\documentclass{article}
\usepackage[plain]{algorithm}
\usepackage[noend]{algorithmic}
\usepackage{amssymb}
\usepackage{amsfonts}
\usepackage{amsmath}
\usepackage{amsthm}
\usepackage{bm}
\usepackage{latexsym}
\usepackage{epsfig}
\usepackage{sgame,color}
\usepackage{graphicx}
\usepackage{natbib}
\usepackage{url}

\newcommand{\veca}{\vec{a}}

\newcommand{\vecv}{\vec{v}}

\renewcommand{\vec}{\mathbf}

\newtheorem{lemma}{Lemma}
\newtheorem{theorem}{Theorem}

\newsavebox\pdbox
\begin{lrbox}{\pdbox}
  \begin{game}{2}{2}[Player~1][Player~2]
    & {\small $C$} & {\small $D$}\\
    {\small $C$} &$2,2$ & $-1,3$\\
    {\small $D$} &$3,-1$ & $0,0$
  \end{game}
\end{lrbox}

\newsavebox\abreubox
\begin{lrbox}{\abreubox}
  \begin{game}{3}{3}
    & {\small $L$} & {\small $M$} & {\small $H$}\\
    {\small $L$} &$10,10$ & $3,15$ & $0,7$\\
    {\small $M$} &$15,3$ & $7,7$ & $-4,5$\\
    {\small $H$} &$7,0$ & $5,-4$ & $-15,-15$
  \end{game}
\end{lrbox}

\newsavebox\rpcbox
\begin{lrbox}{\rpcbox}
  \begin{game}{3}{3}
    & {\small $R$} & {\small $P$} & {\small $S$}\\
    {\small $R$} &$0,0$ & $-1,1$ & $1,-1$\\
    {\small $P$} &$1,-1$ & $0,0$ & $-1,1$\\
    {\small $S$} &$-1,1$ & $1,-1$ & $0,0$
  \end{game}
\end{lrbox}

\newsavebox\bosbox
\begin{lrbox}{\bosbox}
  \begin{game}{2}{2}
    & {\small $O$} & {\small $F$}\\
    {\small $O$} &$1,2$ & $0,0$\\
    {\small $F$} &$0,0$ & $2,1$
  \end{game}
\end{lrbox}

\newsavebox\nopurebox
\begin{lrbox}{\nopurebox}
  \begin{game}{2}{2}
    & {\small $A$} & {\small $B$}\\
    {\small $A$} &$1,0$ & $2,1$\\
    {\small $B$} &$0,3$ & $3,2$
  \end{game}
\end{lrbox}

\begin{document}

\title{An Approximate Subgame-Perfect Equilibrium Computation Technique for Repeated Games}

\author{Andriy Burkov and Brahim Chaib-draa \\
       DAMAS Laboratory, Laval University,\\
       Quebec, Canada G1K 7P4,\\
       \{burkov,chaib\}@damas.ift.ulaval.ca}

\maketitle

\begin{abstract}
This paper presents a technique for approximating, up to any
precision, the set of subgame-perfect equilibria (SPE) in discounted
repeated games. The process starts with a single hypercube
approximation of the set of SPE. Then the initial hypercube is
gradually partitioned on to a set of smaller adjacent hypercubes,
while those hypercubes that cannot contain any point belonging to
the set of SPE are simultaneously withdrawn.

Whether a given hypercube can contain an equilibrium point is
verified by an appropriate mathematical program. Three different
formulations of the algorithm for both approximately computing the
set of SPE payoffs and extracting players' strategies are then
proposed: the first two that do not assume the presence of an
external coordination between players, and the third one that
assumes a certain level of coordination during game play for
convexifying the set of continuation payoffs after any repeated game
history.

A special attention is paid to the question of extracting players'
strategies and their representability in form of finite automata, an
important feature for artificial agent systems.
\end{abstract}

\section{Introduction}
\label{Introduction} In multiagent systems (MAS) the notion of
optimality cannot usually be applied to each agent separately. In a
MAS, each agent's strategy (i.e., a plan specifying its behavior for
every possible situation) can only be considered optimal if it
maximizes that agent's utility function, subject to the constraints
induced by the respective strategies of the other agents -- members
of the same MAS. When each agent's strategy is optimal in this
interdependent sense, the combination of agents' strategies is
called an equilibrium: as long as no agent can individually improve
its utility, all agents prefer to keep their strategies constant.

Given a MAS, a first problem consists of finding a compact yet
sufficiently rich form of representing such strategic interactions.
Game theory provides a powerful framework for this. Repeated
games~\citep{fudenberg-tirole:1991,osborne1999course,mailath2006repeated}
are an important game theoretic formalism permitting modeling and
studying the long-term strategic interactions between multiple
selfish optimizers.

Probably the most known example of a repeated game is Prisoner's
Dilemma whose example is shown in Figure~\ref{fig:PD}.
\begin{figure}[h]
\center
  \begin{game}{2}{2}[Player~1][Player~2]
    & {\small $C$} & {\small $D$}\\
    {\small $C$} &$2,2$ & $-1,3$\\
    {\small $D$} &$3,-1$ & $0,0$
  \end{game}
  \caption{The payoff matrix of Prisoner's Dilemma.}
    \label{fig:PD}
\end{figure}
In this game, there are two players, and each of them can make two
actions: $C$ or $D$. When those players simultaneously perform their
actions, the pair of actions induces a numerical payoff obtained by
each player. The game then passes to the next stage, where it can be
played again by the same pair of players.

Game theory assumes that the goal of each player is to play
optimally, i.e., to maximize its utility function given the
strategies of the other players. When the \emph{a priori}
information about all players' strategies and their real strategic
preferences coincide, we talk about equilibria.

A pair of ``Tit-For-Tat'' (TFT) strategies is a well-known example
of equilibrium in Repeated Prisoner's dilemma. TFT consists of
starting by playing~$C$. Then, each player should play the same
action as the very recent action played by its opponent. Indeed,
such history dependent equilibrium brings to each player a higher
average payoff, than that of another, stationary, equilibrium of the
repeated game (a pair of strategies that prescribe to play $D$ at
every stage). However, an algorithmic construction of such
strategies, given an arbitrary repeated game, is challenging. For
the case where the utility function is given by the average payoff,
\citet{littman2005ptn} propose a simple and efficient algorithm that
constructs equilibrium strategies in two-player repeated games. On
the other hand, when the players discount their future payoffs with
a discount factor, a pair of TFT strategies constitute an
equilibrium only for certain values of the discount factor.
\citet{judd2003computing} propose an approach for computing
equilibria for different discount factors, but their approach is
limited to pure strategies, and, as we will discuss below, has
several other important limitations.

In this paper, we present an algorithmic approach to the problem of
computing equilibria in repeated games when the future payoffs are
discounted. Our approach is more general than that of
\citet{littman2005ptn}, because it allows an arbitrary discounting,
and is free of four major limitations of the algorithm of
\citet{judd2003computing}. Furthermore, our algorithm finds only
those strategies that can be adopted by artificial agents. The
latter are usually characterized by a finite time to compute their
strategies and a finite memory to implement them. To the best of our
knowledge, this is the first time when all these goals are achieved
simultaneously.

The remainder of this paper is structured as follows. In the next
section, we present all necessary formal notions and definitions,
and we formally state the problem. In Section~\ref{sec:previous}, we
survey the previous work, by pointing out its limitations.
Section~\ref{sec:algorithms} is the principal part of this paper. In
this section, we describe our algorithms for approximately solving
repeated games with discounting and for extracting equilibrium
strategies. In Section~\ref{sec:theory}, we investigate the
theoretical properties of the proposed algorithms.
Section~\ref{sec:results} contains an overview of some experimental
results. We conclude in Section~\ref{sec:discussion} with a short
discussion and summary remarks.

\section{Problem Statement}
\label{sec:problem}
\subsection{Stage-Game}
A \emph{stage-game} is a tuple $(N,\{A_i\}_{i\in N}, \{r_i\}_{i\in
N})$. In a stage-game, there is a finite set~$N$, $|N|\equiv n$, of
individual players that act (play, or make their moves in the game)
simultaneously. Player~$i \in N$ has a finite set $A_i$ of
\emph{pure actions} (or, simply, actions) in its disposal. When each
player~$i$ among $N$ chooses a certain action $a_i \in A_i$, the
resulting vector $a\equiv(a_1,\ldots,a_n)$ forms an \emph{action
profile}, which is then played, and the corresponding stage-game
\emph{outcome} is realized. Each action profile belongs to the set
of action profiles $A \equiv \times_{i\in N} A_i$. A player specific
\emph{payoff function} $r_i$ specifies player $i$'s numerical reward
for different game outcomes. In a standard stage-game formulation, a
bijection is typically assumed between the set of action profiles
and the set of game outcomes. In this case, a player's payoff
function can be defined as the mapping $r_i:A \mapsto \mathbb{R}$;
also, this assumption permits, with no ambiguity, to interchangeably
use the notions of action profile and game outcome.

Given an action profile $a$, $r(a)\equiv \times_{i\in N} r_i(a)$ is
called a \emph{payoff profile}. A \emph{mixed action} $\alpha_i$ of
player~$i$ is a probability distribution over its actions, i.e.,
$\alpha_i \in \operatorname{\Delta}(A_i)$. A \emph{mixed action
profile} is a vector $\alpha \equiv (\alpha_i)_{i\in N}$. We denote
by $\alpha^{a_i}_i$ and $\alpha^{a}$ respectively the probability to
play action~$a_i$ by player~$i$ and the probability that the
outcome~$a$ will be realized by~$\alpha$, i.e.,
$\alpha^{a}\equiv\prod_i\alpha^{a_i}_i$. The payoff function can be
extended to mixed action profiles by taking expectations.

The set of players' stage-game payoffs that can be generated by pure
action profiles is denoted as
$$
F \equiv \{ v\in\mathbb{R}^n: \exists a \in A\ s.t.\ v = r(a) \}.
$$
The set $F^{\dag}$ of \emph{feasible payoffs} is the convex hull of
the set $F$, i.e., $F^{\dag} = \operatorname{co}F$.

Let $-i$ stand for ``all players except~$i$''. An \emph{equilibrium}
(or a Nash equilibrium) in a stage-game is a mixed action profile
$\alpha$ with the property that for each player~$i$ and for all
$\alpha'_i\in \Delta(A_i)$, the following inequality holds:
$$
r_i(\alpha) \geq r_i(\alpha'_i,\alpha_{-i}),
$$
\noindent where $\alpha\equiv (\alpha_i,\alpha_{-i})$.

\subsection{Repeated Game}
\label{subsec:repeated-game} In a repeated game, the same stage-game
is played in \emph{periods} $t=0,1,2,\ldots$, also called stages. At
the beginning of each stage, the players choose their actions that
consequently form an action profile. Then they simultaneously play
this action profile, and collect the stage-game payoffs
corresponding to the resulting stage-game outcome. Then the repeated
game passes to the next stage. When the number of game periods is
not known in advance and can be infinite, the repeated game is
called \emph{infinite}. This is the scope of the present paper.

The set of the repeated game \emph{histories up to period}~$t$ is
given by $H^t \equiv \times_t A$. The set of \emph{all possible
histories} is given by $H \equiv \bigcup_{t=0}^{\infty} H^t$. For
instance, a history $h^t \in H^t$ is a stream of outcomes realized
in the repeated game starting from period~$0$ up to period~$t-1$:
$$
h^t \equiv (a^0,a^1,a^2,\ldots,a^{t-1}).
$$
A \emph{pure strategy of player $i$} in the repeated game,
$\sigma_i$, is a mapping from the set of all possible histories to
the set of player $i$'s actions, i.e., $\sigma_i: H \mapsto A_i$. A
\emph{mixed strategy of player $i$} is a mapping $\sigma_i: H
\mapsto \operatorname{\Delta}(A_i)$. $\Sigma_i$ denotes
\emph{player~$i$'s strategy space} and $\Sigma \equiv \times_{i\in
N}\Sigma_i$ denotes the \emph{set of strategy profiles}.

A \emph{subgame} of an original repeated game is a repeated game
based on the same stage-game as the original repeated game but
started from a given history $h^t$. Let a subgame be induced by a
history $h^t$. The behavior of players in that subgame after a
history $h^{\tau}$ is identical to the behavior of players in the
original repeated game after the history $h^t\cdot h^{\tau}$, where
$h^t\cdot h^{\tau} \equiv (h^t,h^{\tau})$ is a concatenation of two
histories. Given a strategy profile $\sigma\in\Sigma$ and a history
$h \in H$, we denote the \emph{subgame strategy profile induced
by~$h$} as~$\sigma|_{h}$.

An \emph{outcome path} in the repeated game is a possibly infinite
stream of action profiles $\veca \equiv (a^0,a^1,\ldots)$. A finite
prefix of length~$t$ of an outcome path corresponds to a history in
$H^{t+1}$. A strategy profile $\sigma$ induces an outcome path
$\veca(\sigma) \equiv (a^0(\sigma),a^1(\sigma),a^2(\sigma),\ldots)$
in the following way:
\begin{equation*}
\begin{array}{l}
  a^0(\sigma) \sim \sigma(\varnothing),\\
  a^1(\sigma) \sim \sigma(a^0(\sigma)),\\
  a^2(\sigma) \sim \sigma(a^0(\sigma),a^1(\sigma)),\\
  \ldots,
\end{array}
\end{equation*}
\noindent where the notation~$a^t(\sigma) \sim \sigma(h^t)$ means
that the outcome $a^t$ is realized at stage~$t$ when the players
were playing according to the (mixed) action profile~$\sigma(h^t)$.
Obviously, in any two independent runs of the same repeated game,
the same pure strategy profile induces two identical outcome paths.
On the contrary, at each period~$t$, the action
profile~$a^t(\sigma)$ belonging to the outcome path induced by a
mixed strategy profile~$\sigma$ is a realization of the random
process~$\sigma(h^t)$.

In order to compare two repeated game strategies in terms of the
utility induced by each strategy, one needs a criterion that permits
comparing infinite payoff streams. Given an infinite sequence of
payoff profiles $\vecv = (v^0,v^1,\ldots)$, the \emph{discounted
average payoff}~$u^{\gamma}_i(\vecv)$ of this sequence for
player~$i$ is given by
\begin{equation}
\label{eq:discounted_avg_payoff} u^{\gamma}_i(\vecv) \equiv
(1-\gamma)\sum_{t=0}^{\infty}\gamma^t v_i^t,
\end{equation}
\noindent where $\gamma\in[0,1)$ is the \emph{discount
factor}\footnote{In the notation $\gamma^t$, $t$ is the power of
$\gamma$ and not a superscript.}. One way to interpret the discount
factor is to view it as a probability that the repeated game will
continue at the next stage (similarly, $(1-\gamma)$ can be viewed as
the probability that the repeated game stops after the current
stage). This interpretation is especially convenient for artificial
agents, because a machine has a non-zero probability of fault at any
moment of time.

Notice that in Equation~(\ref{eq:discounted_avg_payoff}), the sum of
discounted payoffs is normalized by the factor $(1-\gamma)$. This
ensures that $u^{\gamma}_i(\vecv)\in F^{\dag}$ for any instance
of~$\vecv$ or $\gamma$. In other words, after the normalization, the
player's discounted average payoffs can be compared both between
them and with the payoffs of the stage-game. Notice that because a
sequence of payoff profiles,~$\vecv$, always corresponds to an
outcome path,~$\veca$, one can interchangeably and with no ambiguity
write $u^{\gamma}_i(\vecv)$ and $u^{\gamma}_i(\veca)$ referring to
the same quantity.

To compare strategy profiles, a similar criterion can be defined.
Let $\sigma$ be a pure strategy profile and $\gamma$ be a discount
factor. Then the \emph{utility of the strategy profile~$\sigma$ for
player $i$} can be defined as
\begin{equation}
\label{eqn:dap-strategy} u^{\gamma}_i(\sigma) =
(1-\gamma)\sum_{t=0}^{\infty}\gamma^t r_i(a^t(\sigma)).
\end{equation}
As usually, when the players' strategies are mixed, one should take
an expectation over the realized outcome paths.

We define a \emph{utility profile induced by strategy
profile~$\sigma$} as $u^{\gamma}(\sigma) \equiv
(u^{\gamma}_i(\sigma))_{i\in N}$. As previously, due to the
normalization by the factor $(1-\gamma)$, for any $\sigma\in\Sigma$
and for any $\gamma\in [0,1)$, $u^{\gamma}(\sigma)\in F^{\dag}$.
Therefore, when the meaning will be clear from the context, we will
use the terms ``payoff'' and ``payoff profile'' to refer to,
respectively, utility and utility profile.

\subsection{Subgame-Perfect Equilibrium}
In order to act effectively in a given environment, any agent should
have a strategy. When we talk about a \emph{rational agent}, this
strategy has to be optimal in the sense that it should maximize that
agent's expected payoff with respect to the known properties of the
environment. In a single agent case, it can often be assumed that
the properties of the environment do not change in response to the
actions executed by the agent. In this case, it is said that the
environment is \emph{stationary}~\citep{sutton-barto:1998}. In order
to act optimally in a stationary environment, the agent has to solve
the following optimization problem:
$$
\sigma_i = \max_{a_i \in A_i}
\operatorname{E}_{a_j\sim\alpha_j}\left[r_i(a_i,a_j)\right],
$$
\noindent where~$j$ denotes the environment as if it was a player
repeatedly playing a mixed action~$\alpha_j$.

When a rational agent plays a game with other rational agents, it
has to optimize in the presence of the other optimizing players.
This makes the problem non-trivial, since an optimal strategy for
one player depends on the strategies chosen by the other players. In
this context, if the opponents change their strategies, the player's
strategy cannot generally retain optimality.

The concept of equilibrium describes strategies, in which all
players' strategic choices simultaneously optimize with respect to
each other. The strategy profile $\sigma$ is an \emph{equilibrium}
(or a Nash equilibrium) if, for each player $i$ and its strategies
$\sigma'_i\in\Sigma_i$,
$$
u^{\gamma}_i(\sigma) \geq u^{\gamma}_i(\sigma'_i,\sigma_{-i}),
$$
\noindent where $\sigma\equiv(\sigma_i,\sigma_{-i})$. In other
words, in the equilibrium, no player can unilaterally change its
strategy so as to augment its own payoff.

Another notion is important when we consider strategies in repeated
games. This is the notion of \emph{sequential rationality} or, if
applied to the strategy profiles, of \emph{subgame-perfection}. A
strategy profile $\sigma$ is a \emph{subgame-perfect equilibrium}
(SPE) in the repeated game, if for all histories $h\in H$, the
subgame strategy profile $\sigma|_{h}$ is an equilibrium in the
subgame.

Let us first informally explain why, in the repeated games, the
notion of subgame-perfection is of such a high importance. Consider
a grim trigger strategy. This strategy is similar to TFT in that the
two players start by playing~$C$ at the first period. Then grim
trigger prescribes playing~$C$ until any player plays~$D$, in which
case the strategy prescribes playing~$D$ forever. Let the game be as
shown in Figure~\ref{fig:subgame}.
\begin{figure}[h]
\center
  \begin{game}{2}{2}[Player~1][Player~2]
    & {\small $C$} & {\small $D$}\\
    {\small $C$} &$2,2$ & $-1,3$\\
    {\small $D$} &$3,-1$ & $0,-2$
  \end{game}
  \caption{A game in which a profile of two grim trigger strategies is not a subgame-perfect equilibrium.}
    \label{fig:subgame}
\end{figure}
Observe that in this game, the reason why each player would prefer
to play the cooperative action~$C$ while its opponent plays~$C$ is
that the profile of two grim trigger strategies is an equilibrium
when $\gamma$ is close enough to~$1$. Indeed, let Player~$1$
consider a possibility of deviation to the action~$D$ whenever
Player~$2$ is supposed to play~$C$. Player~$1$ is informed that
according to the strategy profile~$\sigma$ (which is a profile of
two grim trigger strategies) starting from the next period,
Player~$2$ will play~$D$ infinitely often. Thus, when $\gamma$ is
sufficiently close to $1$, after only one stage, at which the
profile~$(D,D)$ is played following the deviation, Player~$1$ looses
all the additional gain it obtains owing to the deviation.

Now, let us suppose that Player~$1$ still decides to deviate after a
certain history $h^t$. It plays $D$ whenever Player~$2$ plays~$C$
and collects the payoff of~$3$ instead of~$2$. The repeated game
enters into the subgame induced by the history $h^{t+1} \equiv
(h^t,(D,C))$. Now, according to the strategy profile
$\sigma|_{h^{t+1}}$, Player~$2$ is supposed to play~$D$ forever and
``let the punishment happen''. However, observe the payoffs of
Player~$2$. If Player~$2$ plays~$D$ forever, as prescribed by the
Nash equilibrium, it certainly obtains the average payoff of $-2$ in
the subgame, because the rational opponent (Player~$1$) will
optimize with respect to this strategy. But if Player~$2$ continues
playing~$C$, it obtains the average playoff of $-1$ in the subgame,
while its opponent, the deviator, will continue enjoying the payoff
of~$3$ at each subsequent period. As one can see, even if after the
equilibrium histories the profile of two grim trigger strategies
constitutes an equilibrium in the game shown in
Figure~\ref{fig:subgame}, it is a non-equilibrium in an
\emph{out-of-equilibrium subgame}. Thus, due to this simple example,
it becomes clear why, in order to implement equilibria in practice,
one needs to have recourse to subgame-perfect equilibria: while one
rational player should have no incentive to deviate being informed
about the strategy prescribed to the opponents (the property of Nash
equilibrium), the rational opponents, in turn, need to have
incentives to follow their prescribed strategies \emph{after} that
player's eventual deviation (the property of subgame-perfection).

A subgame-perfect equilibrium always exists. To see this, observe
first that according to~\citet{nash:1950} in any stage-game, there
exists an equilibrium. It is then sufficient to notice that any
strategy profile that prescribes playing, after any history, a
certain Nash equilibrium of the stage-game is a subgame-perfect
equilibrium.

\subsection{Strategy Profile Automata}
By its definition, a player's strategy is a mapping from an infinite
set of histories into the set of player's actions. In order to
construct a strategy for an artificial agent (which is usually
bounded in terms of memory and performance) one needs a way to
specify strategies by means of finite representations.

Intuitively, one can see that, given a strategy profile $\sigma$,
two different histories $h^t$ and $h^{\tau}$ can induce identical
continuation strategy profiles, i.e., $\sigma|_{h^t} =
\sigma|_{h^{\tau}}$. For example, in the case of TFT strategy,
agents will have the same continuation strategy both after the
history $((C,C),(C,C))$ and after the history $((D,C),(C,D),(C,C))$.
One can put all such histories into the same equivalence class. If
one views these equivalence classes of histories as players' states,
then a strategy profile can be viewed as an automaton.

Let $M \equiv (Q,q^0,f,\tau)$ be an \emph{automaton implementation
of a strategy profile}~$\sigma$. It consists of a set of states~$Q$,
with the initial state~$q^0\in Q$; of a profile of decision
functions $f \equiv \times_{i\in N} f_i$, where the decision
function of player~$i$, $f_i: Q \mapsto \operatorname{\Delta}(A_i)$,
associates mixed actions with states; and of a transition function
$\tau: Q \times A \mapsto Q$, which identifies the next state of the
automaton given the current state and the action profile played in
the current state.

Let~$M$ be an automaton. In order to demonstrate how $M$ induces a
strategy profile, one can first recursively define $\tau(q,h^t)$,
the transition function specifying the next state of the automaton
given its initial state $q$ and a history $h^t$ that starts in $q$,
as
$$
\left\{
  \begin{array}{l}
    \tau(q,h^t) \equiv \tau(\tau(q,h^{t-1}),a^{t-1}),\\
    \tau(q,h^1) \equiv \tau (q,a^0).
  \end{array}
\right.
$$
With the above definition in hand, one can define $\sigma_i$, the
strategy of player $i$ induced by the automaton $M$, as
$$
\left\{
  \begin{array}{l}
    \sigma_i(\varnothing) \equiv f_i(q^0),\\
    \sigma_i(h^t) \equiv f_i(\tau(q^0,h^t)).
  \end{array}
\right.
$$

\begin{figure}[tbh]
    \begin{center}
        \leavevmode
        \includegraphics[scale=0.8]{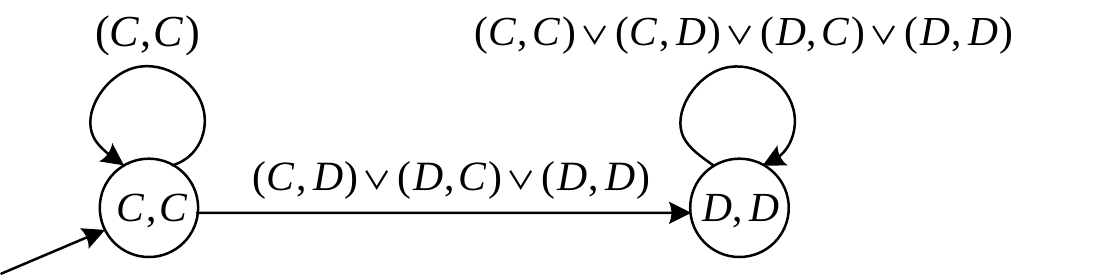}\\
    \end{center}
    \caption{An example of an automaton implementing a profile of two grim trigger strategies. The circles are the states of the automaton; they are labeled with the action profiles prescribed by the profiles of decision functions. The arrows are the transitions between the corresponding states; they are labeled with outcomes.}
    \label{fig:PD_Grim_Trigger_Automaton}
\end{figure}

An example of a strategy profile implemented as an automaton is
shown in Figure~\ref{fig:PD_Grim_Trigger_Automaton}. This automaton
implements the profile of two grim trigger strategies. The circles
are the states of the automaton. The arrows are the transitions
between the corresponding states; they are labeled with outcomes.
The states are labeled with the action profiles prescribed by the
profiles of decision functions.

Since any automaton induces a strategy profile, any two automata can
be compared in terms of the utility they bring to the players. Let
an automaton~$M$ induce a strategy profile~$\sigma$. The
utility~$u^{\gamma}_i(M)$ of the automaton $M$ for player~$i$ is
then equal to $u^{\gamma}_i(\sigma)$, where $u^{\gamma}_i(\sigma)$
is given by Equation (\ref{eqn:dap-strategy}).

Let $|M|$ denote the number of states of automaton $M$. If the value
$|M|$ is finite, such automaton is called a \emph{finite automaton};
otherwise the automaton is called \emph{infinite}. In MAS, most of
the time, we are interested in finite automata, because artificial
agents always have a finite memory to stock their strategies and a
finite processing power to construct them.

Any finite automaton induces a strategy profile, however not any
strategy profile can be represented using finite automata.
\citet{kalai1988finite} demonstrated that any SPE can be
\emph{approximated} with a finite automaton. First of all, they
defined the notion of an approximate SPE. For an \emph{approximation
factor}~$\epsilon > 0$, a strategy profile $\sigma\in\Sigma$ is an
\emph{$\epsilon$-equilibrium} in a repeated game, if for each player
$i$ and for all $\sigma'_i\in\Sigma_i$, $u^{\gamma}_i(\sigma)\geq
u^{\gamma}_i(\sigma'_i,\sigma_{-i})-\epsilon$, where $\sigma\equiv
(\sigma_i,\sigma_{-i})$. A strategy profile $\sigma\in\Sigma$ is a
\emph{subgame-perfect $\epsilon$-equilibrium} ($SP{\epsilon}E$) in
the repeated game, if for all histories $h\in H$, the subgame
strategy profile $\sigma|_{h}$ is an $\epsilon$-equilibrium in the
subgame induced by~$h$. \citet{kalai1988finite} then proved the
following theorem:

\begin{theorem}[\citet{kalai1988finite}]
\label{theorem:approximate-subgame-perfect-discounted} Consider a
repeated game with the discount factor~$\gamma$ and the
approximation factor~$\epsilon$. For any subgame-perfect
equilibrium~$\sigma$, there exists a finite automaton $M$ with the
property that $|u^{\gamma}_i(\sigma) - u^{\gamma}_i(M)| < \epsilon$
for all $i$, and such that~$M$ induces a subgame-perfect
$\epsilon$-equilibrium.
\end{theorem}

\subsection{Problem Statement}
\label{subsec:statement} Let $U^{\gamma}\subset \mathbb{R}^n$ be the
set of all SPE payoff profiles in a repeated game with the discount
factor~$\gamma$. Let $\Sigma^{\gamma,\epsilon}\subseteq\Sigma$ be
the set of all SP$\epsilon$E strategy profiles in a repeated game
with the discount factor~$\gamma$ and the approximation
factor~$\epsilon$.

In this paper, the problem of an approximate subgame-perfect
equilibrium computation is stated as follows: find a set $W
\supseteq U^{\gamma}$ with the property that for any $v \in W$, one
can find a finite automaton~$M$ inducing a strategy profile
$\sigma\in\Sigma^{\gamma,\epsilon}$, such that for all $i$,
$v_i-u^{\gamma}_i(M) \leq \epsilon$.

\section{Previous Work}
\label{sec:previous} The work on equilibrium computation can be
categorized into three main groups. For the algorithms of the first
group, the problem consists in computing one or several stationary
equilibria (or $\epsilon$-equilibria) given a payoff matrix. The
discount factor is implicitly assumed to be equal to
zero~\citep{lemke1964equilibrium,mckelvey1996cef,vonstengel2002cet,chen2006computing,porter2008simple}.
For example, in the repeated Prisoner's Dilemma from
Figure~\ref{fig:PD}, the algorithms of the first group will only
find the stationary equilibrium,
$$
\sigma_i(h)\equiv D,\ \forall i,\ \forall h,
$$
\noindent whose payoff profile is $(0,0)$.

The algorithms belonging to the second group represent the other
extremity. They assume the discount factor to be arbitrarily close
to~$1$. For instance, in two-player repeated games, this permits
obtaining a polynomially fast algorithm for constructing automata
inducing equilibrium strategy profiles~\citep{littman2005ptn}.
Indeed, when $\gamma$ tends to~$1$, the set of SPE payoff profiles
$U^{\gamma}$ converges to the following set:
$$
F^* \equiv \{v\in F^{\dag}: v_i \geq \underline{v}_i,\ \forall i\},
$$
\noindent where the \emph{minmax payoff} $\underline{v}_i$ of
player~$i$ is defined as
$$
\underline{v}_i \equiv \min_{\alpha_{-i}\in\times_{j \neq
i}\operatorname{\Delta}(A_j)}\max_{a_i \in A_i}r_i(a_i,\alpha_{-i}).
$$
The set $F^*$ is called the \emph{set of feasible and individually
rational payoff profiles}. It is the smallest possible set that can
be guaranteed to entirely contain the set of all SPE payoff profiles
in any repeated game. Having in hand the set of SPE payoff profiles,
in order to construct an SPE strategy profile, it is remaining to
choose any point $v\in F^*$ and to construct an automaton having a
structure similar to TFT. More precisely, in this automaton, there
will be one ``in-equilibrium'' (or, ``cooperative'') cycle that
generates~$v$ as an average payoff profile, and two
out-of-equilibrium (or, ``punishment'') cycles, one for each player,
where the deviator obtains at most its minmax payoff during a finite
number of periods (see \citet{littman2005ptn} for more details).

If the discount factor is viewed as the probability that the
repeated game will be continued by the same set of players, it
usually cannot be arbitrarily modified (e.g., moved closer to 1).
The third group of algorithms for computing SPE payoffs and
strategies aims at finding a solution by assuming that the discount
factor~$\gamma$ is a fixed given value between $0$ and
$1$~\citep{cronshaw1994strongly,cronshaw1997algorithms,judd2003computing}.
These algorithms are based on the concept of \emph{self-generating
sets}, introduced by~\citet{abreu1990toward}.

Let us formally develop the idea of self-generation in application
to the problem of computing the set of pure SPE payoff profiles.
Given a strategy profile~$\sigma$, one can rewrite
Equation~(\ref{eqn:dap-strategy}) as follows:
    \begin{eqnarray*}
        u^{\gamma}_i(\sigma) &\equiv&(1-\gamma)\sum_{t=0}^{\infty}\gamma^t r_i(a^t(\sigma))\\
                   &=&(1-\gamma)r_i(a^0(\sigma))+\gamma\left[\sum_{t=1}^{\infty}\gamma^{t-1}
        r_i(a^t(\sigma))\right]\\
                   &=&(1-\gamma)r_i(a^0(\sigma))+\gamma
                   u^{\gamma}_i(\sigma|_{a^0(\sigma)}).
    \end{eqnarray*}
Let $u^{\gamma}_i(a_i,\sigma|_{h^t})$ denote player~$i$'s utility
for playing action~$a_i$ at history~$h^t$ given the strategy
profile~$\sigma$. Let $\bar{a} \equiv (\bar{a}_i,\bar{a}_{-i})$ be
the action profile prescribed by strategy profile $\sigma$ at
history~$h^t$, i.e.,
$\bar{a}\equiv\sigma(h^t)\equiv\sigma|_{h^t}(\varnothing)$. For
all~$a_i\in A_i$ one can write,
\begin{equation}
\label{eqn:continuation-promise} u^{\gamma}_i(a_i,\sigma|_{h^t}) =
(1-\gamma)r_i(a_i,\bar{a}_{-i})+\gamma
                   u^{\gamma}_i(\sigma|_{h^{t+1}}),
\end{equation}
\noindent where $h^{t+1}\equiv h^t\cdot a$ is obtained as a
concatenation of the history $h^t$ and the action profile~$a \equiv
(a_i,\bar{a}_{-i})$; and $u^{\gamma}_i(\sigma|_{h^{t+1}})$
represents the so-called \emph{continuation promise} of the strategy
$\sigma$ after the history $(h^t\cdot a)$.

Therefore, at each period of the repeated game, player~$i$ has a
choice between different actions~$a_i\in A_i$, each having a
particular utility~$v_i$. Consequently, each period of the repeated
game can be represented as a certain stage-game, whose payoffs are
equal to the original stage-game payoffs augmented by the
corresponding continuation promises. Let us call such new stage-game
an \emph{augmented game}. For instance, let the stage-game of the
repeated game be as shown in Figure~\ref{fig:stage-game}.
\begin{figure}[h]
\center
  \begin{game}{2}{2}[Player~1][Player~2]
    & {\small $C$} & {\small $D$}\\
    {\small $C$} &$r(C,C)$ & $r(C,D)$\\
    {\small $D$} &$r(D,C)$ & $r(D,D)$
  \end{game}
  \caption{A generic stage-game.}
    \label{fig:stage-game}
\end{figure}
Given a strategy profile~$\sigma$ and a history $h^t$, the augmented
game corresponding to this stage-game is shown in
Figure~\ref{fig:augmented-game}.
\begin{figure}[h]
\center
  \begin{game}{2}{2}[Player~1][Player~2]
    & {\small $C$} & {\small $D$}\\
    {\small $C$} &$(1-\gamma)r(C,C)+\gamma u^{\gamma}(\sigma|_{h^t \cdot (C,C)})$ & $(1-\gamma)r(C,D)+\gamma u^{\gamma}(\sigma|_{h^t \cdot (C,D)})$\\
    {\small $D$} &$(1-\gamma)r(D,C)+\gamma u^{\gamma}(\sigma|_{h^t \cdot (D,C)})$ & $(1-\gamma)r(D,D)+\gamma u^{\gamma}(\sigma|_{h^t \cdot (D,D)})$
  \end{game}
  \caption{An augmented game for the generic stage-game from Figure~\ref{fig:stage-game}.}
    \label{fig:augmented-game}
\end{figure}

By reformulating the definition of subgame-perfect equilibrium, the
strategy profile~$\sigma$ is an SPE, if and only if it induces an
equilibrium mixed action profile in the augmented game after any
history.

Let $V^{\gamma}$ denote the set of pure action subgame-perfect
equilibrium payoff profiles one wants to identify. Recall
Equation~(\ref{eqn:continuation-promise}): at a history~$h^t$, in
order to make part of a subgame-perfect equilibrium strategy,
action~$a_i$ has to be ``supported'' by a certain continuation
promise~$u^{\gamma}_i(\sigma|_{h^{t+1}})$, where $h^{t+1}\equiv
h^t\cdot(a^i,\bar{a}_{-i})$ and
$(\bar{a}_i,\bar{a}_{-i})\equiv\sigma(h^t)$. By the property of
subgame-perfection, this must hold after any history. Therefore,
if~$a_i$ does make part of a certain subgame-perfect
equilibrium~$\sigma$ at the history~$h^t$, then
$u^{\gamma}_i(\sigma|_{h^{t+1}})$ has to belong to~$V^{\gamma}$, as
well as~$u^{\gamma}_i(a_i,\sigma|_{h^t})$. This self-referential
property of subgame-perfect equilibrium suggests a way by which one
can find~$V^{\gamma}$.

Let $BR_i(\alpha)$ denote a stationary best response of player~$i$
to the mixed action profile~$\alpha\equiv(\alpha_i,\alpha_{-i})$,
i.e.,
$$
BR_i(\alpha)\equiv\max_{a_i\in A_i}r_i(a_i,\alpha_{-i}).
$$
The analysis focuses on the map~$B^{\gamma}$ defined on a
set~$W\subset \mathbb{R}^n$:
$$
B^{\gamma}(W) \equiv \bigcup_{(a,w)\in A\times W}(1-\gamma)r(a) +
\gamma w,
$$
\noindent where $w$ has to verify for all $i$:
$$
(1-\gamma)r_i(a) + \gamma w_i - (1-\gamma)r_i(BR_i(a),a_{-i}) -
\gamma \underline{w}_i\geq 0,
$$
\noindent and $\underline{w}_i \equiv \inf_{w\in W}w_i$.
\citet{abreu1990toward} show that the largest fixed point of
$B^{\gamma}(W)$ is $V^{\gamma}$.

Any numerical implementation of $B^{\gamma}(W)$ requires an
efficient representation of the set $W$ in a machine.
\citet{judd2003computing} use convex sets in order to approximate
both $W$ and $B^{\gamma}(W)$ as an intersection of a finite number
of hyperplanes. Each application of $B^{\gamma}(W)$ is then reduced
to solving a simple linear program. The algorithm starts with a set
$W\in\mathbb{R}^n$ that is guaranteed to entirely
contain~$V^{\gamma}$. Then it iteratively modifies $W$ as
$W\leftarrow B^{\gamma}(W)$ until convergence. We omit further
details: the interested reader can refer
to~\citet{judd2003computing}.

The approach of~\citet{judd2003computing} has, however, several
important limitations:
\begin{enumerate}
  \item It assumes the existence of at least one pure action equilibrium in the
  stage-game;
  \item It permits computing only pure action SPE strategy profiles;
  \item It cannot find SPE strategy profiles implementable by finite automata with given
  precision;
  \item It can only be naturally applicable if the set of SPE payoff profiles is convex. In practice, this is often not the case. This means that in order to be capable of adopting strategies computed by the algorithm, the players need to have a way to convexify the set of continuation promises by randomizing between them. This can be done, for example, by means of a special communication protocol (e.g., \emph{jointly controlled lotteries} by~\citet{aumann1995repeated}) or by using a \emph{public correlating device}~\citep{mailath2006repeated}.
\end{enumerate}
In the next section, we present three different formulations of our
algorithm for solving the problem of an approximate SPE computation,
as it was stated in Section~\ref{sec:problem}. The first formulation
is only free of the last two limitations of the approach
of~\citet{judd2003computing}. The second formulation is free of all
four limitations, but it is not guaranteed to find a set containing
\emph{all} mixed strategy SPE payoff profiles. The third version of
the algorithm, in turn, finds a set containing all (pure and mixed)
SPE payoff profiles. However, it accomplishes this for the sake of
convexifying the set of continuation promises, i.e., has the fourth
limitation.

\section{The Algorithms}
\label{sec:algorithms} The fixed point property of the
map~$B^{\gamma}$ and its relation to the set of SPE payoff profiles
can be used to approximate the latter. Indeed, according
to~\citet{abreu1990toward}, if (\emph{i})~for a certain set $W$ we
have $B^{\gamma}(W)=W$ and (\emph{ii})~$W$ is the largest such set,
then $W = U^{\gamma}$. The idea is to start by a certain set~$W$
that is guaranteed to entirely contain~$U^{\gamma}$, and then to
iteratively eliminate all those points $w' \in W$ for which
$\nexists (w,\alpha) \in
W\times\Delta(A_1)\times\ldots\times\Delta(A_n)$, such that,
\begin{equation}
\label{eqn:verifyCube}
\begin{array}{rl}
  (1) & w' = (1-\gamma)r(\alpha) + \gamma w\hbox{, and}, \\
  (2) & (1-\gamma)r_i(\alpha) + \gamma w_i - (1-\gamma)r_i(BR_i(\alpha),\alpha_{-i}) - \gamma \underline{w}_i\geq 0,\ \forall i.
\end{array}
\end{equation}

Algorithm~\ref{alg:basic} outlines the basic structure for three
different formulations that will be defined in the following
subsections. The algorithm starts with an initial approximation~$W$
of the set of SPE payoff profiles~$U^{\gamma}$. The set~$W$ is
represented by a union of disjoint hypercubes belonging to the
set~$C$. Each hypercube $c \in C$ is identified by its origin
$o^c\in \mathbb{R}^n$ and by the side length~$l$, the same for all
hypercubes. Initially, $C$ contains only one hypercube~$c$, whose
origin $o^c$ is set to be a vector $(\underline{r})_{i\in N}$; the
side length~$l$ is set to be $l=\bar{r}-\underline{r}$, where
$\underline{r}\equiv\min_{a,i}r_i(a)$ and
$\bar{r}\equiv\max_{a,i}r_i(a)$. I.e., $W$ entirely contains
$U^{\gamma}$.
\begin{algorithm}[htb]
\caption{The basic structure for all proposed algorithms.}
\begin{algorithmic}[1]
\REQUIRE $r$, a payoff matrix; $\gamma$ a discount factor;
$\epsilon$, an approximation factor.%
\STATE Let $l\equiv\bar{r}-\underline{r}$ and
$o^c\equiv(\underline{r})_{i\in N}$;%
\STATE Set $C\leftarrow\{(o^c,l)\}$;%
\LOOP%
    \STATE Set $\textsc{AllCubesCompleted} \leftarrow \textsc{True}$;%
    \STATE Set $\textsc{NoCubeWithdrawn} \leftarrow \textsc{True}$;%
    \FOR{\textbf{each} $c\equiv(o^c,l) \in C$}%
        \STATE Let $\underline{w}_i \equiv \min_{c\in C} o^c_i$;%
        \STATE Set $\underline{w} \leftarrow (\underline{w}_i)_{i\in N}$;%
        \IF{\textsc{CubeSupported}($c$, $C$, $\underline{w}$) is \textsc{False}}%
            \STATE Set $C \leftarrow C\backslash\{c\}$;%
            \IF{$C = \varnothing$}%
                \RETURN{$\textsc{False}$;}%
            \ENDIF%
            \STATE Set $\textsc{NoCubeWithdrawn} \leftarrow \textsc{False}$;%
        \ELSE%
            \IF{$\textsc{CubeCompleted($c$)}$ is $\textsc{False}$}%
                \STATE Set $\textsc{AllCubesCompleted} \leftarrow \textsc{False}$;%
            \ENDIF%
        \ENDIF%
    \ENDFOR%
    \IF{$\textsc{NoCubeWithdrawn}$ is $\textsc{True}$}%
        \IF{$\textsc{AllCubesCompleted}$ is $\textsc{False}$}%
            \STATE Set {$C\leftarrow\textsc{SplitCubes}(C)$;}%
        \ELSE%
            \RETURN{$C$.}%
        \ENDIF%
    \ENDIF%
\ENDLOOP%
\end{algorithmic}
\label{alg:basic}
\end{algorithm}

Each iteration of Algorithm~\ref{alg:basic} consists of verifying,
for each hypercube $c\in C$, whether it has to be eliminated from
the set $C$ (procedure \textsc{CubeSupported}). If~$c$ does not
contain any point $w'$ satisfying the conditions of
Equation~(\ref{eqn:verifyCube}), this hypercube is withdrawn from
the set~$C$. If, by the end of a certain iteration, no hypercube was
withdrawn, each remaining hypercube is split into $2^n$ disjoint
hypercubes with side $l/2$ (procedure $\textsc{SplitCubes}$). The
process continues until, for each remaining hypercube, a certain
stopping criterion is satisfied (procedure \textsc{CubeCompleted}).

\subsection{Pure Strategy Equilibria}
For the case where the goal is to only approximate pure action
equilibria, the definition of the procedure \textsc{CubeSupported}
is given in Algorithm~\ref{alg:CubeSupportedPure}.
\begin{algorithm}[htb]
\caption{\textsc{CubeSupported} for pure strategies. The procedure
verifies whether a given hypercube~$c$ has to be kept in the set of
hypercubes~$C$. If yes,~\textsc{CubeSupported} returns a pure action
profile and a continuation promise. Otherwise the procedure returns
\textsc{False}.}
\begin{algorithmic}[1]
\REQUIRE $c\equiv(o^c,l)$, a hypercube; $C$, a set of hypercubes;
$\underline{w}$ a vector of payoffs. \STATE $S\leftarrow
\textsc{GetClusters}($C$)$ \FOR{\textbf{each} $s\equiv(o^s,l^s) \in
S$}
    \FOR{\textbf{each} $a \in A$}
    \STATE Solve the following linear constraint satisfaction problem:
\begin{itemize}
  \item[] Decision variables: $w\in \mathbb{R}^n$ and $w'\in
  \mathbb{R}^n$;
  \item[] Subject to constraints:
$$
\begin{array}{rl}
  (1) & w' = (1-\gamma)r(a) + \gamma w;\\
  \multicolumn{2}{l}{\hbox{For all }i \hbox{:}}\\
  (2) & (1-\gamma)r_i(a) + \gamma w_i - (1-\gamma)r_i(BR_i(a),a_{-i}) - \gamma \underline{w}_i\geq 0,\\
  (3) & o^s_i \leq w_i \leq o^s + l^s_i,\\
  (4) & o^c_i \leq w'_i \leq o^c_i + l;
\end{array}
$$
\end{itemize}
    \IF{a pair $(w,w')$ satisfying the constraints is found}
        \RETURN{$(a,w)$};
    \ENDIF
    \ENDFOR
\ENDFOR \RETURN{\textsc{False}.}
\end{algorithmic}
\label{alg:CubeSupportedPure}
\end{algorithm}
In this algorithm, the set of hyperrectangular clusters,~$S$, is
obtained from the set of hypercubes,~$C$, by finding a smaller set,
such that the union of its elements is equal to the union of the
elements of~$C$ (procedure \textsc{GetClusters}). Each~$s\in S$ is
identified by its origin $o^s\in \mathbb{R}^n$ and by the vector of
side lengths~$l^s\in \mathbb{R}^n$. The clusterization of the
set~$C$ is done to speed up the algorithm in practice. In our
experiments, we used a simple greedy algorithm to identify
hyperrectangles. Of course, one can always define $S \equiv C$. In
this case, for each $c\equiv(o^c,l)\in C$, there will be exactly one
$s\in S$, such that $s\equiv(o^c,(l)_{i\in N})$.

The linear constraint satisfaction program of
Algorithm~\ref{alg:CubeSupportedPure} can be solved by any linear
program solver. We used CPLEX~\citep{cplex200811} together with
OptimJ~\citep{optimj2009} for solving all mathematical programs
defined in this paper.

If the conditions of Equation~(\ref{eqn:verifyCube}) are verified,
and~$c$ has to be kept in~$C$, the \textsc{CubeSupported} procedure
of Algorithm~\ref{alg:CubeSupportedPure} returns a pure action
profile~$a$ and a payoff profile~$w$ such that
$(1-\gamma)r(a)+\gamma w\equiv w'$ belongs to the hypercube~$c$.
Otherwise, the procedure returns~\textsc{False}.

\subsection{Mixed Strategy Equilibria}
Computing the set of all equilibria (i.e., pure action and mixed
action, stationary and non stationary) is a more challenging task.
To the best of our knowledge, there is no algorithm capable of at
least approximately solving this problem. The previous pure strategy
case was greatly simplified by two circumstances:
\begin{enumerate}
  \item It is possible to enumerate pure action profiles one by one in order to test all possibilities to satisfy the two conditions of Equation~(\ref{eqn:verifyCube}).
  \item Any deviation of player~$i$ from the recommended (by the equilibrium strategy profile) action profile~$a\equiv(a_i,a_{-i})$ in a case, where $a_i\notin \{a'_i:r(a'_i,a_{-i})=r_i(BR_i(a),a_{-i})\}$ is immediately detected by the other players. This makes possible to enforce the condition~(2) of Equation~(\ref{eqn:verifyCube}) in practice.
\end{enumerate}

When the action profiles, which strategy profiles can recommend, are
allowed to be mixed, their one by one enumeration is impossible.
Furthermore, deviations from mixed actions can only be detected if
the deviation is done in favor of an out-of-the-support action. In
game theory, the \emph{support of a mixed action} $\alpha_i$ is a
set $A^{\alpha_i}_i\subseteq A_i$, which contains all pure actions
to which $\alpha_i$ assigns a non-zero probability. Therefore, if
player~$i$ plays an action $a_i\notin A^{\alpha_i}_i$, whenever it
is supposed to play a mixed action~$\alpha_i$, only in this case the
other players can immediately detect the deviation. Deviations that
only involve the actions in the support of~$\alpha_i$ cannot be
detected.

We solve the two aforementioned problems in the following way. We
first define a special mixed integer program (MIP). We then let the
solver decide on which actions to be included into the mixed action
support of each player, and what probability has to be assigned to
those actions. The MIP has to be solved for all agents
simultaneously. Because the goal is to only satisfy the constraints
of Equation~(\ref{eqn:verifyCube}), the presence of a particular
objective function in the MIP is not generally necessary. In our
implementation, we have chosen an objective to minimize the sum of
the cardinalities of the supports. This means that, when possible,
the preference is given to pure action strategies.

Player~$i$ is only willing to randomize according to a
mixture~$\alpha_i$, if it is indifferent over the pure actions in
the support of the mixture. The technique is to specify different
continuation promises for different actions in the support of the
mixture, such that the utility of each action remains bounded by the
dimensions of the hypercube.
Algorithm~\ref{alg:CubeSupportedMixedClusters} defines the procedure
\textsc{CubeSupported} for the case, where the set of continuation
promises is represented by the union of hyperrectangular clusters.

For each hyperrectangular cluster,~$s$, containing possible
continuations, Algorithm~\ref{alg:CubeSupportedMixedClusters}
verifies whether, for the given hypercube~$c$, one can find a mixed
action profile $\alpha\equiv(\alpha_i,\alpha_{-i})$, such that for
all~$i$ and for all $a_i\in A^{\alpha_i}_i$, $(1-\gamma)r_i(\alpha)
+ \gamma w_i(a_i)\equiv w'_i(a_i)$ lies between $o^s_i$ and $o^s_i +
l^s_i$. This will satisfy the first condition of
Equation~(\ref{eqn:verifyCube}). To satisfy the second condition,
the choice of the mixed action and of the continuation payoffs has
to be such that, for all $i\in N$ and for all $a_i\notin
A^{\alpha_i}_i$, $\underline{w}_i \leq w_i(a_i) \leq \underline{w}_i
+ l$, to make any out-of-the-support deviation \emph{approximately
unprofitable}.

Observe that in the MIP of
Algorithm~\ref{alg:CubeSupportedMixedClusters}, we assign the
continuation payoffs $w_i(a_i)$ to pure actions and not to pure
action profiles (as, for example, can follow from the definition of
an augmented game). This permits avoiding the non-linear term
$\alpha_{-i}^{a_{-i}}{w_i(a_i,a_{-i})}$ in the constraint~(3) of the
MIP. One can do this, without missing any continuation promise
belonging to the cluster, thanks to the rectangular structure of the
latter: for any action profile $a\equiv(a_1,\ldots,a_n)$ realized
during the repeated game play, the corresponding continuation payoff
profile $(w_1(a_1),\ldots,w_n(a_n))$ will always be found inside, or
on the boundary of, a certain hyperrectangle. This assures that the
continuation payoff profiles belong to~$W$.

In Algorithm~\ref{alg:CubeSupportedMixedClusters}, the required
indifference of player~$i$ between the actions in the support of the
mixed action $\alpha_i$ is (approximately) secured by the
constraint~(4) of the MIP. Observe that in an optimal solution of
the MIP, the binary variables~$y^{a_i}_i$, known as \emph{indicator
variables}, can only be equal to~$1$ if $a_i$ is in the support of
$\alpha_i$. Therefore, according to the constraint~$(4)$, each
$w'_i(a_i)$ is either bounded by the dimensions of the hypercube, if
$a_i\in A^{\alpha_i}_i$, or is below the origin of the hypercube,
otherwise.

Notice that the MIP of
Algorithm~\ref{alg:CubeSupportedMixedClusters} is only linear in the
case of two players. For more than two players, the problem becomes
non-linear due to the fact that~$\alpha_{-i}$ is now given by a
product of decision variables~$\alpha_{j}$, for all $j\in
N\backslash\{i\}$. For three players, for example, such optimization
problem becomes a mixed integer quadratically constrained program
(MIQCP); such optimization problems are generally very difficult as
they combine two kinds of non-convexities: integer variables and
non-convex quadratic constraints~\citep{saxena2008disjunctive}.

The fact that all continuation payoff profiles are contained within
one cluster makes the optimization problem easier to solve; however,
such an approach also restricts the set of equilibria by allowing
only those SPE, for which the continuation payoffs are always
contained within a certain cluster. Nevertheless, the solutions that
can be computed by Algorithm~\ref{alg:CubeSupportedMixedClusters}
include, among others, all pure strategy SPE (because, in this case,
the continuation payoff profile is a unique point belonging to a
certain cluster) as well as all stationary mixed strategy SPE
(because for any~$i$ and any stationary SPE payoff~$w'_i(a_i)$,
belonging to a certain hypercube, the continuation payoff~$w_i(a_i)$
belongs to the same hypercube). A more general formulation of the
MIP could, for example, allow the continuations for different action
profiles to belong to different clusters. The task of selecting a
particular cluster for~$w_i(a)$, for all $i$, can also be left to
the solver. This would, however, again result in a non-linear MIP,
because, now, the constraint~(3) would look as follows,
$$
w'_i(a_i) =
\sum_{a_{-i}}{{\alpha_{-i}^{a_{-i}}}\left((1-\gamma)r_i(a_i,a_{-i})
+ \gamma w_i(a_i,a_{-i})\right)},
$$
\noindent where $w_i(a_i,a_{-i})$, the continuation payoff assigned
to an action profile~$a\equiv(a_i,a_{-i})$, is bounded by the
dimensions of a certain cluster.

There is a way to modify
Algorithm~\ref{alg:CubeSupportedMixedClusters} so as to keep in~$W$
all SPE payoff profiles while preserving the linearity of the MIP,
at least for two-player repeated games. This can be achieved by
assuming a certain level of coordination between players during the
game play. This is the subject of the next subsection.

\begin{algorithm}[htb]
\caption{\textsc{CubeSupported} for mixed strategies. The procedure
verifies whether a given hypercube~$c$ has to be kept in the set of
hypercubes~$C$. If yes, \textsc{CubeSupported} returns a mixed
action profile~$\alpha$ and the corresponding continuation promise
payoffs for each pure action in the support of~$\alpha_i$. Otherwise
the procedure returns \textsc{False}.}
\begin{algorithmic}[1]
\REQUIRE $c\equiv(o^c,l)$, a hypercube; $C$, a set of hypercubes;
$\underline{w}$ a vector of payoffs.%
\STATE $S\leftarrow \textsc{GetClusters}($C$)$%
\FOR{\textbf{each} $s\equiv(o^s,l^s) \in S$}
    \STATE Solve the following mixed integer program:
    \renewcommand{\labelenumi}{({enumi})}
    \begin{itemize}
      \item[] Decision variables: $w_i(a_i)\in \mathbb{R}$, $w'_i(a_i)\in \mathbb{R}$, $y^{a_i}_i\in\{0,1\}$, $\alpha^{a_i}_i\in [0,1]$ for all $i\in \{1,2\}$ and for all $a_i\in
      A_i$;
      \item[] Objective function: $\min f \equiv
      \sum_i\sum_{a_i}y^{a_i}_i$;
      \item[] Subject to constraints:
$$
\begin{array}{rl}
\multicolumn{2}{l}{\hbox{For all }i \hbox{:}}\\
(1) & \sum_{a_i}{\alpha^{a_i}_i} = 1;\\
\multicolumn{2}{l}{\hbox{For all }i \hbox{ and for all }a_i\in A_i\hbox{:}}\\
(2) & \alpha^{a_i}_i \leq y^{a_i}_i,\\
(3) & w'_i(a_i) = (1-\gamma)\sum_{a_{-i}}{{\alpha_{-i}^{a_{-i}}}r_i(a_i,a_{-i})} + \gamma w_i(a_i),\\
(4) & {o^c_i}{y^{a_i}_i} \leq w'_i(a_i) \leq l{y^{a_i}_i} + o^c_i,\\
(5) & \underline{w}_i - {\underline{w}_i}y^{a_i}_i +
{o^s_i}y^{a_i}_i \leq w_i(a_i) \leq (\underline{w}_i + l) -
(\underline{w}_i + l)y^{a_i}_i + (o^s_i + l^s_i)y^{a_i}_i;
\end{array}
$$
    \end{itemize}
    \IF{a solution is found}
        \RETURN{$w_i(a_i)$ and $\alpha^{a_i}_i$ for all $i\in \{1,2\}$ and for all $a_i\in A_i$;}
    \ENDIF
\ENDFOR \RETURN{\textsc{False}.}
\end{algorithmic}
\label{alg:CubeSupportedMixedClusters}
\end{algorithm}

\subsection{Public Correlation}
\label{subsec:publicCorrelation} A mixed SPE strategy
profile~$\sigma$, after each history~$h$, suggests to the players a
certain mixed action profile~$\alpha$ and has a certain value
$w(\alpha)$ associated with it. More precisely, $w(\alpha)$ is an
expected continuation promise for playing mixed action~$\alpha$ at
history~$h$, such that $u^{\gamma}(\sigma|_{h})\equiv w' =
(1-\gamma)r(\alpha)+\gamma w(\alpha)$. Also, $\sigma|_{h}$ induces a
certain continuation payoff profile~$w(a)$ for each outcome~$a$
realized at~$h$. Because~$\sigma$ is an SPE, every such~$w(a)$
belongs to~$U^{\gamma}$, the set of SPE payoff profiles. However,
$w(\alpha)$ does not necessarily belongs to~$U^{\gamma}$,
because~$w(\alpha)$ is a mixture $\sum_{a\in A}\alpha^a w(a)$. On
the other hand, for any~$\alpha$, $w(\alpha)$ does belong
to~$\operatorname{co}U^{\gamma}$, the convex hull of the set of SPE
payoff profiles.

Let us assume that one can select, as a continuation payoff profile,
any payoff profile from $\operatorname{co}W$. In the MIP of the
\textsc{CubeSupported} procedure, one can, therefore, associate
continuation payoffs~$w_i$ with player~$i$'s actions, and not with
action profiles. This would permit avoiding the previously seen
non-linearity when we allowed the continuations to belong to
different hypercubes. To achieve this, one can rewrite~$w(\alpha)$
as $(w_i(\alpha))_{i\in N}$, where
$w_i(\alpha)\equiv\sum_{a_i}\alpha^{a_i}_i w_i(a_i|\alpha)$ and
$w_i(a_i|\alpha)\equiv\sum_{a'\in A:a'_i=a_i}w_i(a')\prod_{j\in
N\backslash \{i\}}\alpha^{a'_j}_j$. Let $w'\equiv(w'_i)_{i\in N}$ be
an SPE payoff profile and let one want to identify $\alpha$ and
$w(\alpha)$ in support of $w'$. If $a_i\in A^{\alpha_i}_i$, then,
for all~$i$,
$$
w'_i = (1-\gamma)r_i(a_i|\alpha) + \gamma w_i(a_i|\alpha),
$$
\noindent where $r_i(a_i|\alpha)\equiv\sum_{a'\in
A:a'_i=a_i}r_i(a')\prod_{j\in N\backslash \{i\}}\alpha^{a'_j}_j$.
For two players, the right-hand expression for $w'_i$ is linear.
Furthermore, for any choice of~$\alpha$, the payoff profile obtained
as $(w_i(a_i|\alpha))_{i\in N}$ is a point in $\mathbb{R}^n$ that
belongs to $\operatorname{co}W$. One can now modify the optimization
problem of Algorithm~\ref{alg:CubeSupportedMixedClusters} so as to
keep inside the convex hull of~$W$ any point $(w_i(a_i))_{i\in N}$,
such that $a_i\in A^{\alpha_i}_i,\ \forall i$. In doing so, we are
guaranteed to keep in~$W$ all possible SPE payoff profiles.

A convexification of the set of continuation payoff profiles can be
done in different ways, one of which is public correlation. A
\emph{repeated game with public correlation} is a repeated game,
such that in every stage-game, a realization $\omega\in(0,1]$ of a
public random variable is first drawn, which is observed by all
players, and then each player chooses an action. The public
signal~$\omega$ can be generated by a certain \emph{public
correlating device}~\citep{mailath2006repeated}. This device has to
be capable of generating instances of a given random variable and to
be unbiased, i.e., indifferent with regard to the repeated game
outcomes. A public correlating device can be simulated by a special
communication protocol, such as a \emph{jointly controlled
lottery}~\citep{aumann1995repeated}.

Let $\sigma$ be an SPE strategy profile that suggests playing a
mixed action profile~$\alpha$ at $h^t$ and promises a continuation
payoff profile~$w(a)$ for each $a\in A$. If, for all $a$, $w(a)\in
U^{\gamma}$, no public correlation is necessary: for any possible
$h^{t+1}\equiv h^t\cdot a$, there exists
$\sigma|_{h^{t+1}}\in\Sigma^{\gamma}$, such that
$u^{\gamma}(\sigma|_{h^{t+1}})=w(a)$. Now, let us suppose that after
playing a mixed action~$\alpha$ at $h^t$, an outcome~$a$ has been
realized, such that $w(a)\in \operatorname{co}{U^{\gamma}}\backslash
U^{\gamma}$. In this case, one cannot find any strategy
$\sigma|_{h^{t+1}}\in\Sigma^{\gamma}$, such that
$u^{\gamma}(\sigma|_{h^{t+1}})=w(a)$. On the other hand, by using a
correlating device during the game play, the players can obtain, in
expectation, the continuation payoff profile~$w(a)$ as a convex
combination of~$K$ points $w_k\in U^{\gamma}$. I.e., there exist~$K$
non-negative real numbers~$\rho_k$ with the property that
$\sum_{k=1}^{K}\rho_k w_k = w(a)$ and $\sum_{k=1}^{K}\rho_k = 1$.

Let a public correlating device be available and capable of
generating a uniformly distributed signal $\omega\in(0,1]$ when
needed. Define $\rho_0 = 0$. If
$\omega\in\left(\sum_{j=0}^{k-1}\rho_j,\sum_{k=0}^{k}\rho_j\right]$
for some $k=1,\ldots,K$, then $\sigma|_{h^{t+1}}$ is set to be a
$\sigma\in\Sigma^{\gamma}$ such that $u^{\gamma}(\sigma)=w_k$. By so
doing, any SPE payoff profile~$v$, computed assuming that the set of
continuation payoff profiles is convex, can in practice be induced
by a certain SPE strategy profile~$\sigma$. To achieve this, the
transition function of the automaton implementation of~$\sigma$ has
to be modified into a mapping $f: Q\times A\times (0,1]\mapsto Q$,
such that $f(q,a,w)$ specifies the next state of the automaton,
given that the outcome~$a\in A$ was first realized in the current
state $q\in Q$ and then~$\omega\in (0,1]$ was drawn.

\subsubsection{The Algorithm}
Algorithm~\ref{alg:CubeSupportedMixedConvex} contains the definition
of the \textsc{CubeSupported} procedure that convexifies the set of
continuation promises. The definition is given for two players,
i.e., $N\equiv\{1,2\}$. The procedure first
identifies~$\operatorname{co}W$, the smallest convex set containing
all hypercubes of the set~$C$ (procedure \textsc{GetHalfplanes}).
This convex set is represented as a set~$P$ of half-planes. Each
element~$p\in P\subset\mathbb{R}^3$ is a vector
$p\equiv(\phi^p,\psi^p,\lambda^p)$, such that the inequality $\phi^p
x + \psi^p y \leq \lambda^p$ identifies a half-plane in a
two-dimensional space. The intersection of these half-planes gives
$\operatorname{co}W$. In our experiments, in order to construct the
set~$P$ from the set~$C$, we used the Graham scan, an efficient
technique to identify the boundary points of the convex hull of a
set~\citep{graham1972efficient}.

The procedure $\textsc{CubeSupported}$ defined in
Algorithm~\ref{alg:CubeSupportedMixedConvex} differs from that of
Algorithm~\ref{alg:CubeSupportedMixedClusters} in the following
aspects. It does not compute clusters and, consequently, does not
iterate. Instead, it convexifies the set~$W$ and searches for
continuation promises for the hypercube~$c$ inside
$\operatorname{co}W$. The definition of the MIP is also different.
New indicator variables, $z^{a_1,a_2}$, for all pairs $(a_1,a_2)\in
A_1\times A_2$, are introduced. The new constraint~(6), jointly with
the modified objective function, verify that $z^{a_1,a_2}$ is only
equal to $1$ whenever both $y^{a_1}_1$ and $y^{a_2}_2$ are equal to
$1$. In other words, $z^{a_1,a_2}=1$, only if $a_1\in
A^{\alpha_1}_1$ and $a_2\in A^{\alpha_2}_2$. Another new
constraint~(7) verifies that $(w_1(a_1),w_2(a_2))$, the continuation
promise payoff profile, belongs to~$\operatorname{co}W$ if and only
if $(a_1,a_2)\in A^{\alpha_1}_1\times A^{\alpha_2}_2$. Notice that
in the constraint~(7), $M$ stands for a sufficiently large number.
In constrained optimization, this is a standard technique for
relaxing a given constraint by using binary indicator variables.

\begin{algorithm}[htb]
\caption{\textsc{CubeSupported} for mixed actions and public
correlation. The procedure verifies whether a given hypercube~$c$
has to be kept in the set of hypercubes~$C$. If~$c$ has to be kept
in~$C$, \textsc{CubeSupported} returns a mixed action profile and
the corresponding continuation promise payoffs for each pure action
in the support of mixed actions. Otherwise the procedure returns
\textsc{False}.}
\begin{algorithmic}[1]
\REQUIRE $c\equiv(o^c,l)$, a hypercube; $C$, a set of hypercubes.
    \STATE $P\leftarrow \textsc{GetHalfplanes}($C$)$
    \STATE Solve the following mixed integer linear optimization problem:
    \begin{itemize}
      \item[] Decision variables: $w_i(a_i)\in \mathbb{R}$, $w'_i(a_i)\in \mathbb{R}$, $y^{a_i}_i\in\{0,1\}$, $\alpha^{a_i}_i\in [0,1]$ for all $i\in \{1,2\}$ and for all $a_i\in A_i$; $z^{a_1,a_2}\in\{0,1\}$ for all pairs $(a_1,a_2)\in A_1\times A_2$.
      \item[] Objective function: $\min f \equiv \sum_{(a_1,a_2)\in A_1\times A_2}z^{a_1,a_2}$.
      \item[] Subject to constraints:
$$
\begin{array}{rl}
\multicolumn{2}{l}{\hbox{For all }i\in\{1,2\} \hbox{:}}\\
(1) & \sum_{a_i}{\alpha^{a_i}_i} = 1;\\
\multicolumn{2}{l}{\hbox{For all }i\in\{1,2\} \hbox{ and for all }a_i\in A_i\hbox{:}}\\
(2) & \alpha^{a_i}_i \leq y^{a_i}_i,\\
(3) & w'_i(a_i) = (1-\gamma)\sum_{a_{-i}}{{\alpha_{-i}(a_{-i})}r_i(a_i,a_{-i})} + \gamma w_i(a_i),\\
(4) & {o^c_i}{y^{a_i}_i} \leq w'_i(a_i) \leq l{y^{a_i}_i} + o^c_i,\\
(5) & \underline{w}_i - {\underline{w}_i}y^{a_i}_i \leq w_i(a_i) \leq (\underline{w}_i + l) - (\underline{w}_i + l)y^{a_i}_i + \bar{r}y^{a_i}_i;\\
\multicolumn{2}{l}{\hbox{For all }a_1\in A_1\hbox{ and for all }a_2\in A_2\hbox{:}}\\
(6) & y^{a_1}_1 + y^{a_2}_2 \leq z^{a_1,a_2} + 1;\\
\multicolumn{2}{l}{\hbox{For all }p\equiv(\phi^p,\psi^p,\lambda^p)\in P\hbox{ and for all pairs }(a_1,a_2)\in A_1\times A_2\hbox{:}}\\
(7) & \phi^p w_1(a_1) + \psi^p w_2(a_2) \leq \lambda^p z^{a_1,a_2} +
M - M z^{a_1,a_2}.
\end{array}
$$
    \end{itemize}
    \IF{a solution is found}
        \RETURN{$w_i(a_i)$ and $\alpha^{a_i}_i$ for all $i\in \{1,2\}$ and for all $a_i\in A_i$.}
    \ENDIF
\RETURN{\textsc{False}}
\end{algorithmic}
\label{alg:CubeSupportedMixedConvex}
\end{algorithm}

\subsection{Computing Strategies}
Algorithm~\ref{alg:basic} returns the set of hypercubes~$C$, such
that the union of these hypercubes gives~$W$, a set that contains
$U^{\gamma}$. Intuitively, each hypercube represents all those
strategy profiles that induce similar payoff profiles. Therefore,
one can view hypercubes as states of an automaton. Pick a point
$v\in W$. Algorithm~\ref{alg:automaton} constructs an automaton~$M$
that implements a strategy profile~$\sigma$ that approximately
induces the payoff profile~$v$.
\begin{algorithm}
\begin{algorithmic}[1]
  \REQUIRE $C$, a set of hypercubes, such that~$W$ is their union; $v\in W$, a payoff profile.
  \STATE Find a hypercube~$c\in C$, which $v$ belongs to; set $Q\leftarrow\{c\}$ and $q^0\leftarrow
  c$;
  \FOR{\textbf{each} player~$i$}
      \STATE Find $\underline{w}^i = \min_{w\in W}w_i$ and a hypercube~$c^i\in C$, which $\underline{w}^i$ belongs
      to;
      \STATE Set $Q\leftarrow Q\cup\{c^i\}$;
      \STATE Set $f\leftarrow\varnothing\mapsto\times_i\operatorname{\Delta}(A_i)$;
      \STATE Set $\tau\leftarrow \varnothing\mapsto C$.
  \ENDFOR
  \LOOP
      \IF{$Q=\varnothing$}
        \RETURN{$M\equiv(Q,q^0,f,\tau)$}.
      \ENDIF
      \STATE Pick a hypercube $q\in Q$, for which $f(q)$ is not
        defined.
      \STATE Apply the procedure $\textsc{CubeSupported}(q)$ and obtain a (mixed) action profile~$\alpha$ and continuation payoff profiles~$w(a)$ for all $a\in \times_i
      A^{\alpha_i}_i$.
      \STATE Define $f(q)\equiv\alpha$.
      \FOR{\textbf{each} $a\in \times_i A^{\alpha_i}_i$}
        \STATE Find a hypercube~$c\in C$, which $w(a)$ belongs to, set $Q\leftarrow
        Q\cup\{c\}$;
        \STATE Define $\tau(q,a)\equiv c$.
      \ENDFOR
      \FOR{\textbf{each} $i$ and \textbf{each} $a^i\in (A\backslash A^{\alpha_i}_i)\times_{j\in N\backslash \{i\}}
      A^{\alpha_i}_i$}
        \STATE Define $\tau(q,a^i)\equiv c^i$.
      \ENDFOR
  \ENDLOOP
\end{algorithmic}
\caption{Algorithm for constructing an automaton~$M$ that
approximately induces the given payoff
profile~$v$.}\label{alg:automaton}
\end{algorithm}

\subsection{Stopping Criterion}
The values of the flags \textsc{NoCubeWithdrawn} and
\textsc{AllCubesCompleted} determine whether the basic algorithm
(Algorithm~\ref{alg:basic}) should stop and return the set~$W$
approximating the set of SPE payoff profiles (and entirely
containing it). At the end of each algorithm's iteration, the flag
\textsc{AllCubesCompleted} is only $\textsc{True}$, if for none of
the remaining hypercubes $c\in C$, \textsc{CubeCompleted($c$)} is
$\textsc{False}$. The procedure \textsc{CubeCompleted}, in turn,
verifies, for hypercube~$c$, that the two conditions of the problem
stated in Subsection~\ref{subsec:statement} are satisfied, namely:
\begin{enumerate}
  \item For any $v\in W$, the strategy profile~$\sigma$, implemented by the automaton~$M$, constructed by Algorithm~\ref{alg:automaton}, induces the payoff profile $u^{\gamma}(\sigma)$, such that, for all~$i$,
  $v_i- u^{\gamma}_i(\sigma)\leq\epsilon$,
  \item[] and
  \item The maximum payoff~$g_i$ that each player~$i$ can achieve by
  unilaterally deviating from~$\sigma$ is such that $g_i - u^{\gamma}_i(\sigma)\leq\epsilon$.
\end{enumerate}
Both conditions can be verified by dynamic programming. For example,
the second condition can be verified by using the value iteration
algorithm~\citep{sutton-barto:1998}. To do this, the deviating
agent~$i$ has to be considered as the only decision maker
(optimizer). The remaining agents' strategy profile~$\sigma_{-i}$
can then be viewed as the decision maker's environment.

\section{Theoretical Analysis}\label{sec:theory} In this section, we examine the theoretical properties of
Algorithm~\ref{alg:basic} for the case of mixed strategies. While
the procedure \textsc{CubeSupported} for pure strategies
(Algorithm~\ref{alg:CubeSupportedPure}) is defined differently,
mixed strategies include pure ones. Therefore, in our theoretical
analysis, we concentrate on two more general cases: mixed strategies
with no external coordination (Algorithm~\ref{alg:basic} with
\textsc{CubeSupported} given by
Algorithm~\ref{alg:CubeSupportedMixedClusters}) and mixed strategies
with public correlation (Algorithm~\ref{alg:basic} with
\textsc{CubeSupported} given by
Algorithm~\ref{alg:CubeSupportedMixedConvex}).
\begin{theorem}\label{theorem:main}
For any repeated game, discount factor~$\gamma$ and approximation
factor~$\epsilon$, (1)~Algorithm~\ref{alg:basic} terminates in
finite time, (2)~$C$ contains at least one hypercube, and (3)~for
all $c\in C$, Algorithm~\ref{alg:automaton} terminates in finite
time and returns a finite automaton~$M$ that satisfies:
\begin{enumerate}
  \item The strategy profile~$\sigma$ implemented by~$M$ induces the payoff profile $v\equiv
  u^{\gamma}(\sigma)$, such that, for all~$i$, $o^c_i -
  v_i\leq\epsilon$,
  \item[] and
  \item The maximum payoff~$g_i$ that each player~$i$ can achieve by
  unilaterally deviating from~$\sigma$ is such that $g_i -
  v_i\leq\epsilon$.
\end{enumerate}
\end{theorem}
\noindent The proof of Theorem~\ref{theorem:main} relies on the
following lemmas.

\begin{lemma}
\label{lemma:noEmptyC} At any point of execution of
Algorithm~\ref{alg:basic},~$C$ contains at least one hypercube.
\begin{proof}
According to~\citet{nash:1950}, any stage-game has at least one
equilibrium. Let~$v$ be a payoff profile of a certain Nash
equilibrium in the stage-game. For the hypercube~$c$ that
contains~$v$, the procedure $\textsc{CubeSupported}$ will always
return $\textsc{True}$, because for any $\gamma$,~$v$ satisfies the
two conditions of Equation~(\ref{eqn:verifyCube}), with $w' = w = v$
and~$\alpha$ being a mixed action profile that induces~$v$.
Therefore, $c$ will never be withdrawn.
\end{proof}
\end{lemma}

\begin{lemma}
\label{lemma:NoCubeWithdrawnTrue} An iteration of
Algorithm~\ref{alg:basic}, such that $\textsc{NoCubeWithdrawn}$ is
$\textsc{True}$, will be reached in finite time.
\begin{proof}
Because the number of hypercubes (and, therefore, the number of
clusters) is finite, the procedure \textsc{CubeSupported} given by
Algorithm~\ref{alg:CubeSupportedMixedClusters} will terminate in
finite time. The same is true for \textsc{CubeSupported} given by
Algorithm~\ref{alg:CubeSupportedMixedConvex}. For a constant~$l$,
the set $C$ is finite and contains at most
$\lceil(\bar{r}-\underline{r})/l\rceil$ elements. Therefore, and
according to Lemma~\ref{lemma:noEmptyC}, after a finite time, there
will be an iteration of Algorithm~\ref{alg:basic}, such that for all
$c\in C$, $\textsc{CubeSupported}(c)$ returns $\textsc{True}$.
\end{proof}
\end{lemma}

\begin{lemma}\label{lemma:AutomatonFiniteTime}
Let~$C$ be the set of hypercubes at the end of a certain iteration
of Algorithm~\ref{alg:basic}, such that $\textsc{NoCubeWithdrawn}$
is $\textsc{True}$. For all $c\in C$, Algorithm~\ref{alg:automaton}
terminates in finite time and returns a complete finite automaton.
\begin{proof}
By observing the definition of Algorithm~\ref{alg:automaton}, the
proof follows from the fact that the number of hypercubes and,
therefore, the possible number of the automaton states is finite.
Furthermore, the definition of the automaton will be complete,
because the fact that $\textsc{NoCubeWithdrawn}$ is $\textsc{True}$
implies that for each hypercube $c\in C$, there is a mixed action
$\alpha$ and a continuation payoff profile~$w$ belonging to a
certain hypercube $c'\in C$. Consequently, for each state~$q$ of the
automaton, the functions $f(q)$ and $\tau(q)$ will be defined.
\end{proof}
\end{lemma}

\begin{lemma}\label{lemma:ErrorAutomaton}
Let~$C$ be the set of hypercubes at the end of a certain iteration
of Algorithm~\ref{alg:basic}, such that $\textsc{NoCubeWithdrawn}$
is $\textsc{True}$. Let~$l$ be the current value of the hypercube
side length. For every $c\in C$, the strategy profile~$\sigma$,
implemented by the automaton~$M$ that starts in $c$, induces the
payoff profile $v\equiv u^{\gamma}(M)$, such that, for all~$i$,
$o^c_i - v_i\leq\frac{\gamma l}{1-\gamma}$.
\begin{proof}
When player~$i$ is following the strategy prescribed by the
automaton constructed by Algorithm~\ref{alg:automaton}, this process
can be reflected by an \emph{equilibrium graph}, as the one shown in
Figure~\ref{fig:AutomatonEquilibrium}.
\begin{figure}[htb]
    \begin{center}
        \leavevmode
        \includegraphics[scale=0.8]{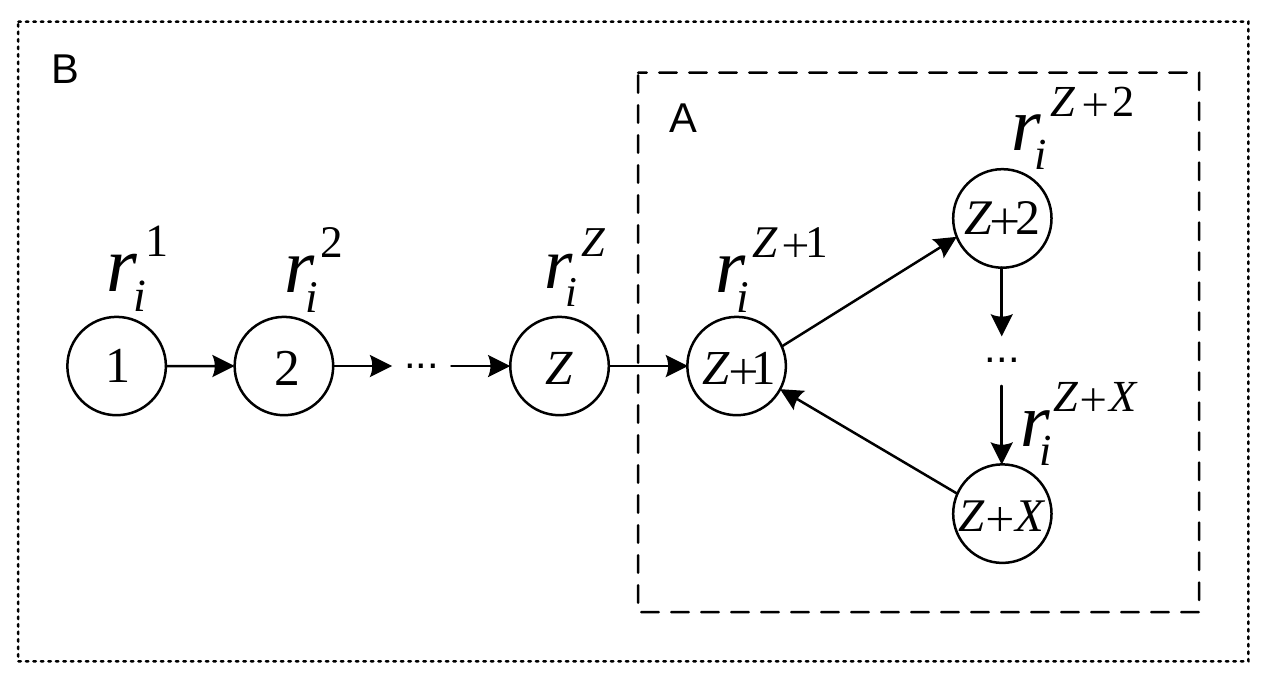}
    \end{center}
    \caption{Equilibrium graph for player~$i$. The graph represents the initial state followed by a non-cyclic sequence of states (nodes $1$ to $Z$) followed by a cycle of $X$ states (nodes~$Z+1$ to $Z+X$). The labels over the nodes are the immediate expected payoffs collected by player~$i$ in the corresponding states.}
    \label{fig:AutomatonEquilibrium}
\end{figure}
Because for all hypercubes~$c$ behind the states of the automaton,
$\textsc{CubeSupported}$ returns $\textsc{True}$, we have:
\begin{equation}
\label{eqn:basicInequalitiesError1}
\begin{array}{cl}
  (1.1) & o^1_i\leq(1-\gamma)r^1_i+\gamma w^1_i\leq o^1_i+l,\\
  (1.2) & o^2_i\leq w^1_i\leq o^2_i+l,\\
  (2.1) & o^2_i\leq(1-\gamma)r^2_i+\gamma w^2_i\leq o^2_i+l,\\
  (2.2) & o^3_i\leq w^2_i\leq o^3_i+l,\\
  \ldots & \\
  (\hbox{Z}.1) & o^Z_i\leq(1-\gamma)r^Z_i+\gamma w^Z_i\leq o^Z_i+l,\\
  (\hbox{Z}.2) & o^{Z+1}_i\leq w^Z_i\leq o^{Z+1}_i+l,\\
  (\hbox{Z+1}.1) & o^{Z+1}_i\leq(1-\gamma)r^{Z+1}_i+\gamma w^{Z+1}_i\leq o^Z_i+l,\\
  (\hbox{Z+1}.2) & o^{Z+2}_i\leq w^{Z+1}_i\leq o^{Z+2}_i+l,\\
  \ldots & \\
  (\hbox{Z+X}.1) & o^{Z+X}_i\leq(1-\gamma)r^{Z+X}_i+\gamma w^{Z+X}_i\leq o^{Z+X}_i+l,\\
  (\hbox{Z+X}.2) & o^{Z+1}_i\leq w^{Z+X}_i\leq o^{Z+1}_i+l,\\
\end{array}
\end{equation}
\noindent where $o^q_i$, $r^q_i$ and $w^q_i$ stand respectively for
(\emph{i})~the payoff of player~$i$ in the origin of the hypercube
behind the state~$q$, (\emph{ii})~the immediate expected payoff of
player~$i$ for playing according to $f_i(q)$ or for deviating inside
the support of~$f_i(q)$, and (\emph{iii})~the continuation promise
payoff of player~$i$ for playing according to the equilibrium
strategy profile in state~$q$.

The following development only uses the inequalities of
Equation~(\ref{eqn:basicInequalitiesError1}), one by one. It starts
with inequality~(Z+1):
\begin{eqnarray}\label{eqn:developmentError1}
     o^{Z+1}_i&\leq&(1-\gamma)r^{Z+1}_i+\gamma w^{Z+1}_i\nonumber\\
     & &\langle\hbox{By inequality (Z+1.2)}\rangle\nonumber\\
     &\leq&(1-\gamma)r^{Z+1}_i+\gamma (o^{Z+2}_i+l)\nonumber\\
     & &\langle\hbox{By inequality (Z+2.1)}\rangle\nonumber\\
     &\leq&(1-\gamma)r^{Z+1}_i+\gamma \left((1-\gamma)r^{Z+2}_i+\gamma w^{Z+2}_i\right)+\gamma l\nonumber\\
     & &\langle\hbox{By inequality (Z+2.2)}\rangle\nonumber\\
     &\leq&(1-\gamma)r^{Z+1}_i+\gamma(1-\gamma)r^{Z+2}_i+\gamma^2 o^{Z+3}_i+\gamma l \nonumber\\
      & & \ldots\\
     & &\langle\hbox{By inequality (Z+X.2)}\rangle\nonumber\\
     &\leq&(1-\gamma)\sum_{x=1}^X{\gamma^{x-1}r^{Z+x}_i}+\gamma^X
o^{Z+1}_i+\gamma\sum_{x=1}^{X}{\gamma^{x-1} l}
\end{eqnarray}
Denote by $g^A_i$ the long-term expected non-normalized payoff for
player~$i$ for passing through the cycle~$A$ of the equilibrium
graph infinitely often.
\begin{eqnarray}\label{eqn:developmentError2}
    g^A_i &=& \sum_{x=1}^{X}{\gamma^{x-1}r^{Z+x}_i} + \gamma^X g^A_i\\
        &=&\frac{\sum_{z=1}^{X}{\gamma^{x-1}r^{Z+x}_i}}{1-\gamma^X}.
\end{eqnarray}
The property of the infinite sum of the geometric series allows us
to write:
\begin{equation}\label{eqn:developmentError3}
\sum_{x=1}^X{\gamma^{x-1}l} =\frac{(1-\gamma^X)l}{1-\gamma}.
\end{equation}
From
Equations~(\ref{eqn:developmentError1}-\ref{eqn:developmentError3})
it follows that,
\begin{equation}\label{eqn:developmentError4}
    o^{Z+1}\geq (1-\gamma)g^A_i + \frac{\gamma l}{1-\gamma}.
\end{equation}
Using inequalities~(1.1 - Z.2) of
Equation~(\ref{eqn:basicInequalitiesError1}), the following
development is possible:
\begin{eqnarray}\label{eqn:developmentError5}
     & &\langle\hbox{By inequality (1.1)}\rangle\nonumber\\
     o^1_i&\leq& (1-\gamma)r^1_i+\gamma w^1_i\nonumber\\
     & &\langle\hbox{By inequality (1.2)}\rangle\nonumber\\
          &\leq& (1-\gamma)r^1_i+\gamma (o^2_i+l)\nonumber\\
          & &\langle\hbox{By inequality (2.1)}\rangle\nonumber\\
          &\leq& (1-\gamma)r^1_i+\gamma \left((1-\gamma)r^2_i+\gamma w^2_i\right)+\gamma l\nonumber\\
          & &\langle\hbox{By inequality (2.2)}\rangle\nonumber\\
          &\leq& (1-\gamma)r^1_i+\gamma (1-\gamma)r^2_i+\gamma^2 (o^3_i + l)+\gamma l\nonumber\\
          & & \ldots\\
          & &\langle\hbox{By inequality (Z.2)}\rangle\nonumber\\
          &\leq& (1-\gamma)\sum_{z=1}^{Z} \gamma^{z-1} r^z_i + \gamma^Z o^{Z+1}_i + \gamma\sum_{z=1}^Z{\gamma^{z-1}l}\nonumber\\
\end{eqnarray}
From Equations~(\ref{eqn:developmentError4}) and
(\ref{eqn:developmentError5}) it follows that,
\begin{equation}\label{eqn:developmentError6}
    o^1_i\leq (1-\gamma)\left(\sum_{z=1}^{Z} \gamma^{z-1} r^z_i + \gamma^Z g^A_i\right) + \frac{\gamma l}{1-\gamma}.
\end{equation}
Denote by $g^B_i$ the long-term expected (normalized) payoff for
player~$i$ for passing through the equilibrium graph (graph $B$ in
Figure~\ref{fig:AutomatonDeviation}a) infinitely often. Observe
that,
$$
g^B_i \equiv (1-\gamma)\left(\sum_{z=1}^{Z} \gamma^{z-1} r^z_i +
\gamma^Z g^A_i\right).
$$
Therefore,
\begin{equation*}\label{eqn:developmentDeviation7}
    o^0_i - g^B_i \leq \frac{\gamma l}{1-\gamma}.
\end{equation*}
\end{proof}
\end{lemma}

\begin{lemma}\label{lemma:DeviationAutomaton}
Let~$C$ be the set of hypercubes at the end of a certain iteration
of Algorithm~\ref{alg:basic}, such that $\textsc{NoCubeWithdrawn}$
is $\textsc{True}$. Let~$l$ be the current value of the hypercube
side length. For every $c\in C$, the maximum gain $g_i$ that each
player~$i$ can achieve by unilaterally deviating from the strategy
profile~$\sigma$ implemented by an automaton~$M$ that starts in $c$
and induces the payoff profile $v\equiv u^{\gamma}(M)$ is such that
$g_i - v_i\leq\frac{2l}{1-\gamma}$.
\begin{proof}
To prove the lemma, one has to bound the maximum gain of a deviation
that starts in an arbitrary state of an automaton. Consider two
deviation graphs for player~$i$ depicted in
Figure~\ref{fig:AutomatonDeviation}.
\begin{figure}[htb]
    \begin{center}
        \leavevmode
        \begin{tabular}{cc}
          \includegraphics[scale=0.8]{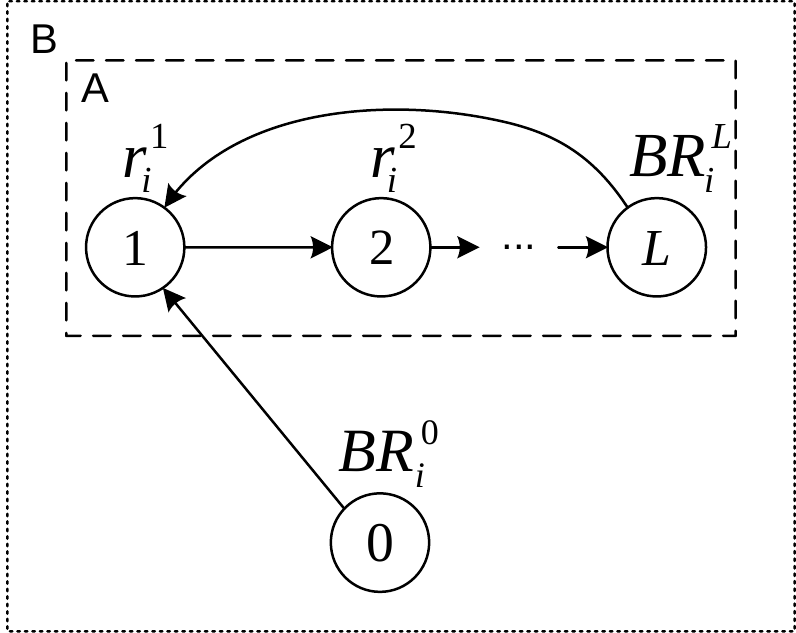} &~~~~~~~~~~~~~~~~~~~\includegraphics[scale=0.8]{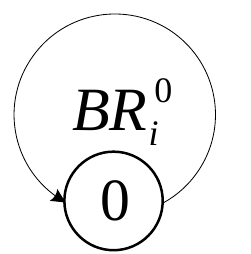} \\
          (a) &~~~~~~~~~~~~~~~(b) \\
        \end{tabular}

    \end{center}
    \caption{Deviation graphs for player~$i$. (a)~A generic deviation graph for player~$i$. The graph represents the initial deviation state (node~$0$) followed by a transition into the punishment state (node~$1$) followed by a number of in-equilibrium (or, otherwise, inside-the-support deviation) states (nodes $1$ to $L-1$) followed by the subsequent out-of-the-support deviation state (node~$L$). (b)~A particular, one state deviation graph, where the only deviation state is the punishment state for player~$i$. The labels over the nodes are the immediate expected payoffs collected by player~$i$ in the corresponding states.}
    \label{fig:AutomatonDeviation}
\end{figure}
A \emph{deviation graph for player~$i$} is a finite graph, which
reflects the optimal behavior for player~$i$ assuming that the
behavior of the other players is fixed and is given by an automaton
returned by Algorithm~\ref{alg:automaton}. The nodes of the
deviation graph correspond to the states of the automaton. The
labels over the nodes are the immediate expected payoffs collected
by player~$i$ in the corresponding states. A \emph{generic deviation
graph for player~$i$} (Figure~\ref{fig:AutomatonDeviation}a) is a
deviation graph that has one cyclic and one non-cyclic part. In the
cyclic part (subgraph $A$), player~$i$ follows the equilibrium
strategy or deviations take place inside the support of the
prescribed mixed actions (nodes $1$ to $L-1$, with node~$1$
corresponding to the punishment state\footnote{The punishment state
for player~$i$ is the automaton state, which is based on the
hypercube that contains a payoff profile~$v$, such that $v_i =
\underline{w}_i$.} for player~$i$). In the last node of the cyclic
part (node $L$), an out-of-the-support deviation takes place. The
non-cyclic part of the generic deviation graph contains a single
node corresponding to the state, where the initial
out-of-the-support deviation of player~$i$ from the SPE strategy
profile occurs. If the state of the initial deviation of player~$i$
is itself the punishment state for player~$i$, then the deviation
graph will look as shown in Figure~\ref{fig:AutomatonDeviation}b.

The present proof only considers the generic deviation graph
(Figure~\ref{fig:AutomatonDeviation}a); the proof for the particular
cases, like that of Figure~\ref{fig:AutomatonDeviation}b, can be
obtained by analogy, and, because they bring the same result, we
omit it here. Consider first the subgraph~$A$ of the generic
deviation graph. Because, for all hypercubes~$c$ behind the states
of the automaton, $\textsc{CubeSupported}$ returns $\textsc{True}$,
we have:
\begin{equation}
\label{eqn:basicInequalitiesDeviation1}
\begin{array}{cl}
  (1.0) & o^1_i\leq \underline{w}_i\leq o^1_i+l,\\
  (1.1) & o^1_i\leq(1-\gamma)r^1_i+\gamma w^1_i\leq o^1_i+l,\\
  (1.2) & o^2_i\leq w^1_i\leq o^2_i+l,\\
  (2.1) & o^2_i\leq(1-\gamma)r^2_i+\gamma w^2_i\leq o^2_i+l,\\
  (2.2) & o^3_i\leq w^2_i\leq o^3_i+l,\\
  \ldots & \\
  (\hbox{Z}.1) & o^Z_i\leq(1-\gamma)r^Z_i+\gamma w^Z_i\leq o^Z_i+l,\\
  (\hbox{Z}.2) & (1-\gamma)r^Z_i+\gamma w^Z_i - (1-\gamma)BR^Z_i-\gamma\underline{w}_i\geq
  0,
\end{array}
\end{equation}
\noindent where $o^q_i$, $r^q_i$ and $w^q_i$ stand respectively for
(\emph{i})~the payoff of player~$i$ in the origin of the hypercube
behind the state~$q$, (\emph{ii})~the immediate expected payoff of
player~$i$ for playing according to $f_i(q)$ or for deviating inside
the support of~$f_i(q)$, and (\emph{iii})~the continuation promise
payoff of player~$i$ for playing according to the equilibrium
strategy profile in state~$q$.

The following development only uses the inequalities of
Equation~(\ref{eqn:basicInequalitiesDeviation1}), one by one. It
starts with inequality~(1.1):
\begin{eqnarray}\label{eqn:developmentDeviation1}
o^1_i&\geq&(1-\gamma)r^1_i+\gamma w^1_i - l\nonumber\\
     & &\langle\hbox{By inequality (1.2)}\rangle\nonumber\\
     &\geq&(1-\gamma)r^1_i+\gamma o^2_i - l\nonumber\\
     & &\langle\hbox{By inequality (2.1)}\rangle\nonumber\\
     &\geq&(1-\gamma)r^1_i+\gamma \left((1-\gamma)r^2_i+\gamma w^2_i - l\right) - l\nonumber\\
     & &\langle\hbox{By inequality (2.2)}\rangle\nonumber\\
     &\geq&(1-\gamma)r^1_i+\gamma(1-\gamma)r^2_i+\gamma^2 o^3_i - \gamma l - l\nonumber\\
     & &\ldots\nonumber\\
     & &\langle\hbox{By inequalities (3.1) to (Z-1.1)}\rangle\nonumber\\
     &\geq&(1-\gamma)\sum_{z=1}^{Z-1}{\gamma^{z-1}r^z}+\gamma^{Z-1} o^Z - \gamma\sum_{z=1}^{Z-1}{\gamma^{z-1}l}\nonumber\\
     & &\langle\hbox{By inequality (Z.1)}\rangle\nonumber\\
     &\geq&(1-\gamma)\sum_{z=1}^{Z-1}{\gamma^{z-1}r^z}+\gamma^{Z-1} \left((1-\gamma)r^Z_i+\gamma w^Z_i - l\right) - \gamma\sum_{z=1}^{Z-1}{\gamma^{z-1}l}\nonumber\\
     & &\langle\hbox{By inequality (Z.2)}\rangle\nonumber\\
     &\geq&(1-\gamma)\sum_{z=1}^{Z-1}{\gamma^{z-1}r^z}+\gamma^{Z-1} \left((1-\gamma)BR^Z_i+\gamma\underline{w}_i - l\right) - \gamma\sum_{z=1}^{Z-1}{\gamma^{z-1}l}\nonumber\\
     & &\langle\hbox{By inequality (1.0)}\rangle\nonumber\\
     &\geq&(1-\gamma)\left(\sum_{z=1}^{Z-1}{\gamma^{z-1}r^z}+\gamma^{Z-1}BR^Z_i\right)+\gamma^Z o^1_i -
     \gamma\sum_{z=1}^Z{\gamma^{z-1}l}.
\end{eqnarray}
Denote by $g^A_i$ the long-term expected non-normalized payoff of
player~$i$ for passing through the cycle~$A$ of the generic
deviation graph infinitely often:
\begin{eqnarray}\label{eqn:developmentDeviation2}
    g^A_i &=& \sum_{z=1}^{Z-1}{\gamma^{z-1}r^z_i}+\gamma^{Z-1}BR^Z_i +
    \gamma^Z g^A_i\\
        &=&\frac{\sum_{z=1}^{Z-1}{\gamma^{z-1}r^z_i}+\gamma^{Z-1}BR^Z_i}{1-\gamma^Z}.
\end{eqnarray}
The property of the infinite sum of the geometric series permits us
to write:
\begin{equation}\label{eqn:developmentDeviation3}
\sum_{z=1}^Z{\gamma^{z-1}l} =\frac{(1-\gamma^Z)l}{1-\gamma}.
\end{equation}
From
Equations~(\ref{eqn:developmentDeviation1}-\ref{eqn:developmentDeviation3})
it follows that
\begin{equation}\label{eqn:developmentDeviation4}
    o^1\geq (1-\gamma)g^A_i - \frac{\gamma l}{1-\gamma}.
\end{equation}
Furthermore, we have,
\begin{equation}
\label{eqn:basicInequalitiesDeviation2}
\begin{array}{cl}
  (1.1) & (1-\gamma)r^0_i+\gamma w^0_i - (1-\gamma)BR^0_i-\gamma\underline{w}_i\geq 0,\\
  (1.2) & o^0_i\leq(1-\gamma)r^0_i+\gamma w^0_i\leq o^0_i+l.
\end{array}
\end{equation}
The following development is possible:
\begin{eqnarray}\label{eqn:developmentDeviation5}
     & &\langle\hbox{By inequality (1.1) of Equation~(\ref{eqn:basicInequalitiesDeviation2})}\rangle\nonumber\\
     (1-\gamma)r^0_i+\gamma w^0_i &\geq& (1-\gamma)BR^0_i+\gamma\underline{w}_i\nonumber\\
     & &\langle\hbox{By inequality (1.2) of Equation~(\ref{eqn:basicInequalitiesDeviation2})}\rangle\nonumber\\
     o^0_i &\geq& (1-\gamma)BR^0_i+\gamma\underline{w}_i - l\nonumber\\
           & &\langle\hbox{By inequality (1.0) of Equation~(\ref{eqn:basicInequalitiesDeviation1})}\rangle\nonumber\\
           &\geq& (1-\gamma)BR^0_i+\gamma o^1_i + \gamma l - l
\end{eqnarray}
From Equations~(\ref{eqn:developmentDeviation4}) and
(\ref{eqn:developmentDeviation5}) it follows that
\begin{equation}\label{eqn:developmentDeviation6}
    o^0_i\geq (1-\gamma)BR^0_i+\gamma \left((1-\gamma)g^A_i - \frac{\gamma l}{1-\gamma}\right) + \gamma l -
    l.
\end{equation}
Denote by $g^B_i$ the long-term expected (normalized) payoff of
player~$i$ for following the generic deviation graph (graph $B$ in
Figure~\ref{fig:AutomatonDeviation}a). Observe that,
$$
g^B_i \equiv (1-\gamma)BR^0_i+\gamma(1-\gamma)g^A_i.
$$
Therefore,
\begin{equation*}\label{eqn:developmentDeviation7}
    g^B_i - o^0_i \leq \frac{\gamma^2 l}{1-\gamma} - \gamma l + l.
\end{equation*}
Finally, by Lemma~\ref{lemma:ErrorAutomaton}, starting from the
state that corresponds to the node~$0$ of the generic deviation
graph, the payoff profile~$v$, induced by the automaton, satisfies:
$o^0_i\leq\frac{\gamma l}{1-\gamma} + v_i$. Therefore,
\begin{eqnarray}\label{eqn:developmentDeviation7}
    g^B_i - v_i &\leq& \frac{\gamma^2 l}{1-\gamma} + \frac{\gamma l}{1-\gamma} - \gamma l + l\nonumber\\
                &\leq& \frac{2l}{1-\gamma}.\nonumber
\end{eqnarray}

\end{proof}
\end{lemma}

\begin{lemma}
Algorithm~\ref{alg:basic} terminates in finite time.
\begin{proof}
The hypercube side length~$l$ is reduced by half every time that no
hypercube was withdrawn by the end of an iteration of the algorithm.
Therefore, and by Lemma~\ref{lemma:NoCubeWithdrawnTrue}, any given
value of~$l$ will be reached after a finite time. By
Lemmas~\ref{lemma:ErrorAutomaton} and
\ref{lemma:DeviationAutomaton}, Algorithm~\ref{alg:basic}, in the
worst case, terminates whenever~$l$ becomes lower than or equal to
$\frac{\epsilon(1-\gamma)}{2}$.
\end{proof}
\end{lemma}

\section{Experimental Results} \label{sec:results}
In this section, we present several significant experimental results
for a number of well-known games. These are Prisoner's Dilemma
(Figure~\ref{fig:PD}), Duopoly (Figure~\ref{fig:matrix_games}a),
Rock, Paper, Scissors (Figure~\ref{fig:matrix_games}b), Battle of
the Sexes (Figure~\ref{fig:matrix_games}c), and a game with no
stage-game pure action equilibrium (Figure~\ref{fig:matrix_games}d).
For these games, certain equilibrium properties are known or can be
readily analytically verified.

\begin{figure}
\center
\begin{tabular}{cc}
  \usebox{\abreubox}
    &
  \usebox{\rpcbox}\\[8 mm]
    (a) & (b)\\
  \usebox{\bosbox} &
  \usebox{\nopurebox}\\[6 mm]
    (c) & (d)\\
\end{tabular}
  \caption{Four game matrices: (a)~Duopoly game, (b)~Rock, Paper, Scissors, (c)~Battle of the Sexes and (d)~Game with no pure action Nash equilibrium in stage-game.}\label{fig:matrix_games}
\end{figure}

\begin{figure}
\center
\begin{tabular}{cccccc}
\\
\rotatebox{180}{\reflectbox{\includegraphics[scale=0.06]{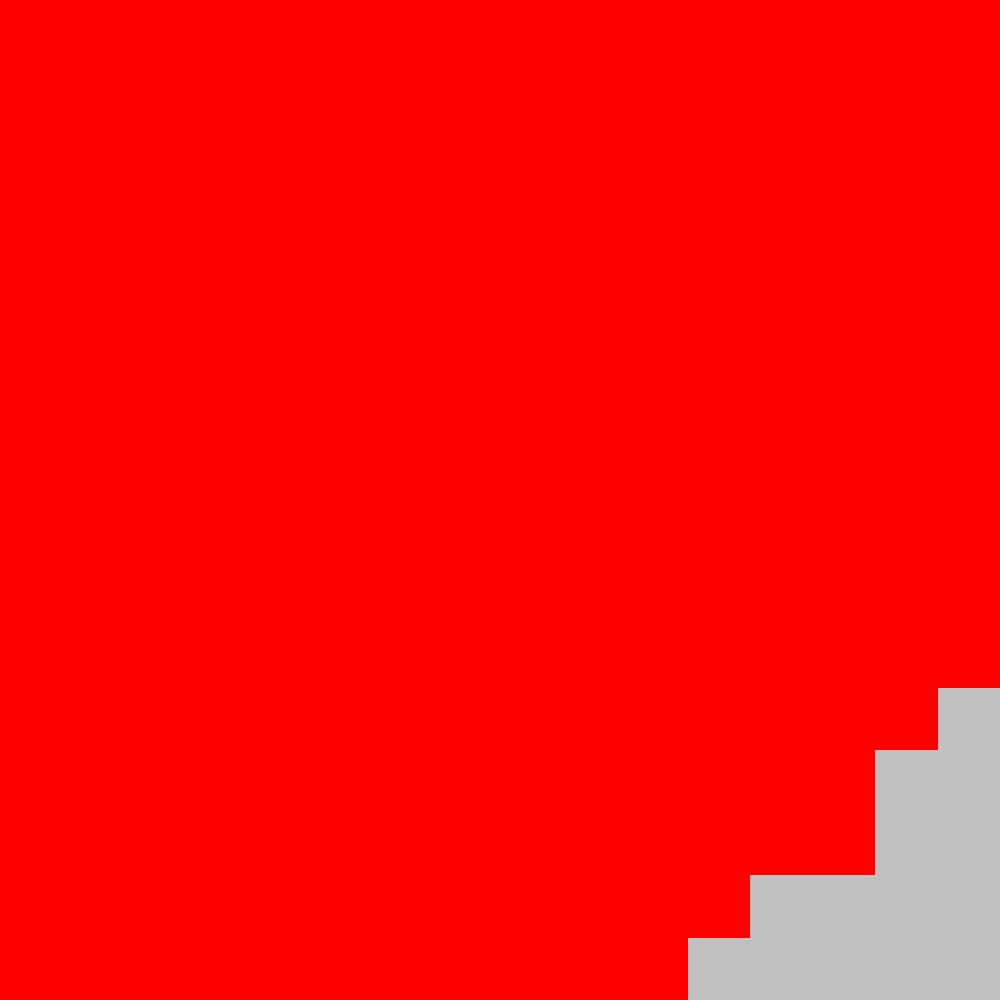}}}
&
\rotatebox{180}{\reflectbox{\includegraphics[scale=0.06]{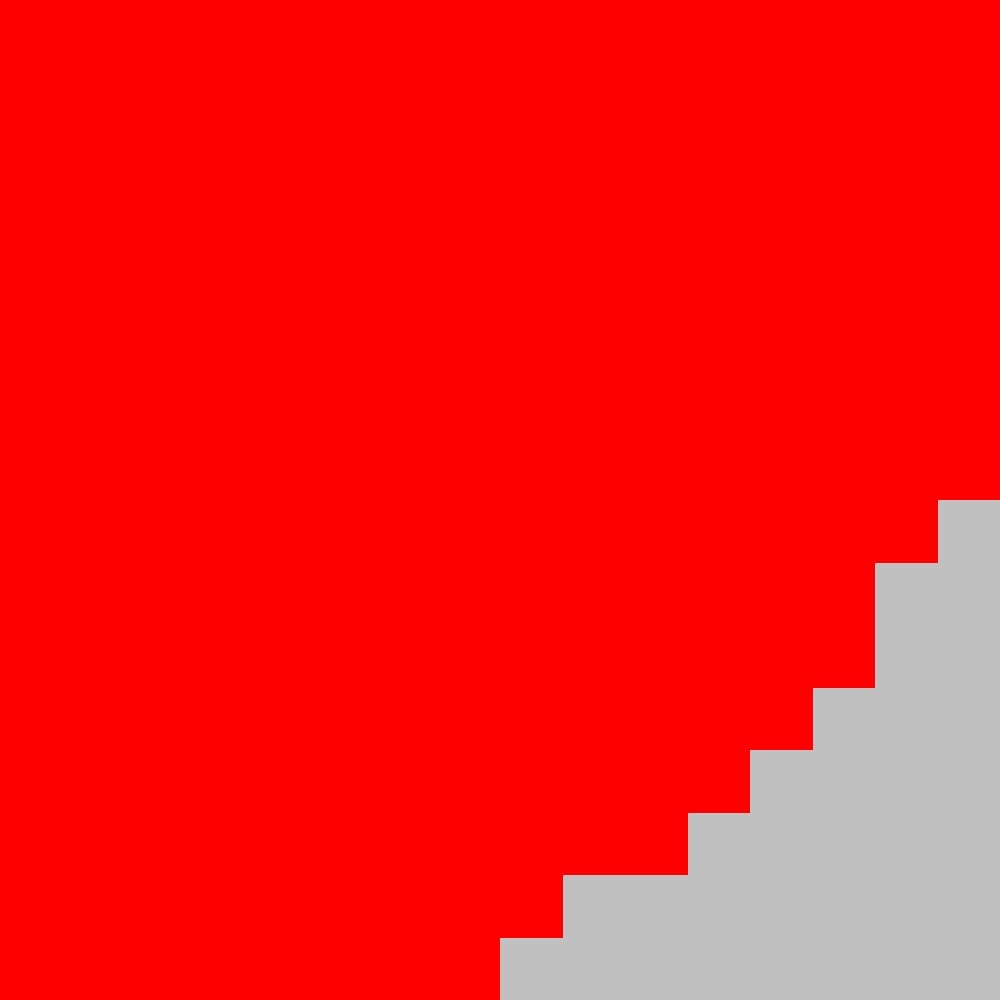}}}
&
\rotatebox{180}{\reflectbox{\includegraphics[scale=0.06]{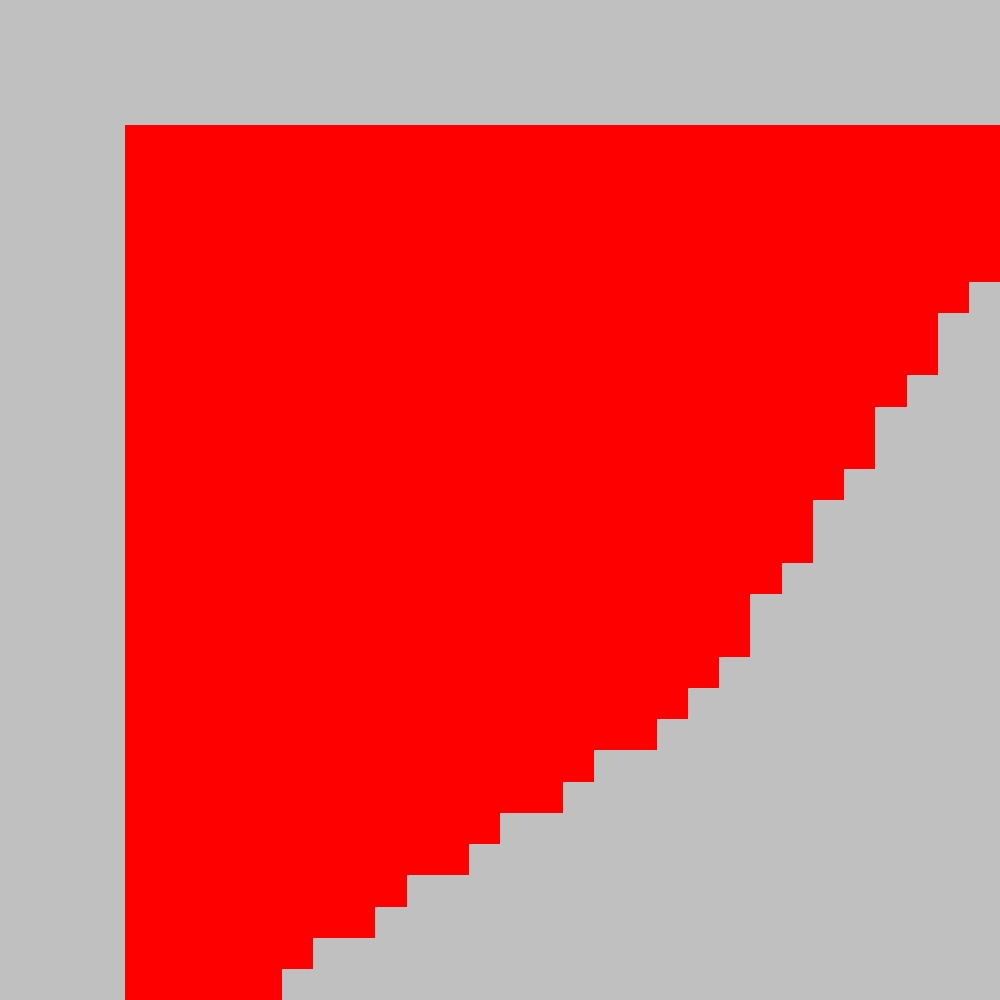}}}
&
\rotatebox{180}{\reflectbox{\includegraphics[scale=0.06]{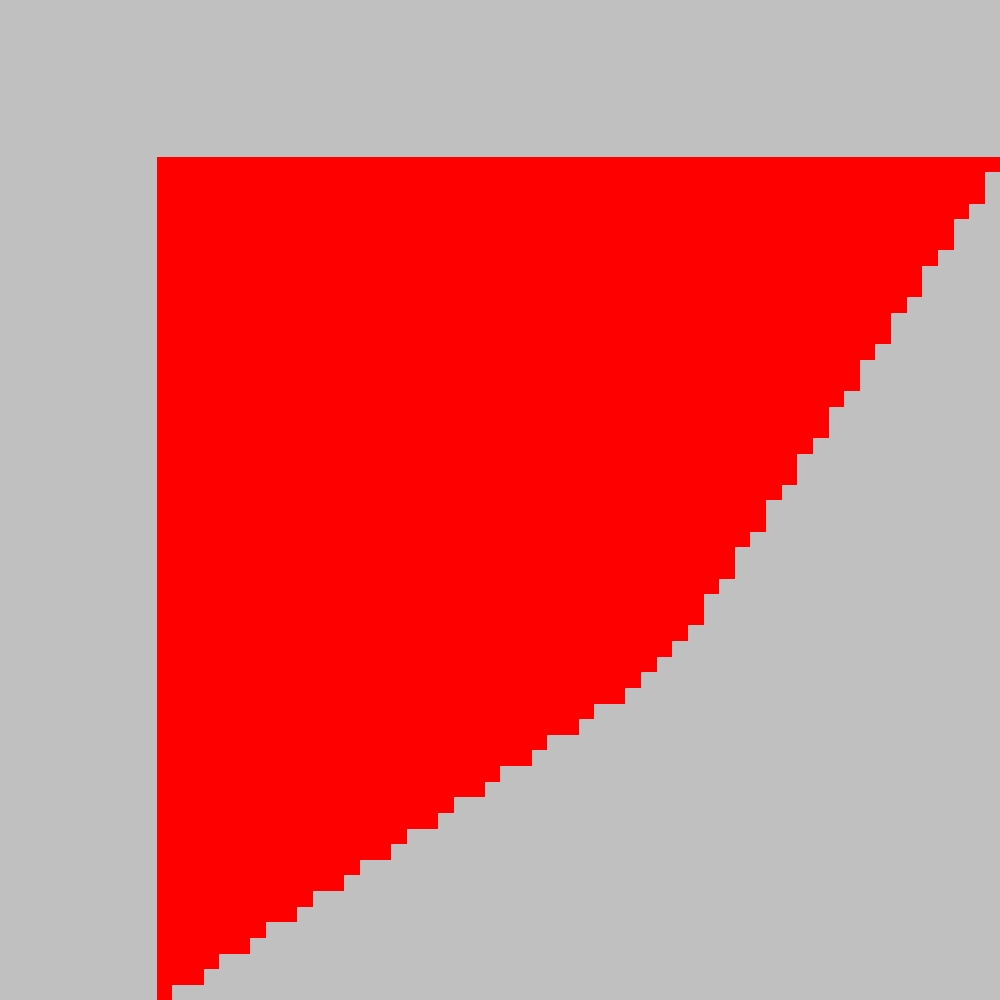}}}
&
\rotatebox{180}{\reflectbox{\includegraphics[scale=0.06]{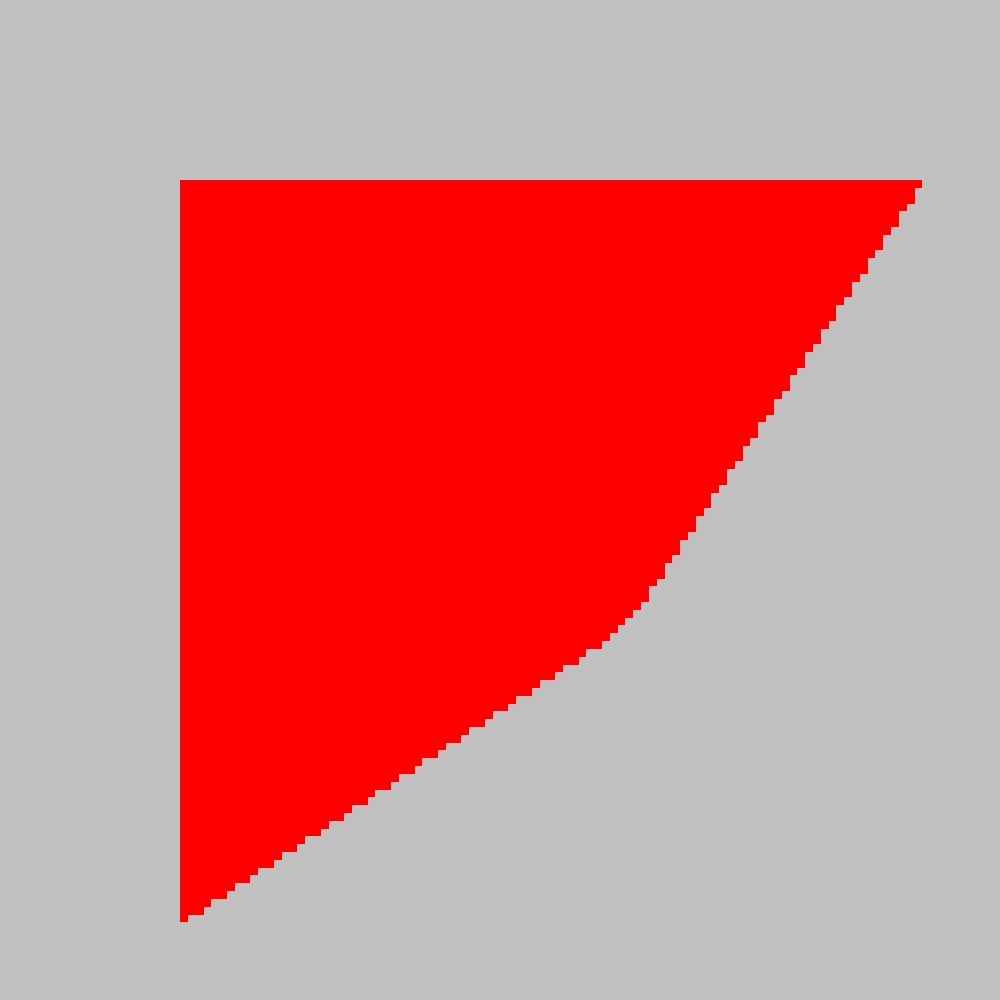}}}
&
\rotatebox{180}{\reflectbox{\includegraphics[scale=0.06]{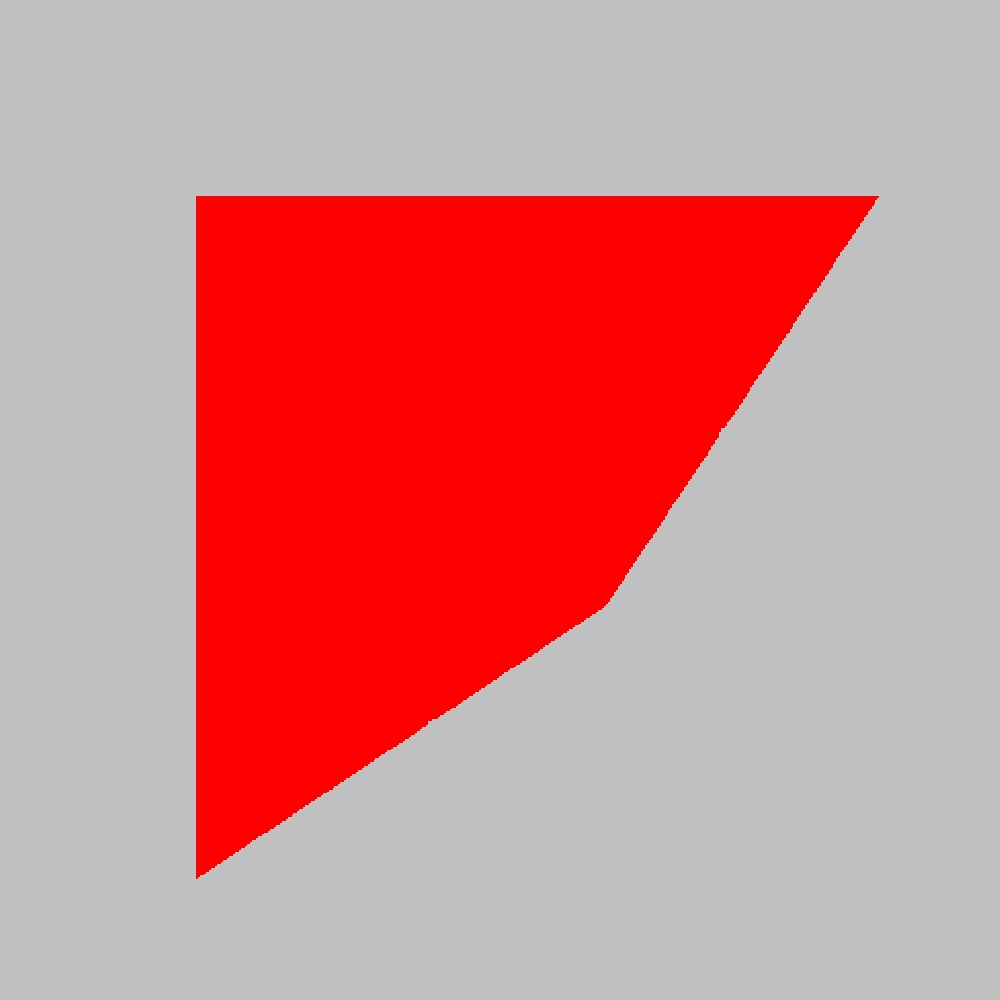}}}\\
  5 & 10 & 15 & 20 & 30 & 50 \\
  \multicolumn{6}{c}{(a)~$\gamma=0.7$, $\epsilon=0.01$}
\\
\rotatebox{180}{\reflectbox{\includegraphics[scale=0.06]{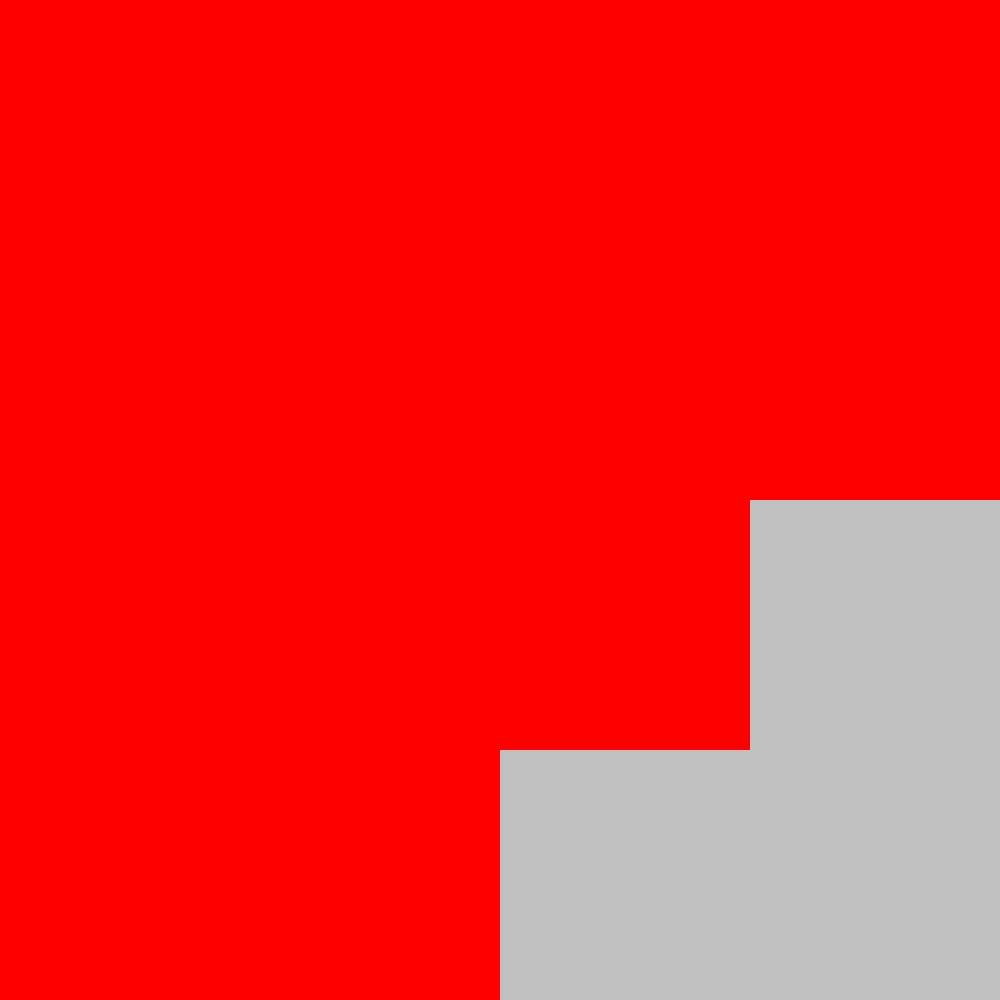}}}
&
\rotatebox{180}{\reflectbox{\includegraphics[scale=0.06]{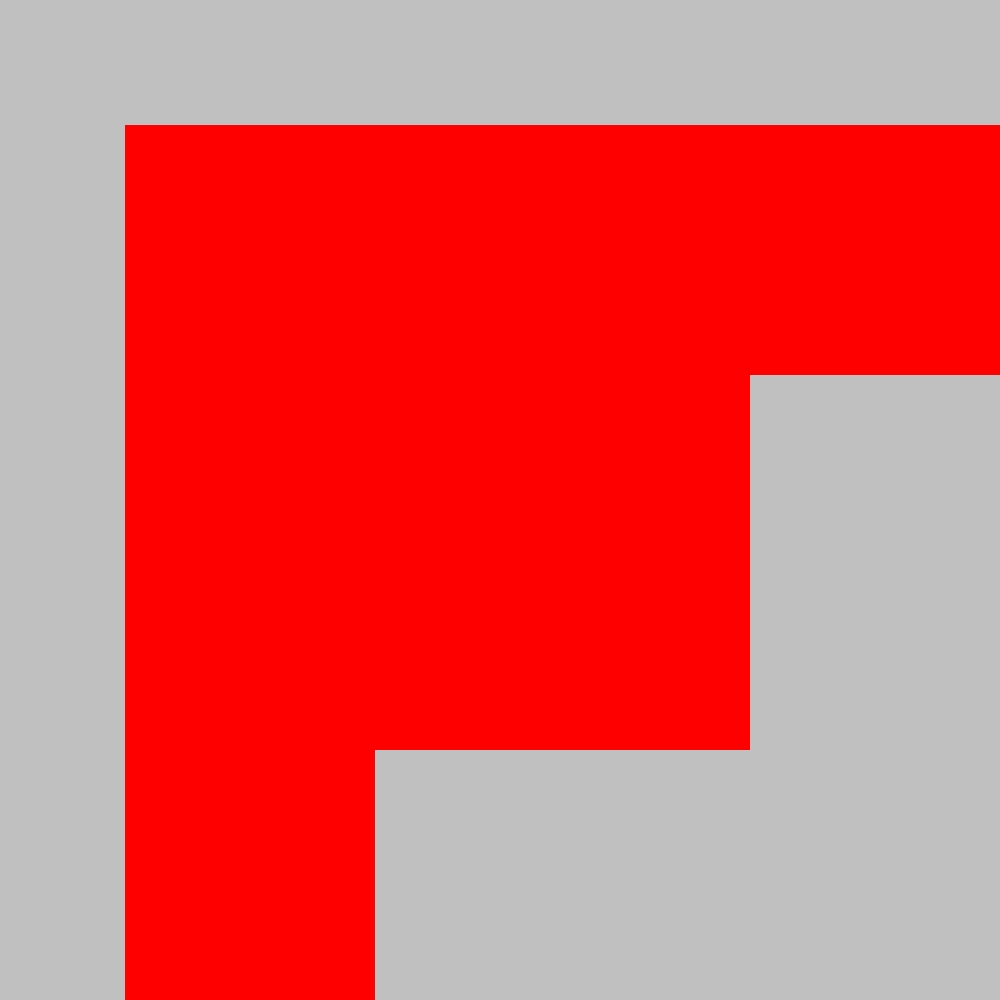}}}
&
\rotatebox{180}{\reflectbox{\includegraphics[scale=0.06]{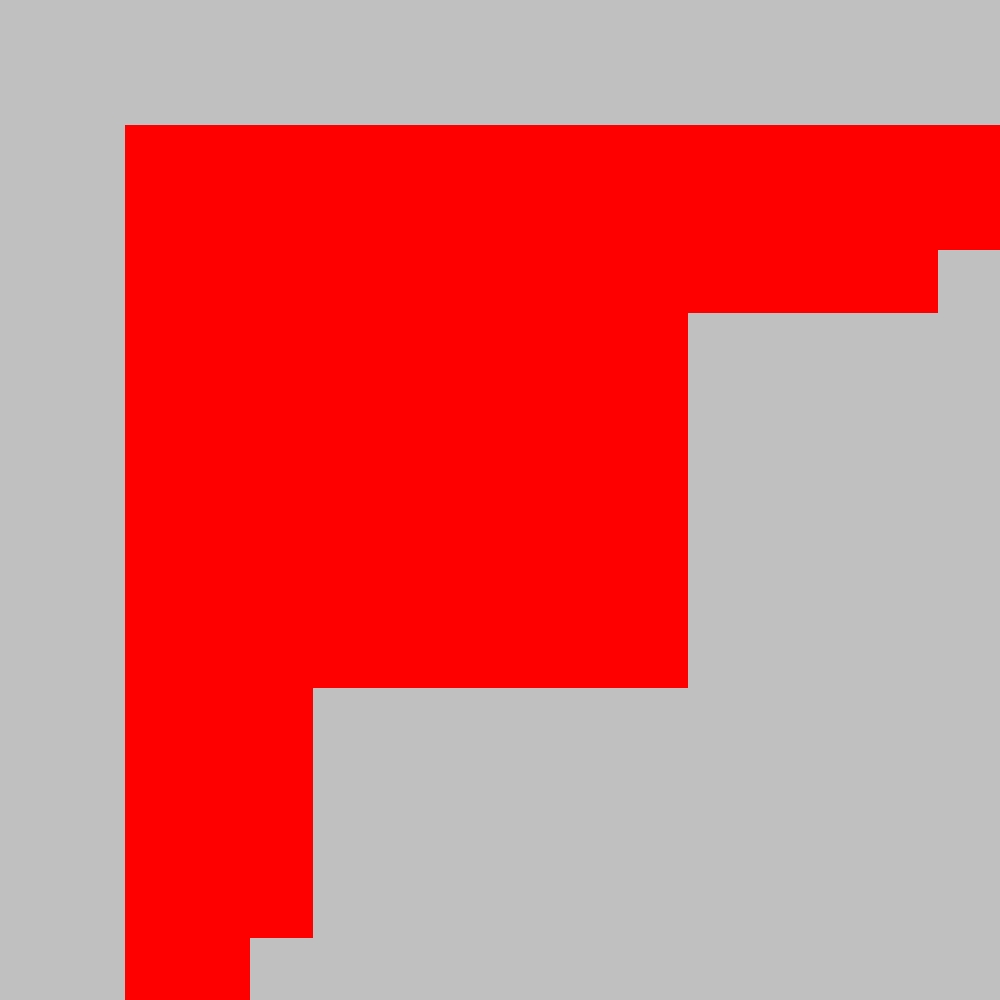}}}
&
\rotatebox{180}{\reflectbox{\includegraphics[scale=0.06]{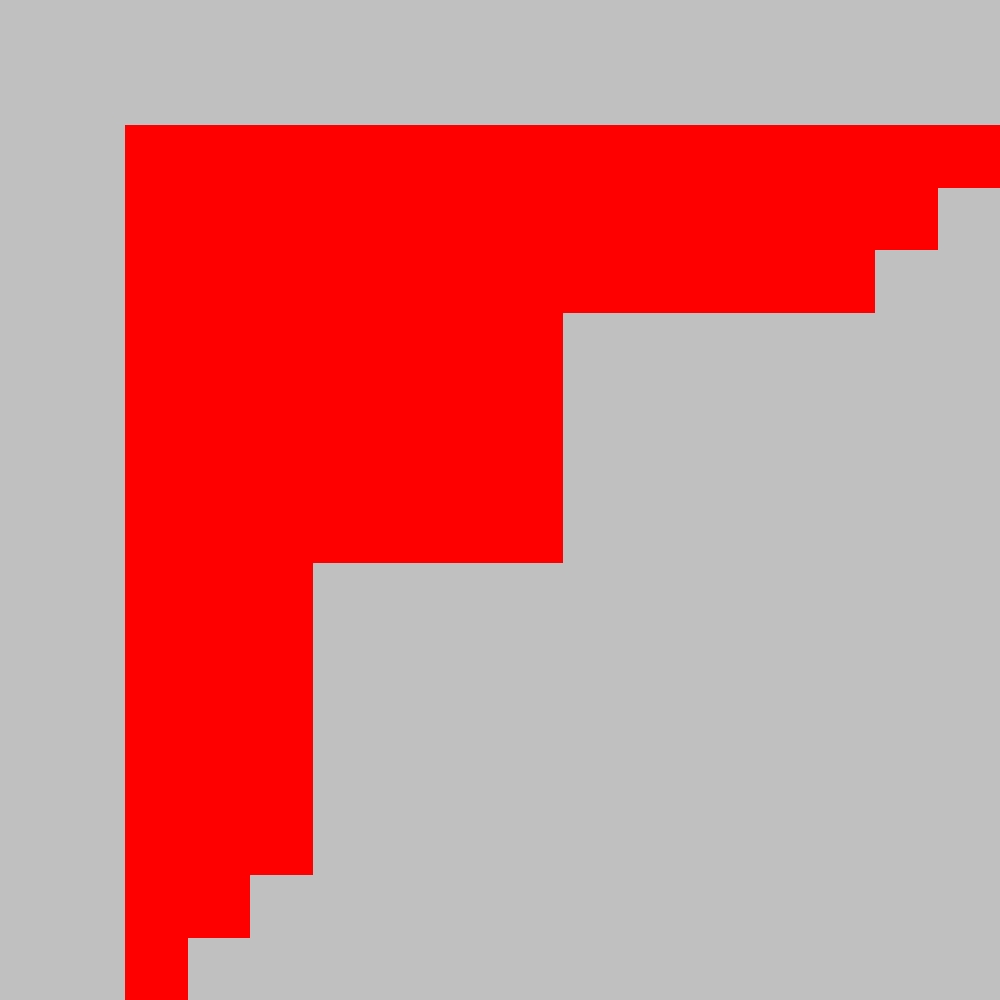}}}
&
\rotatebox{180}{\reflectbox{\includegraphics[scale=0.06]{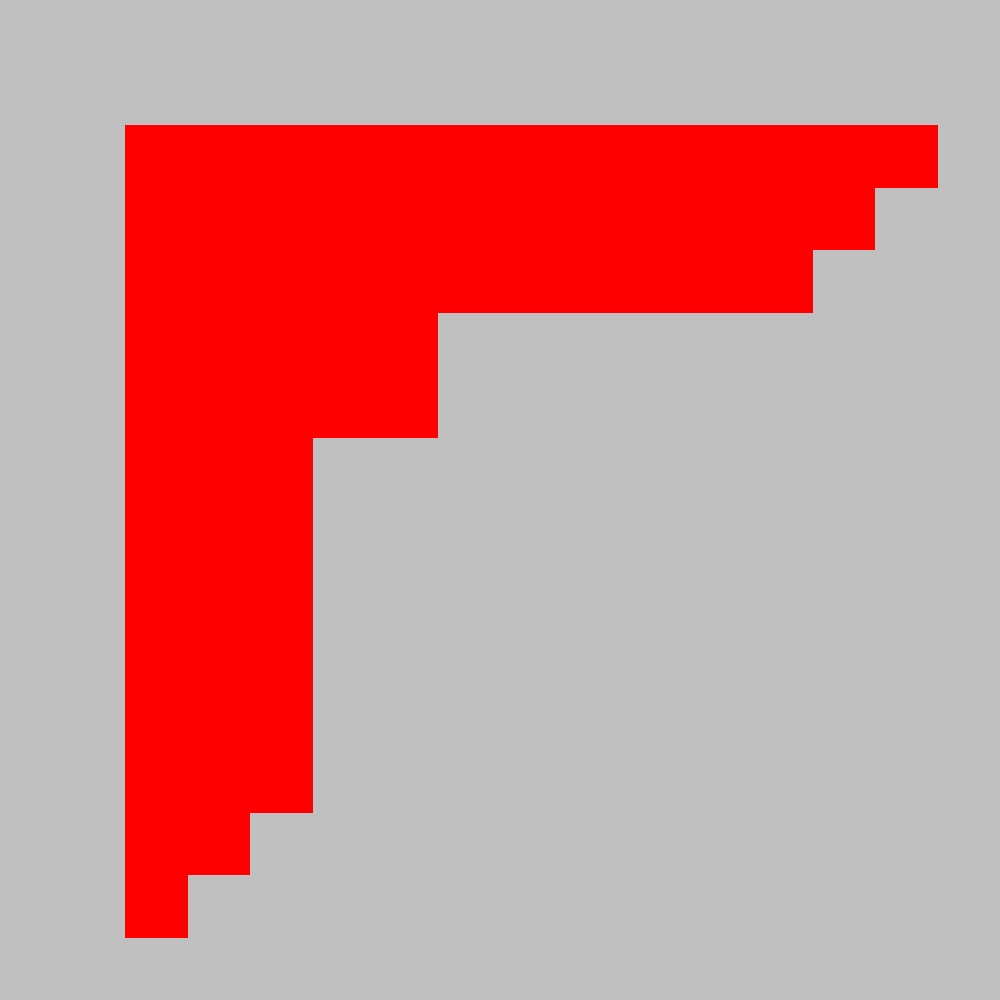}}}
&
\rotatebox{180}{\reflectbox{\includegraphics[scale=0.06]{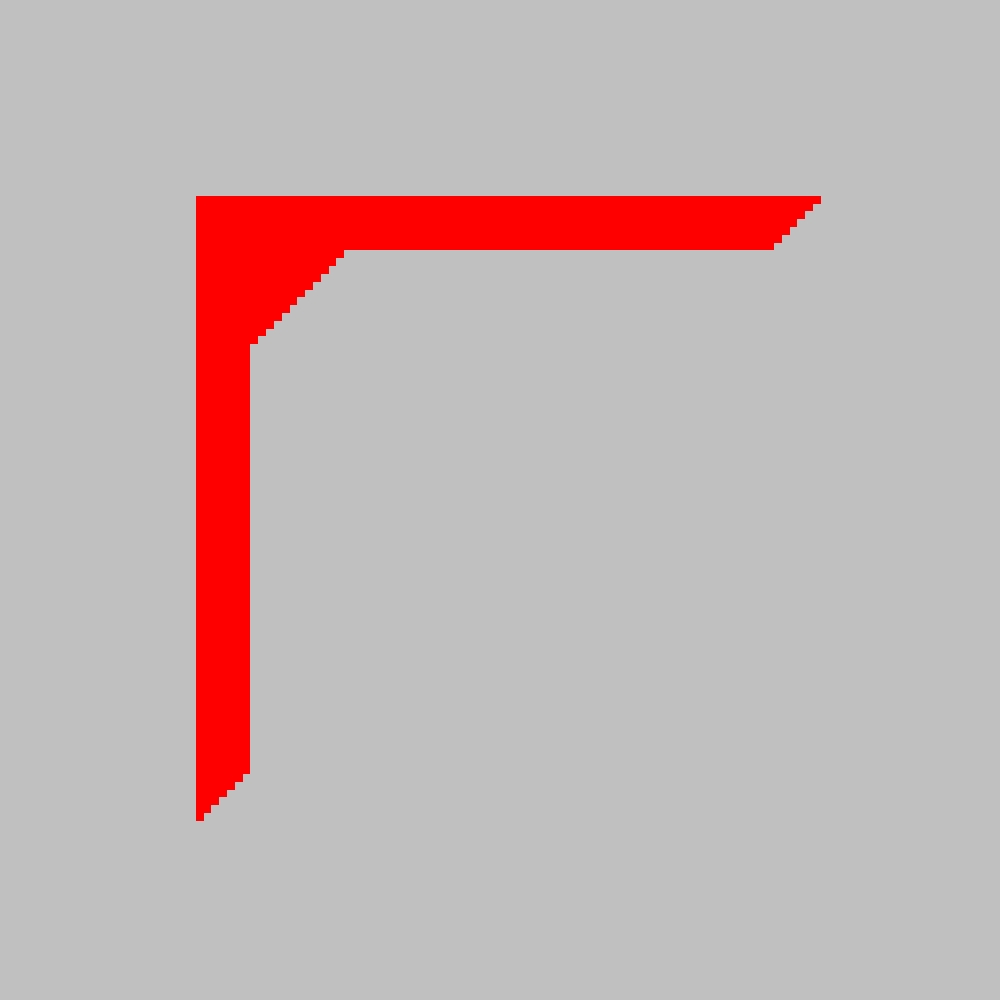}}}\\
  2 & 4 & 6 & 8 & 10 & 20 \\
  \multicolumn{6}{c}{(b)~$\gamma=0.3$, $\epsilon=0.01$}
  \\
\rotatebox{180}{\reflectbox{\includegraphics[scale=0.06]{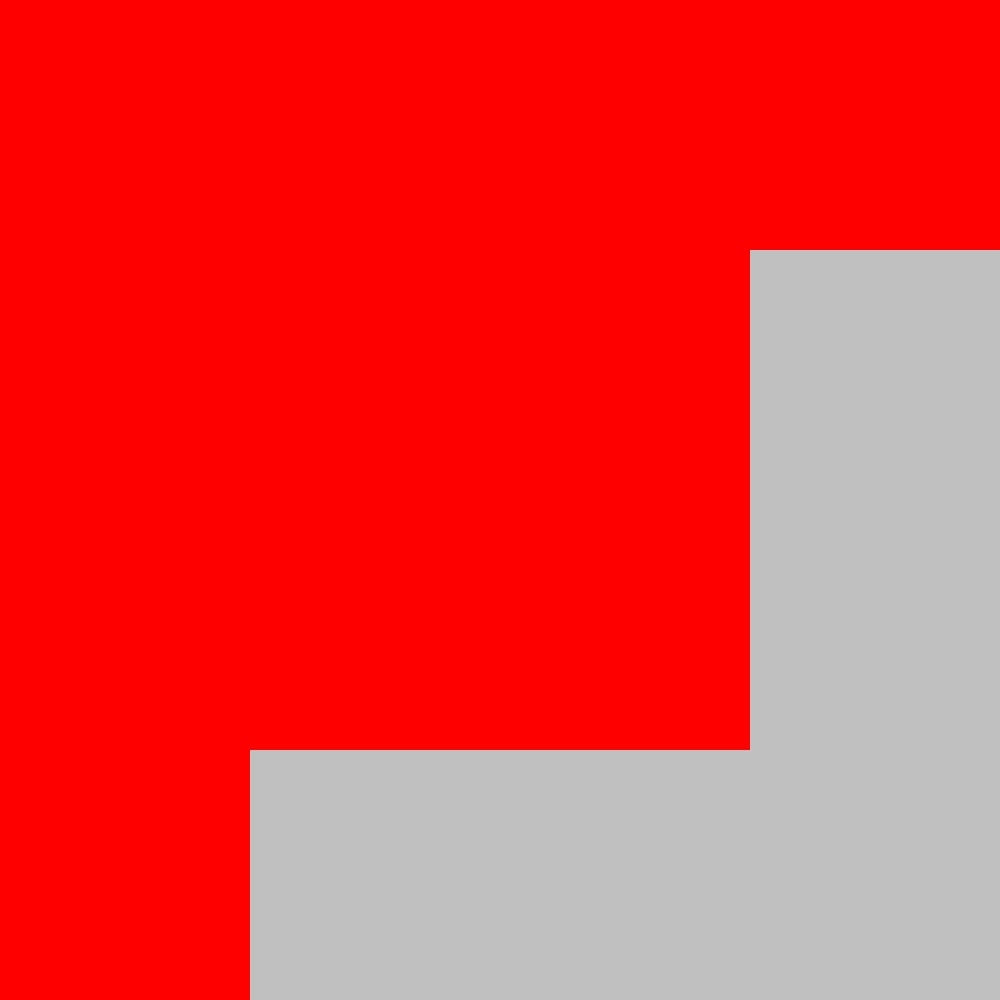}}}
&
\rotatebox{180}{\reflectbox{\includegraphics[scale=0.06]{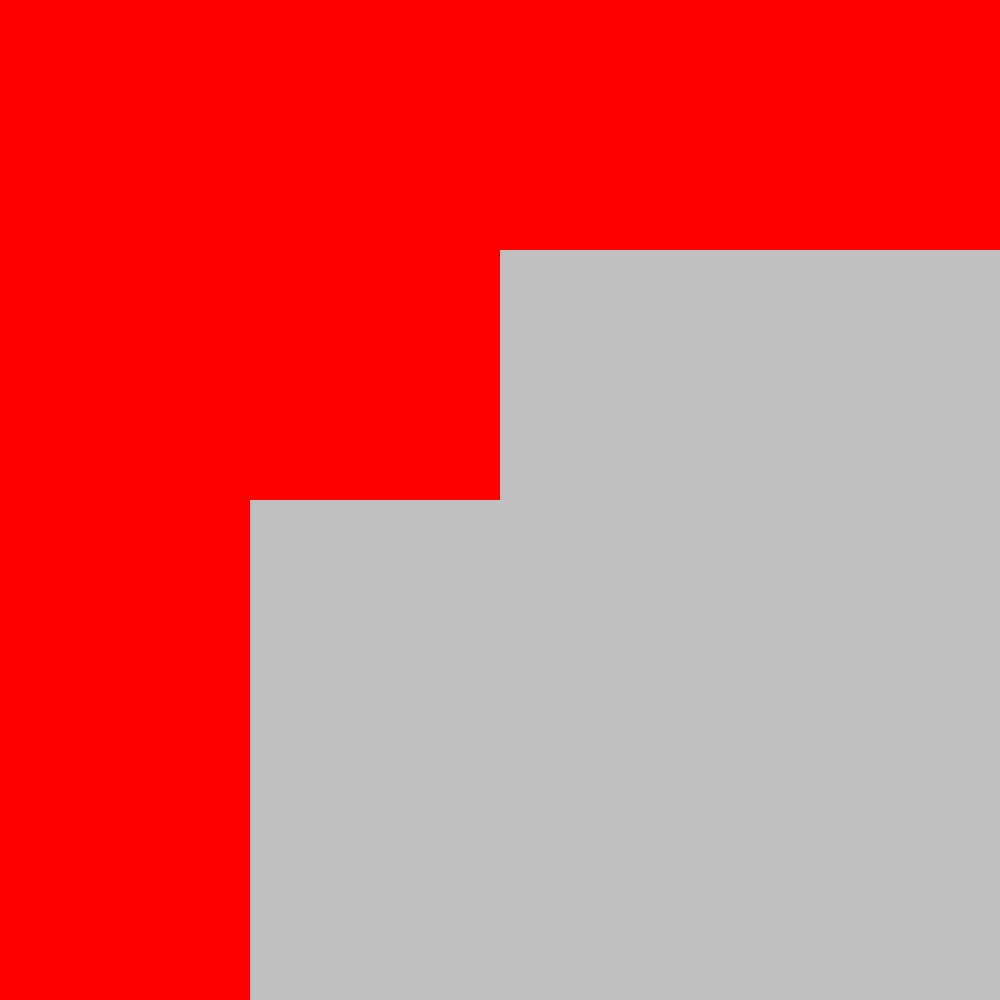}}}
&
\rotatebox{180}{\reflectbox{\includegraphics[scale=0.06]{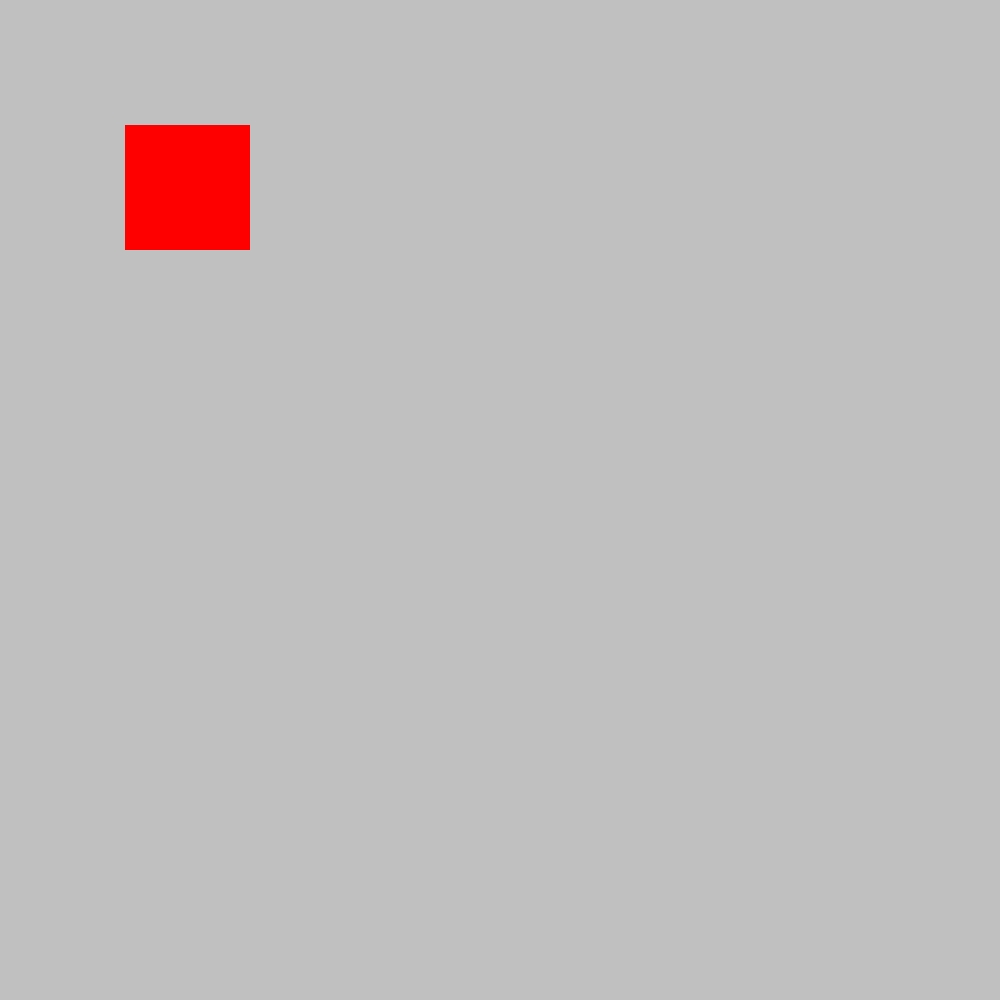}}}
&
\rotatebox{180}{\reflectbox{\includegraphics[scale=0.06]{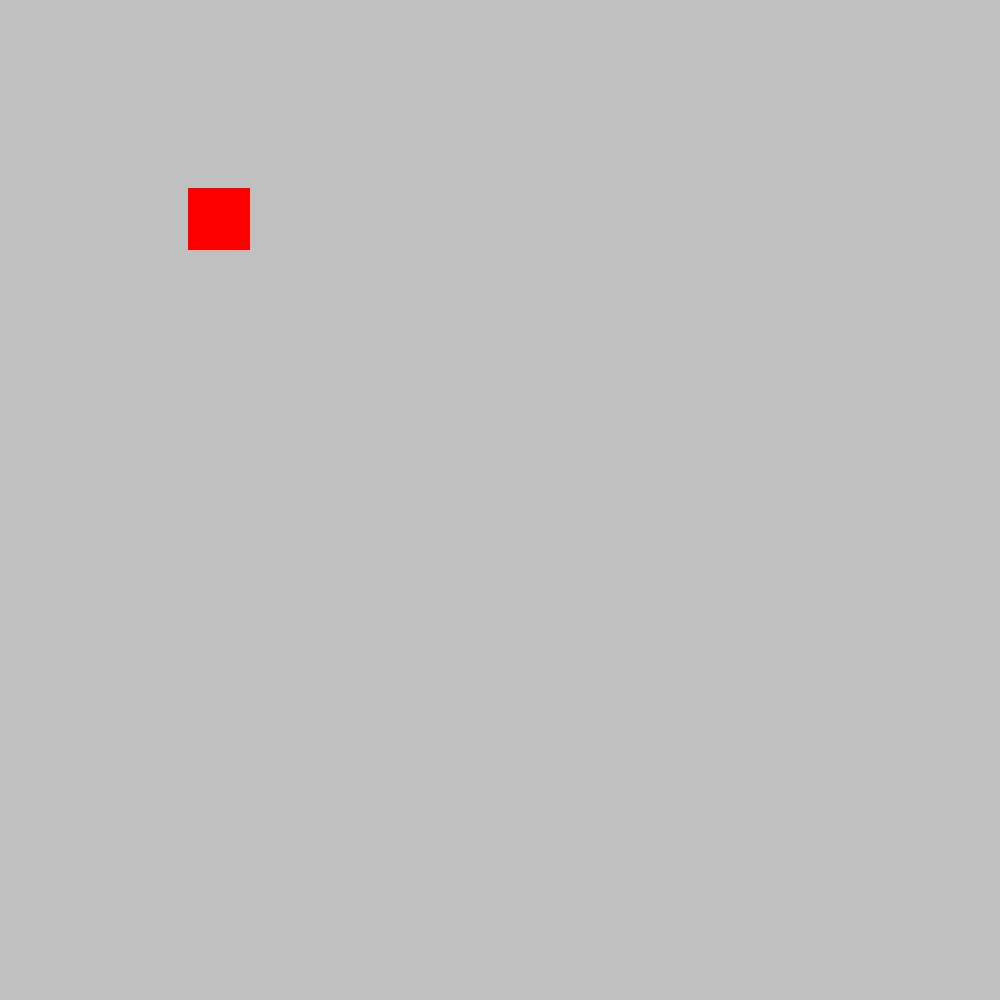}}}
&
\rotatebox{180}{\reflectbox{\includegraphics[scale=0.06]{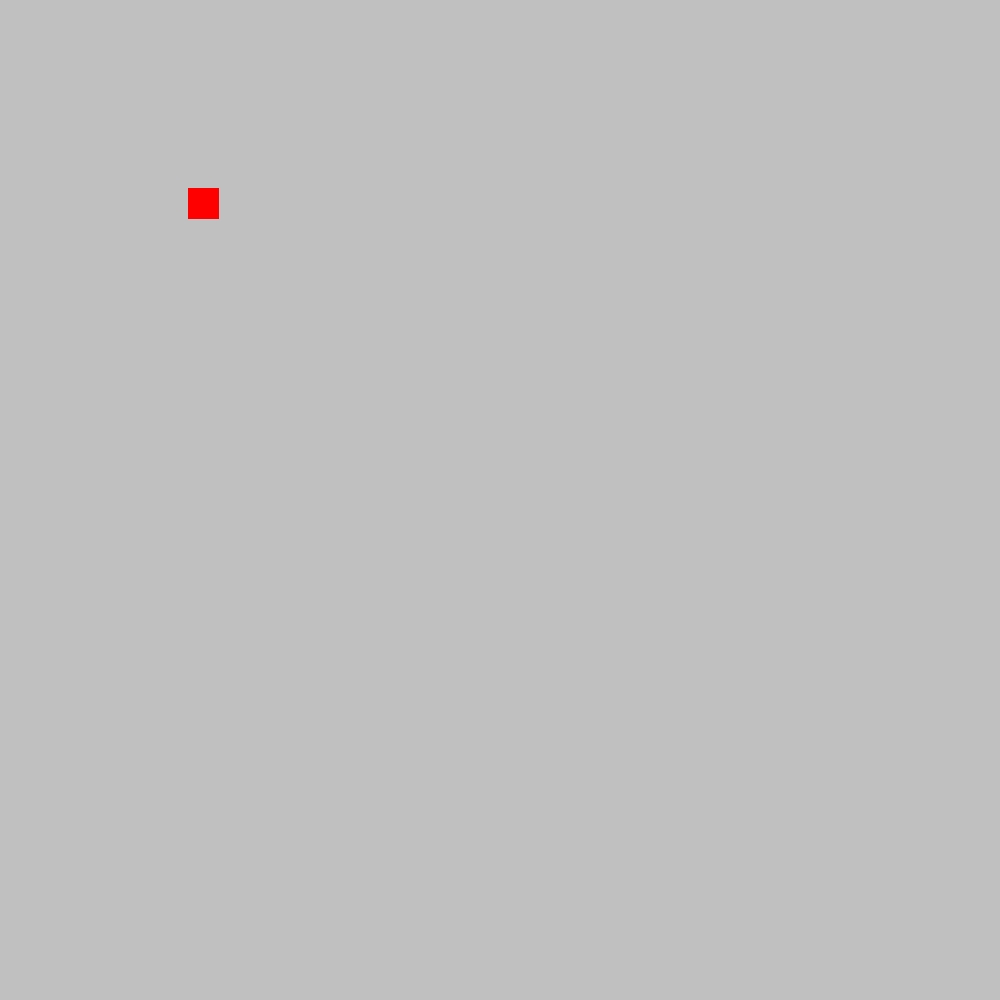}}}
&
\rotatebox{180}{\reflectbox{\includegraphics[scale=0.06]{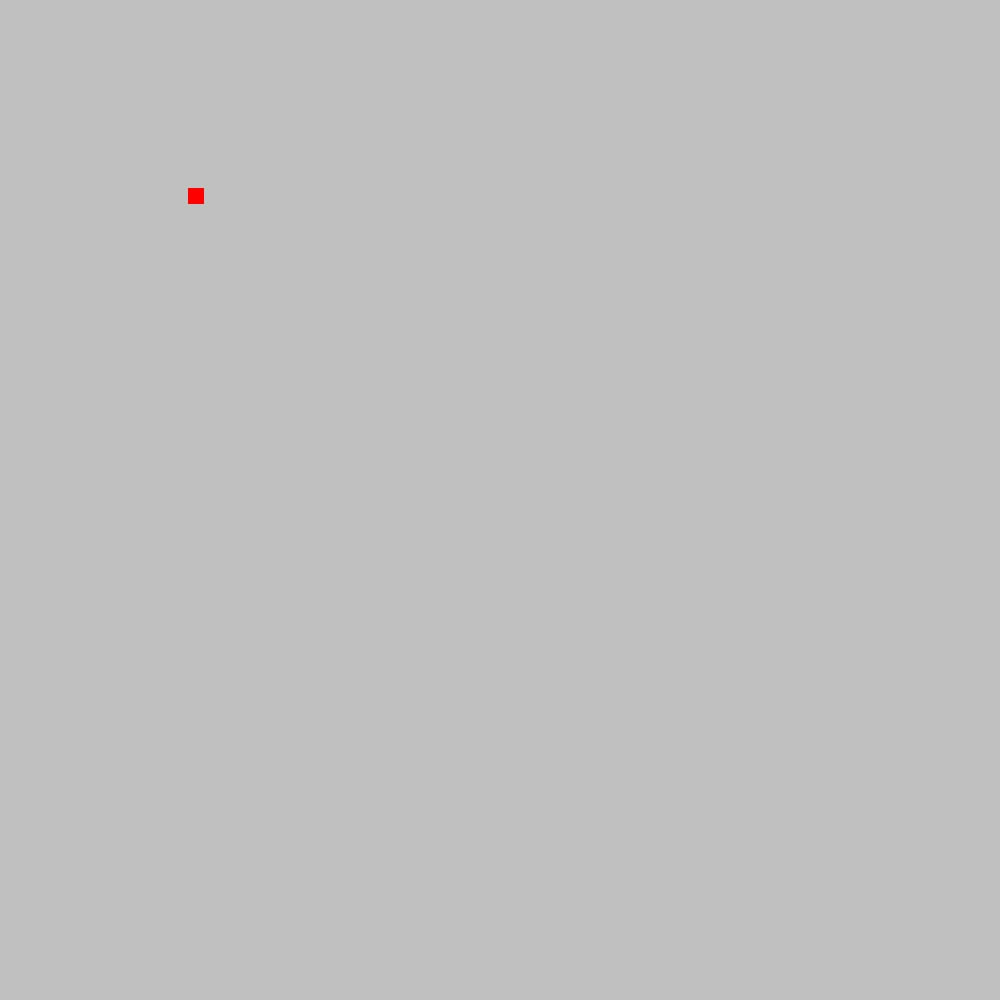}}}\\
  2 & 3 & 5 & 7 & 9 & 12 \\
  \multicolumn{6}{c}{(c)~$\gamma=0.05$, $\epsilon=0.01$}
\end{tabular}
  \caption{The evolution of the set of SPE payoff profiles computed by Algorithm~\ref{alg:basic} for mixed strategies with public correlation in the repeated Prisoner's Dilemma. The numbers under the graphs reflect the algorithm's iterations. The red (darker) regions denote the hypercubes that remain in the set~$C$ by the end of the corresponding iteration.}
  \label{fig:Results_PD_SPE}
\end{figure}

The graphs in Figure~\ref{fig:Results_PD_SPE} reflect, for three
different values of the discount factor, the evolution of the set of
SPE payoff profiles computed by Algorithm~\ref{alg:basic} for the
case of mixed strategies with public correlation in the repeated
Prisoner's Dilemma. Here and below, the vertical and the horizontal
axes of each graph correspond respectively to the payoffs of the
first and the second players. The upper and lower limits of each
axis are given respectively by $\bar{r}$ and $\underline{r}$. The
numbers under the graphs reflect the algorithm's iterations. The red
(darker) regions on a graph reflect the hypercubes that remain in
the set~$C$ by the end of the corresponding iteration. One can see
in Figure~\ref{fig:Results_PD_SPE}a that when $\gamma$ is
sufficiently large, the algorithm maintains a set that converges
towards the set $F^*$ of feasible and individually rational payoff
profiles, the largest possible set of SPE payoff profiles. On the
other hand, in Figure~\ref{fig:Results_PD_SPE}c, one can see that
when $\gamma$ is close enough to~$0$ the set of SPE payoff profiles
converges, as expected, towards the point~$(0,0)$ that corresponds
to the Nash equilibrium of the stage-game: a strategy profile that
prescribes playing~$D$ at every repeated game period.

\begin{figure}
\center
\begin{tabular}{cccccc}
\rotatebox{180}{\reflectbox{\includegraphics[scale=0.06]{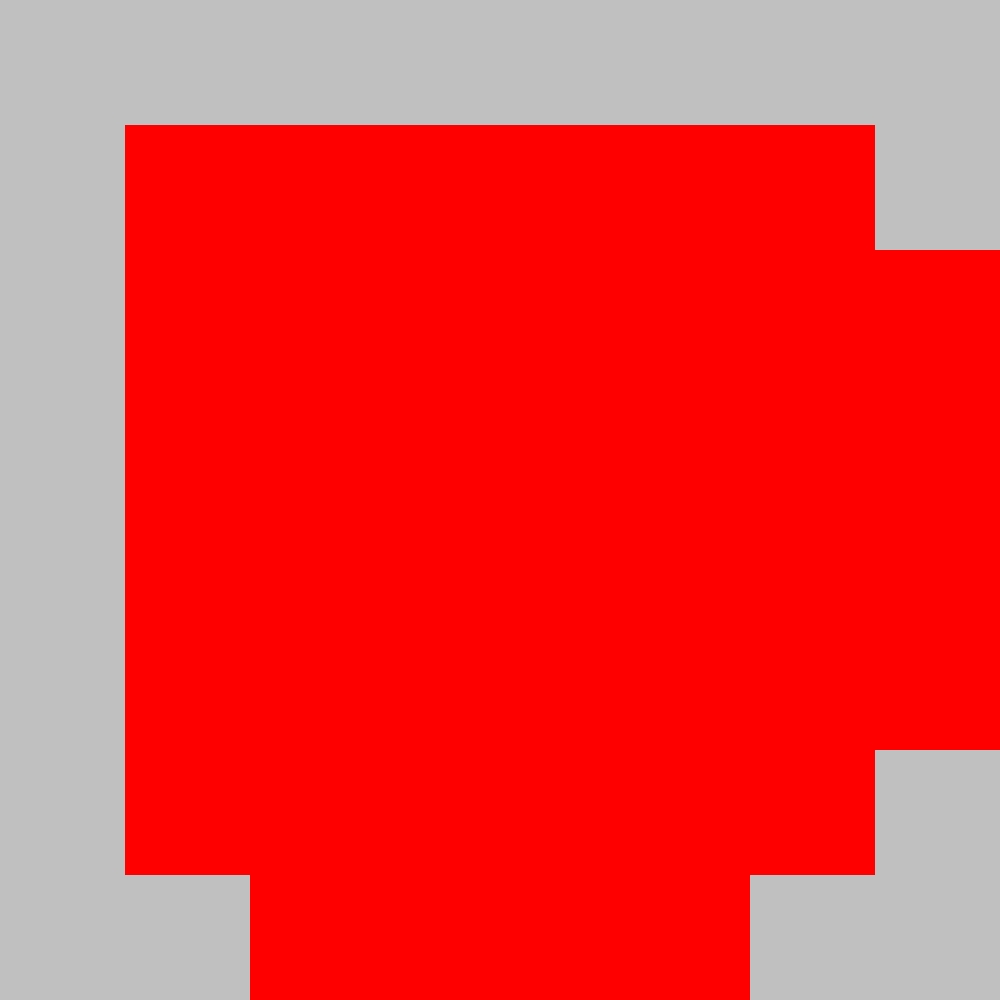}}}
&
\rotatebox{180}{\reflectbox{\includegraphics[scale=0.06]{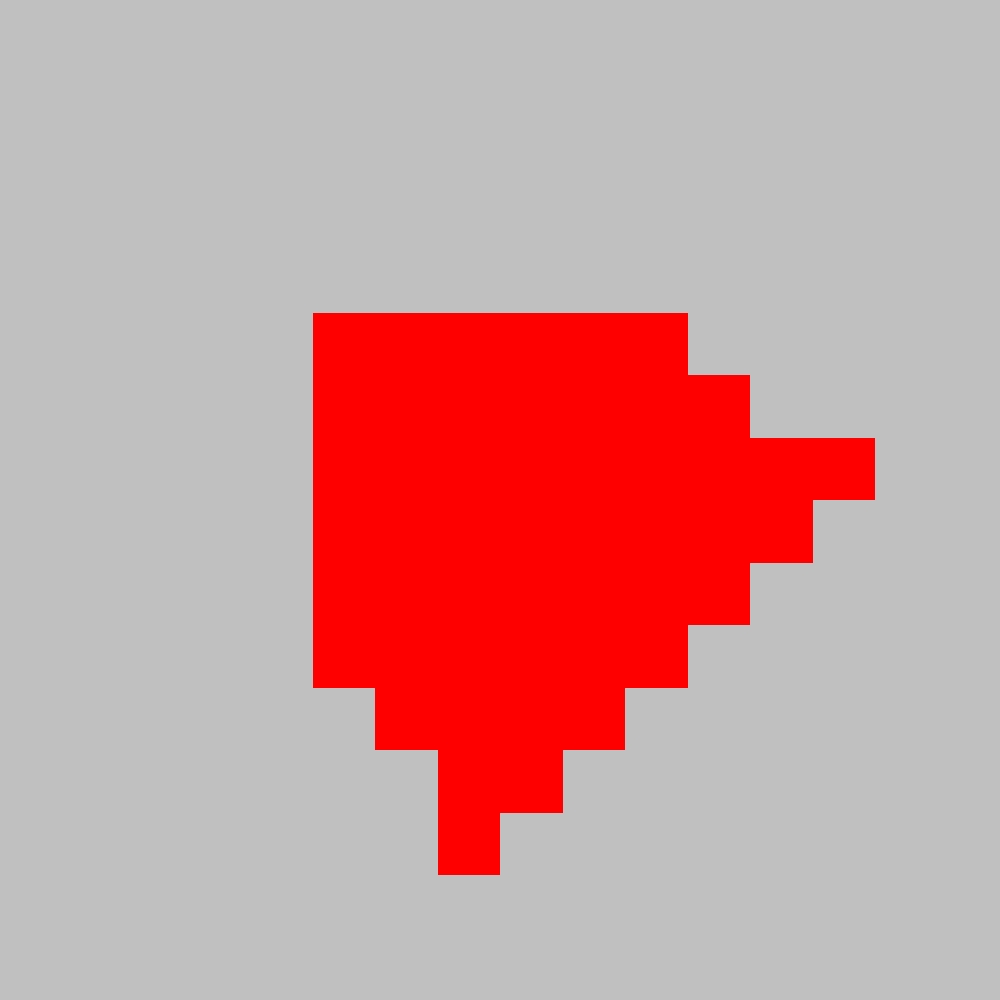}}}
&
\rotatebox{180}{\reflectbox{\includegraphics[scale=0.06]{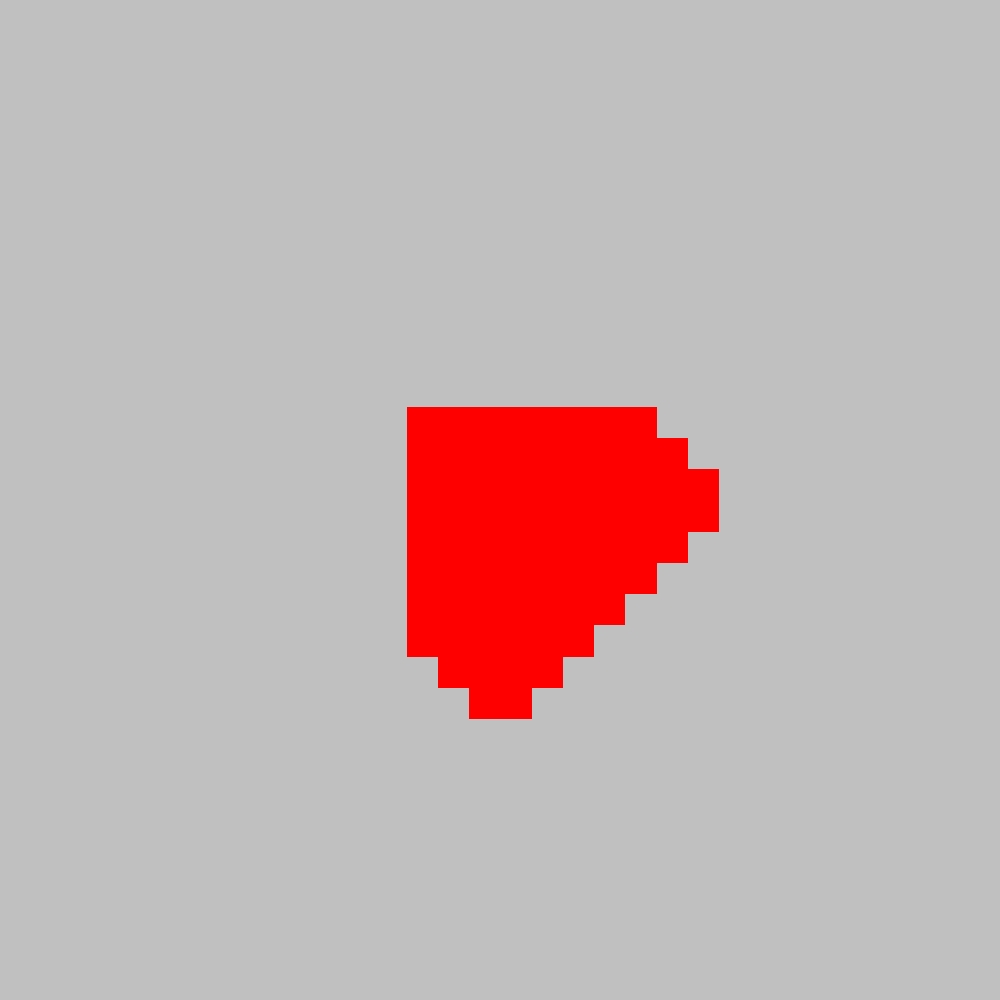}}}
&
\rotatebox{180}{\reflectbox{\includegraphics[scale=0.06]{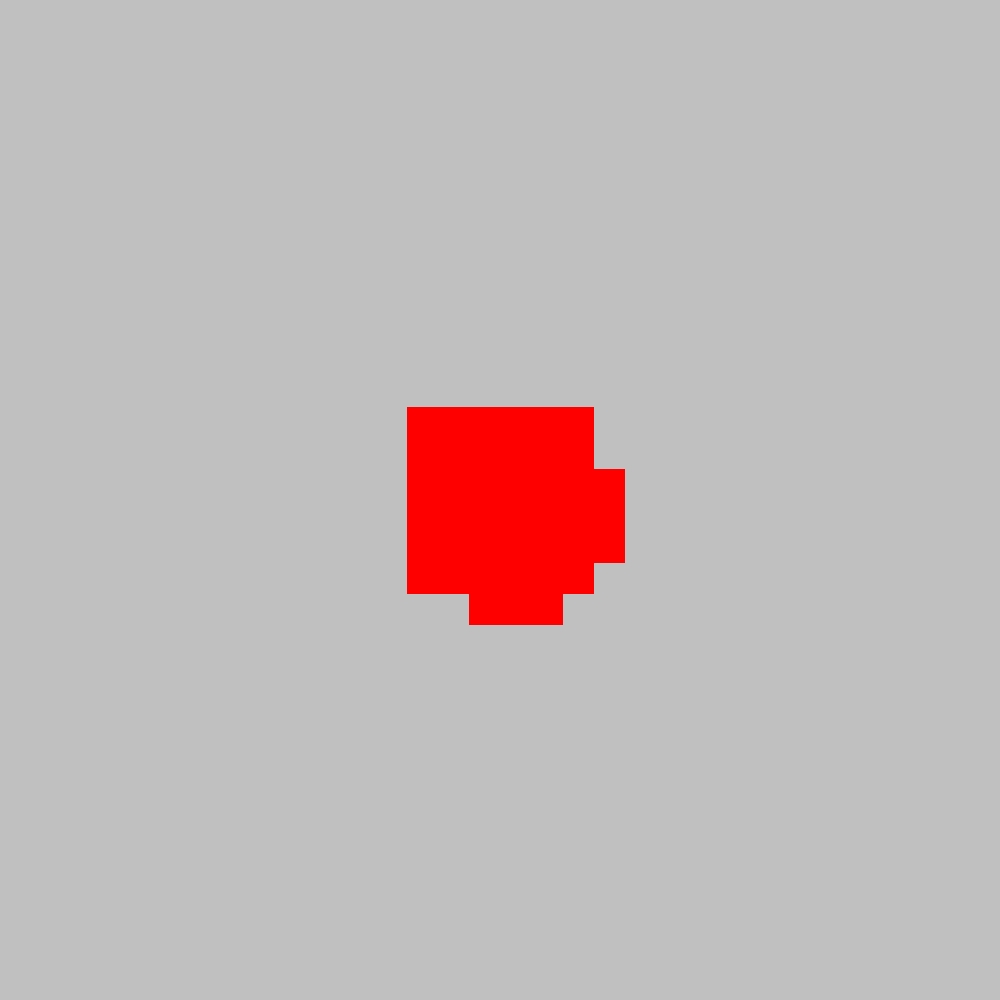}}}
&
\rotatebox{180}{\reflectbox{\includegraphics[scale=0.06]{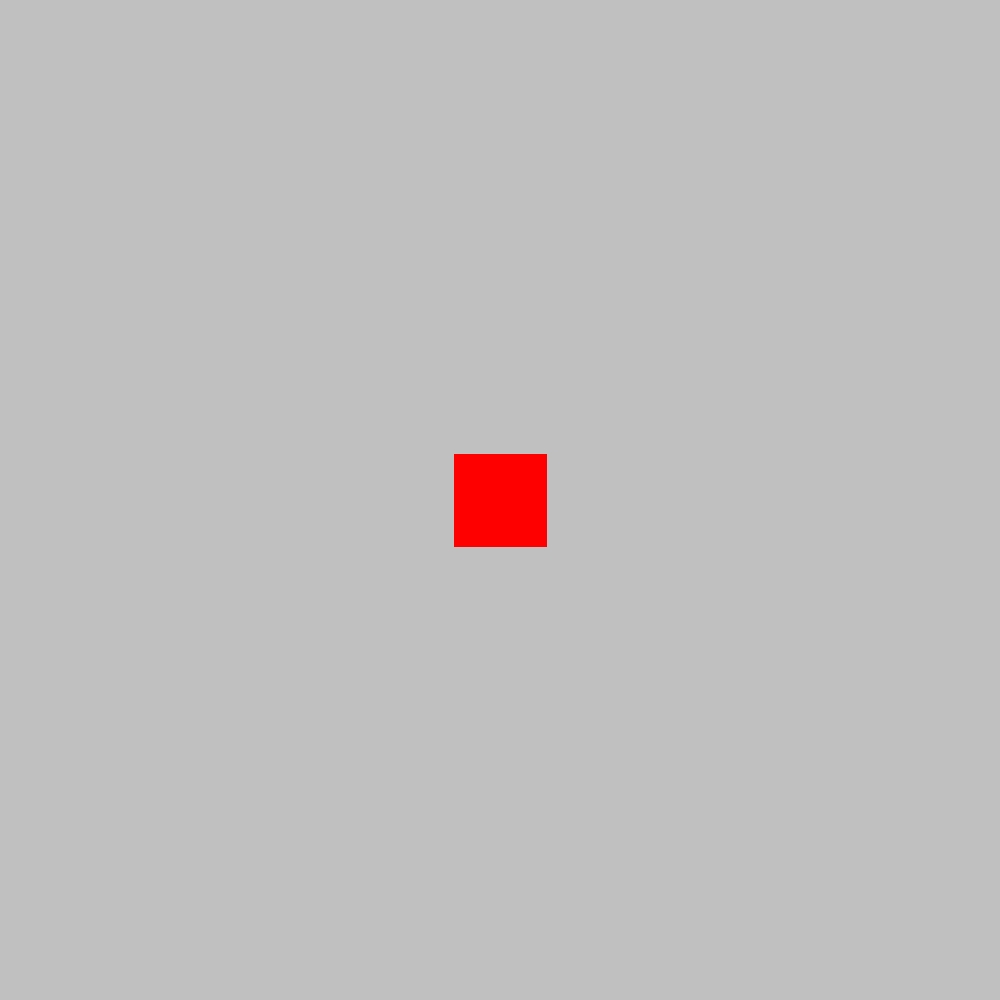}}}
&
\rotatebox{180}{\reflectbox{\includegraphics[scale=0.06]{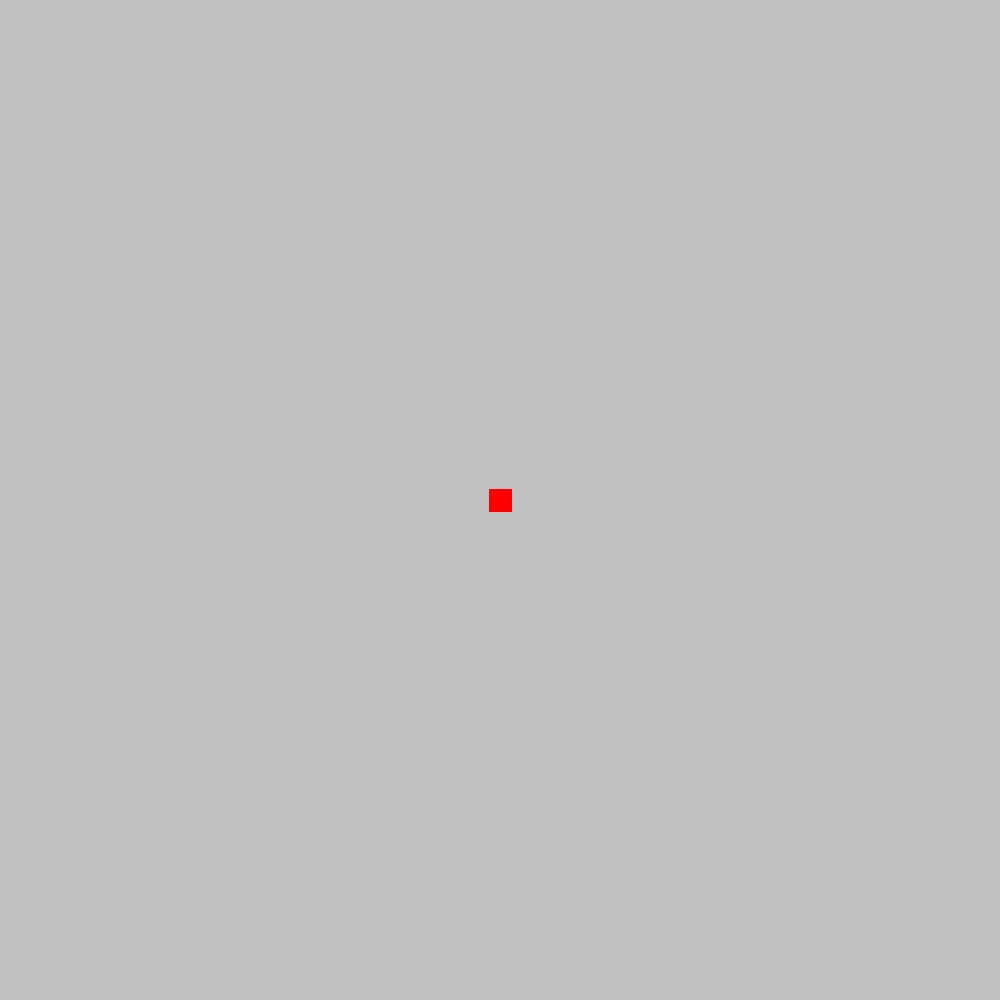}}}\\
  5 & 10 & 15 & 20 & 25 & 33
\end{tabular}
  \caption{The evolution of the set of SPE payoff profiles computed by Algorithm~\ref{alg:basic} for mixed action with public correlation in the repeated Rock, Paper, Scissors with $\gamma=0.7$ and $\epsilon=0.01$.}
  \label{fig:Results_RPC_SPE}
\end{figure}

Rock, Paper, Scissors (RPC) is a symmetrical zero-sum game. In the
repeated RPC game, the point $(0,0)$ is the only possible SPE payoff
profile, regardless of the discount factor. This payoff profile can
be realized by a stationary strategy profile prescribing to each
player to sample actions from a uniform distribution. The graphs in
Figure~\ref{fig:Results_RPC_SPE} certify the correctness of
Algorithm~\ref{alg:basic} in this case.

\begin{figure}
\center
\begin{tabular}{cccccc}
\\
\rotatebox{180}{\reflectbox{\includegraphics[scale=0.06]{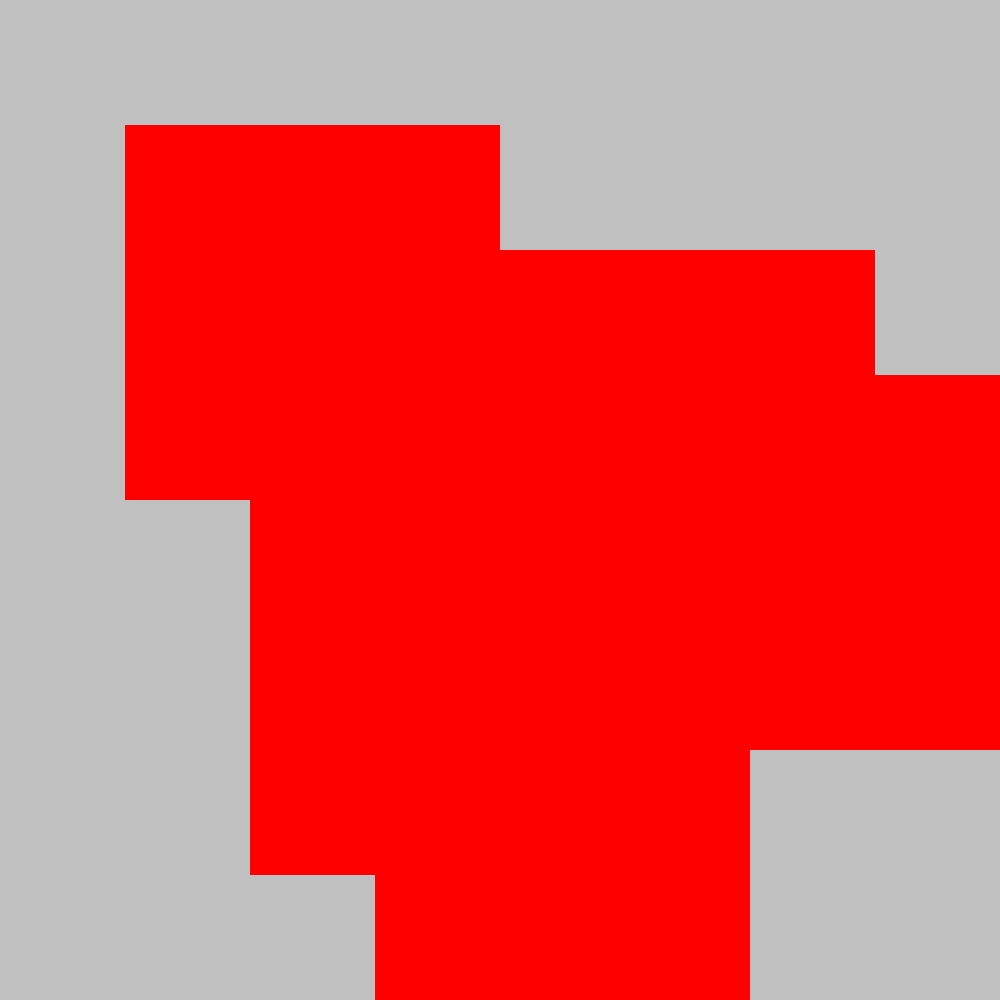}}}
&
\rotatebox{180}{\reflectbox{\includegraphics[scale=0.06]{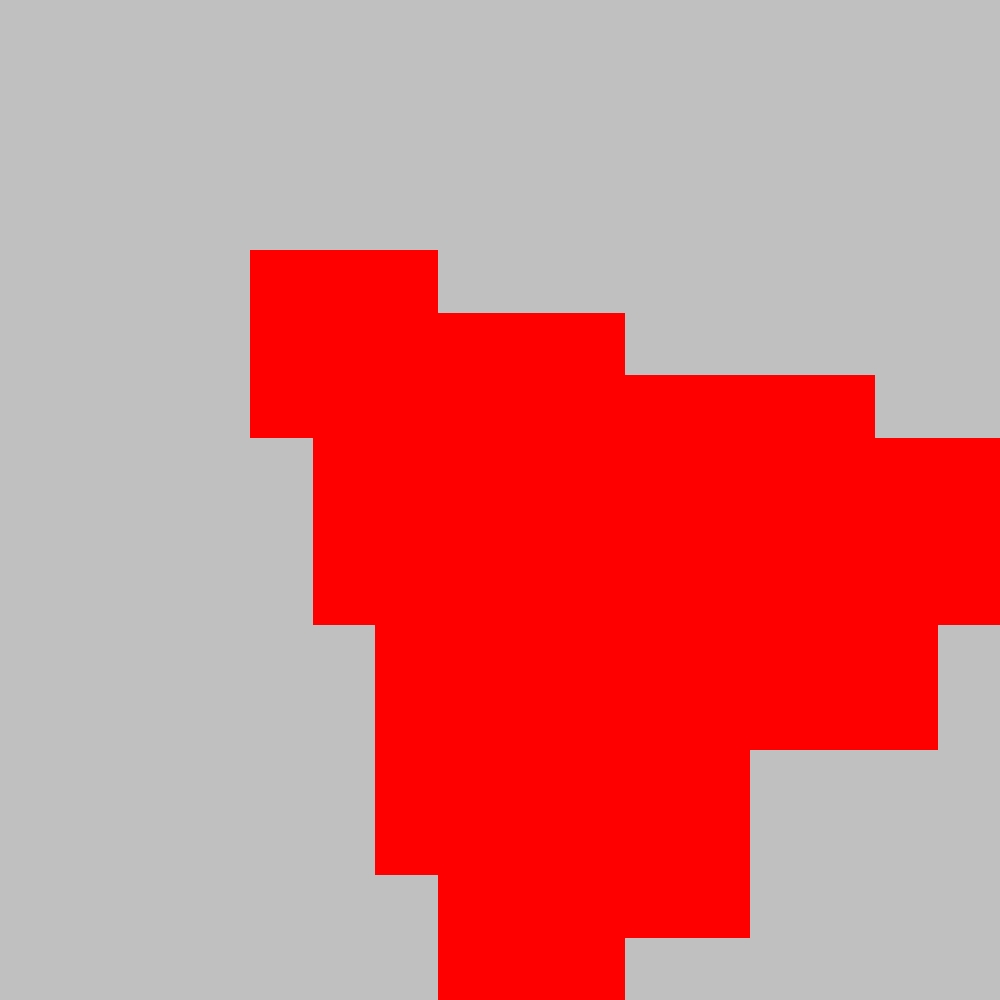}}}
&
\rotatebox{180}{\reflectbox{\includegraphics[scale=0.06]{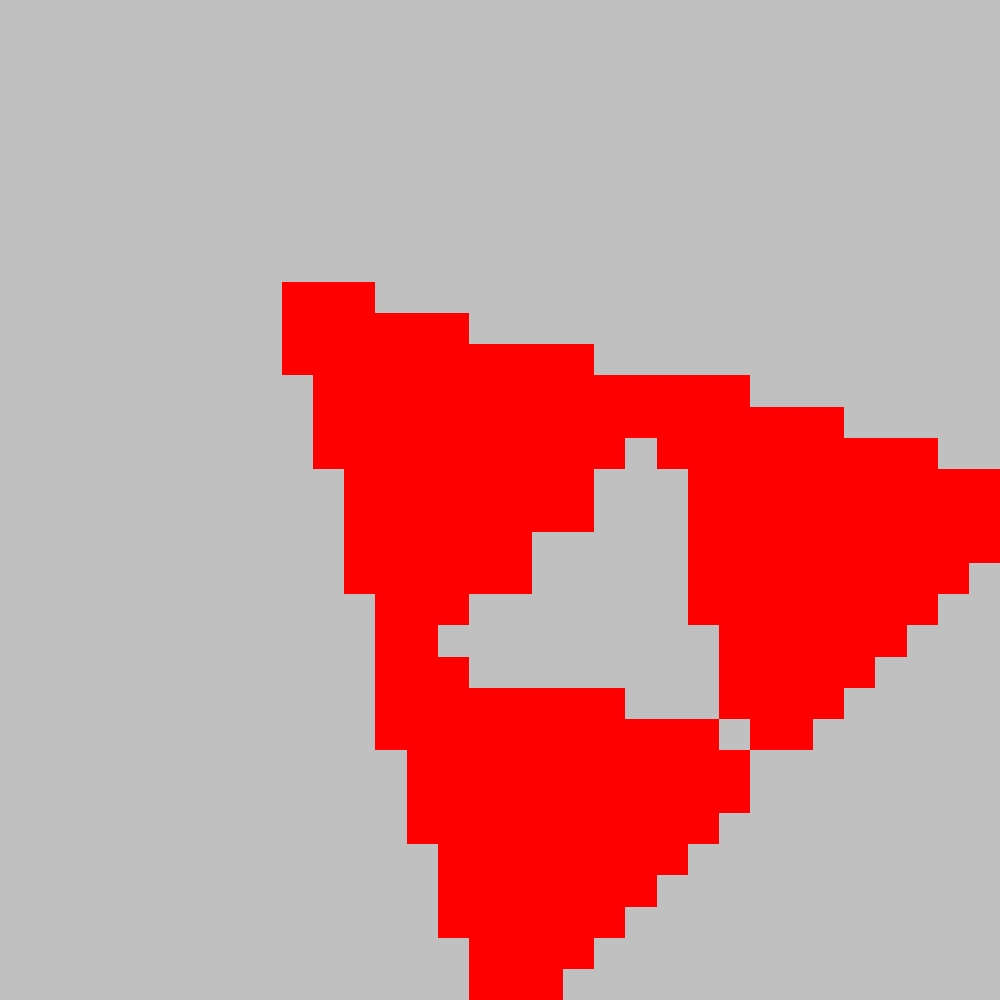}}}
&
\rotatebox{180}{\reflectbox{\includegraphics[scale=0.06]{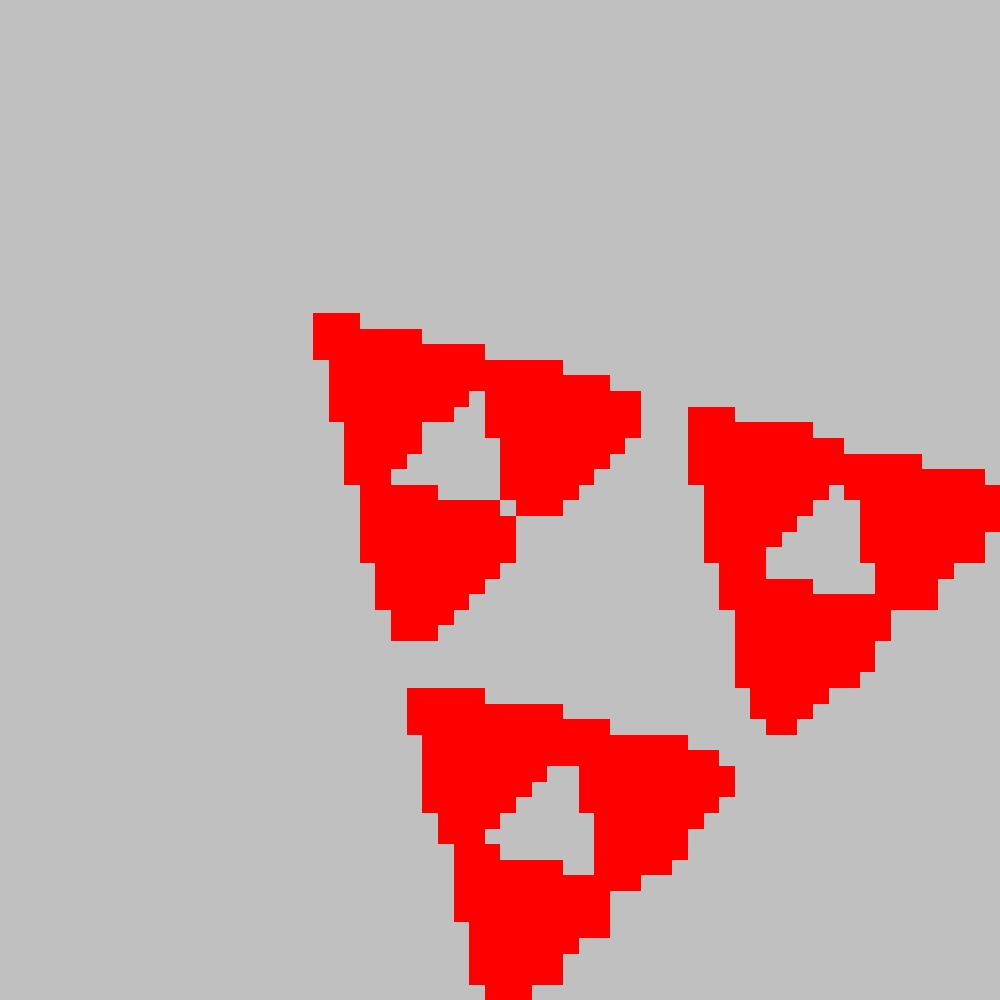}}}
&
\rotatebox{180}{\reflectbox{\includegraphics[scale=0.06]{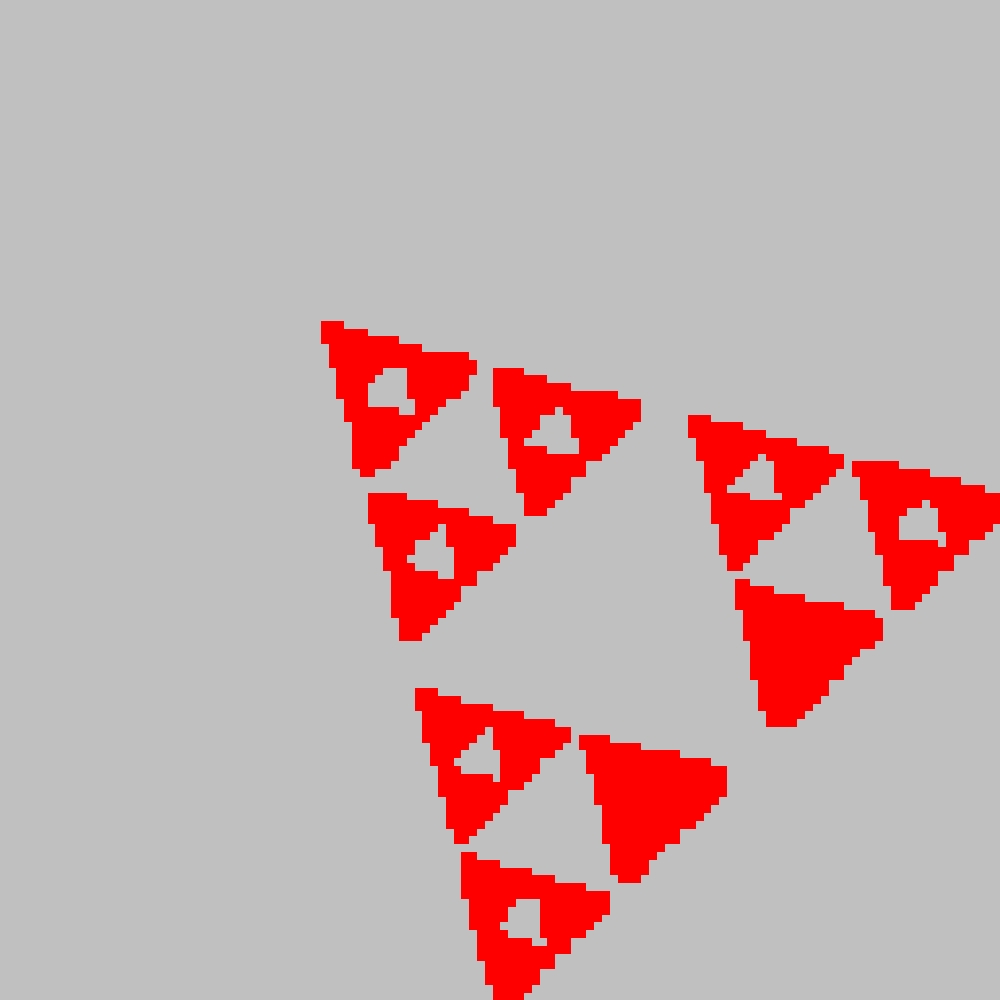}}}
&
\rotatebox{180}{\reflectbox{\includegraphics[scale=0.06]{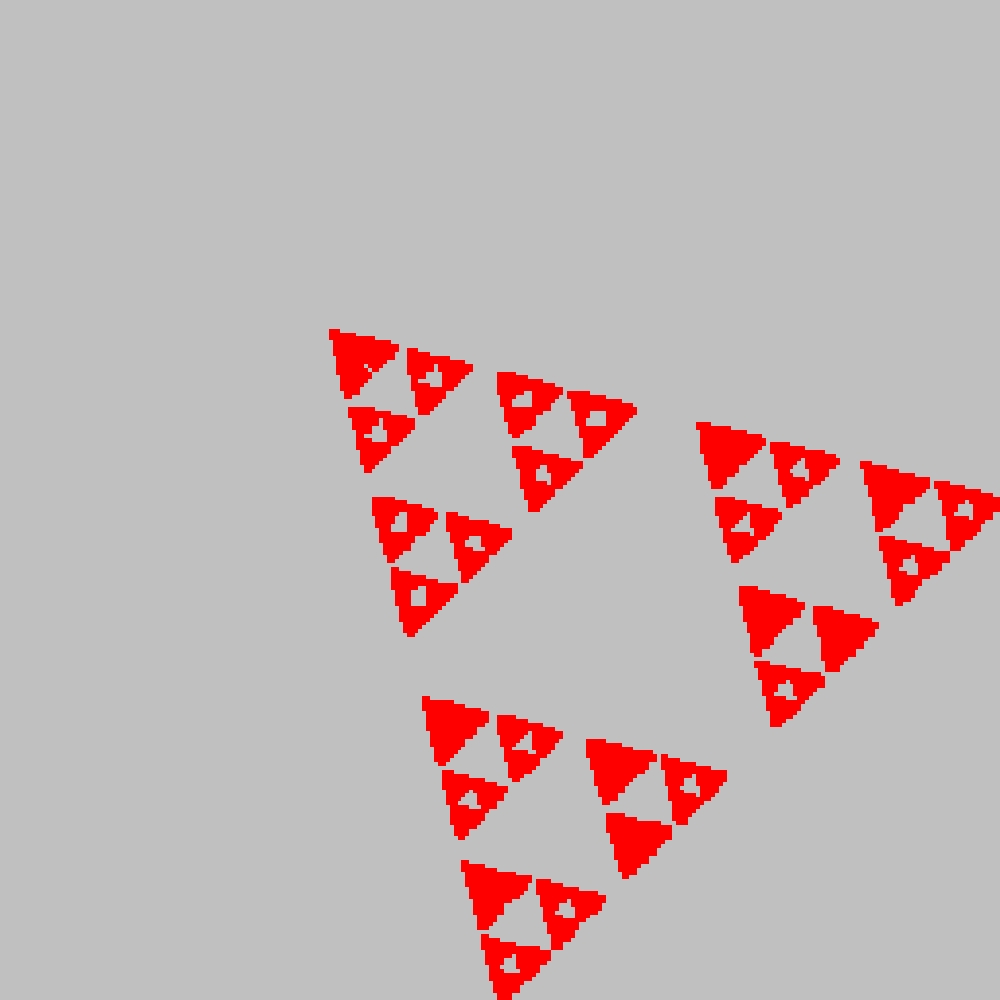}}}\\
  5 & 10 & 15 & 20 & 30 & 40 \\
  \multicolumn{6}{c}{(a) $\gamma=0.45$, $\epsilon=0.01$}
\\
\rotatebox{180}{\reflectbox{\includegraphics[scale=0.06]{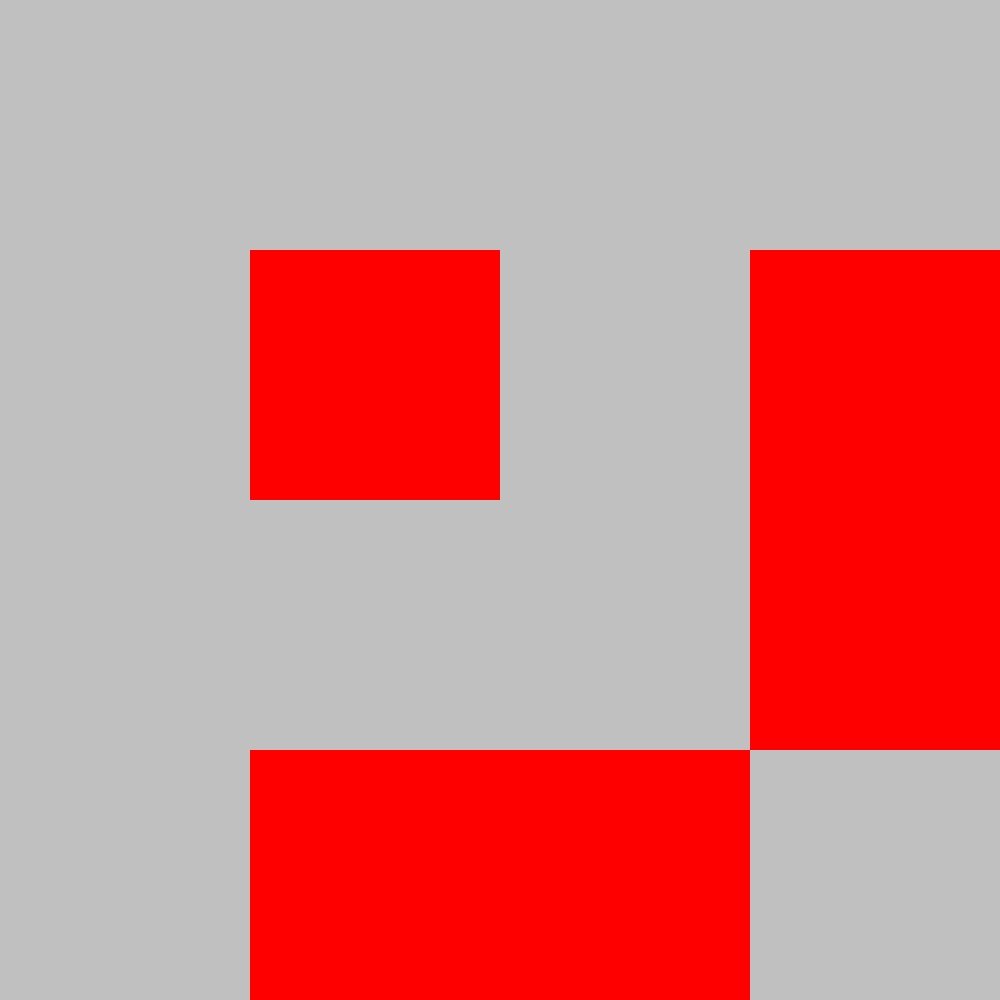}}}
&
\rotatebox{180}{\reflectbox{\includegraphics[scale=0.06]{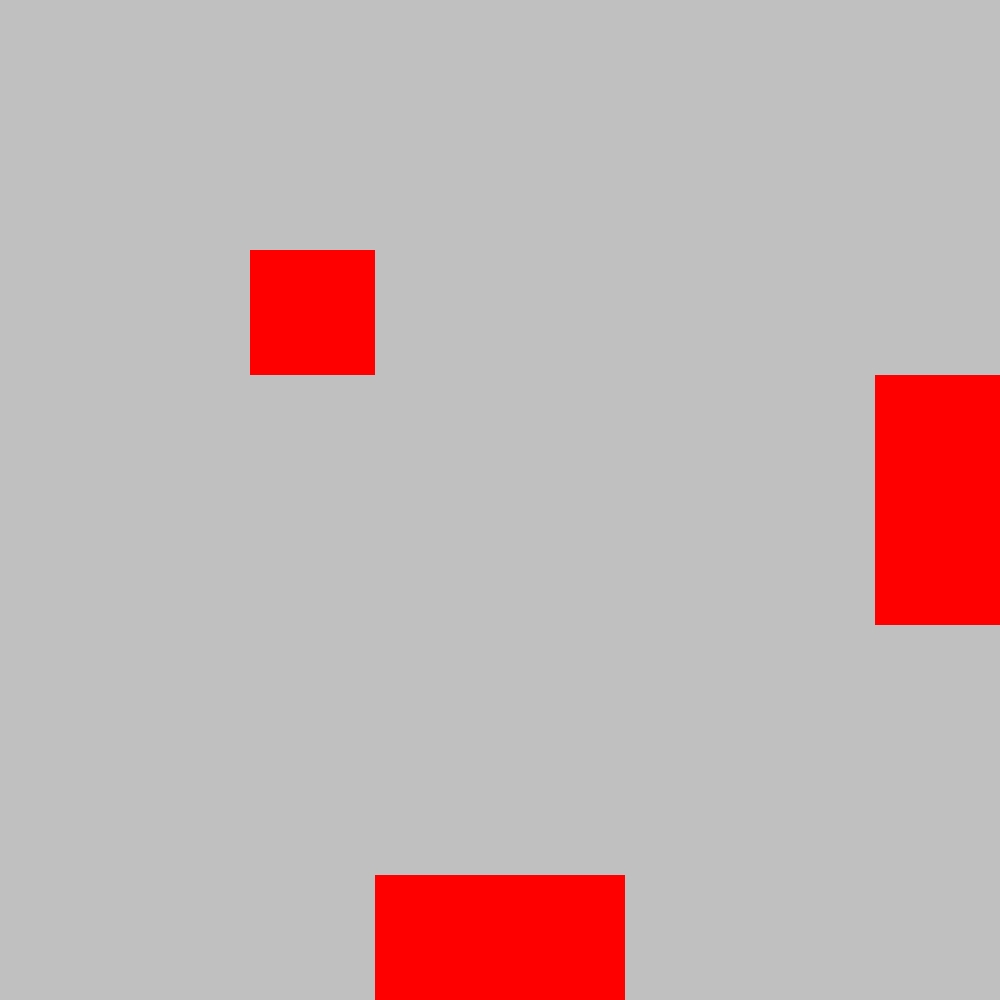}}}
&
\rotatebox{180}{\reflectbox{\includegraphics[scale=0.06]{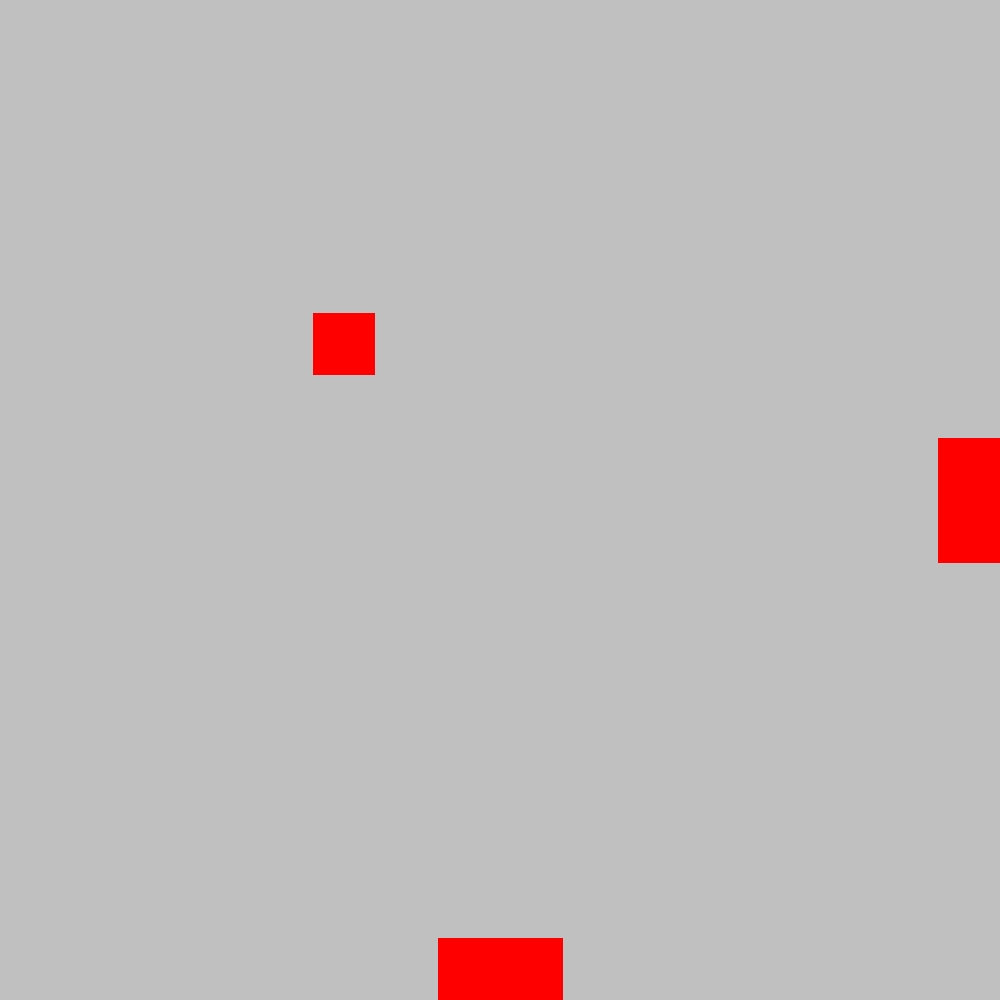}}}
&
\rotatebox{180}{\reflectbox{\includegraphics[scale=0.06]{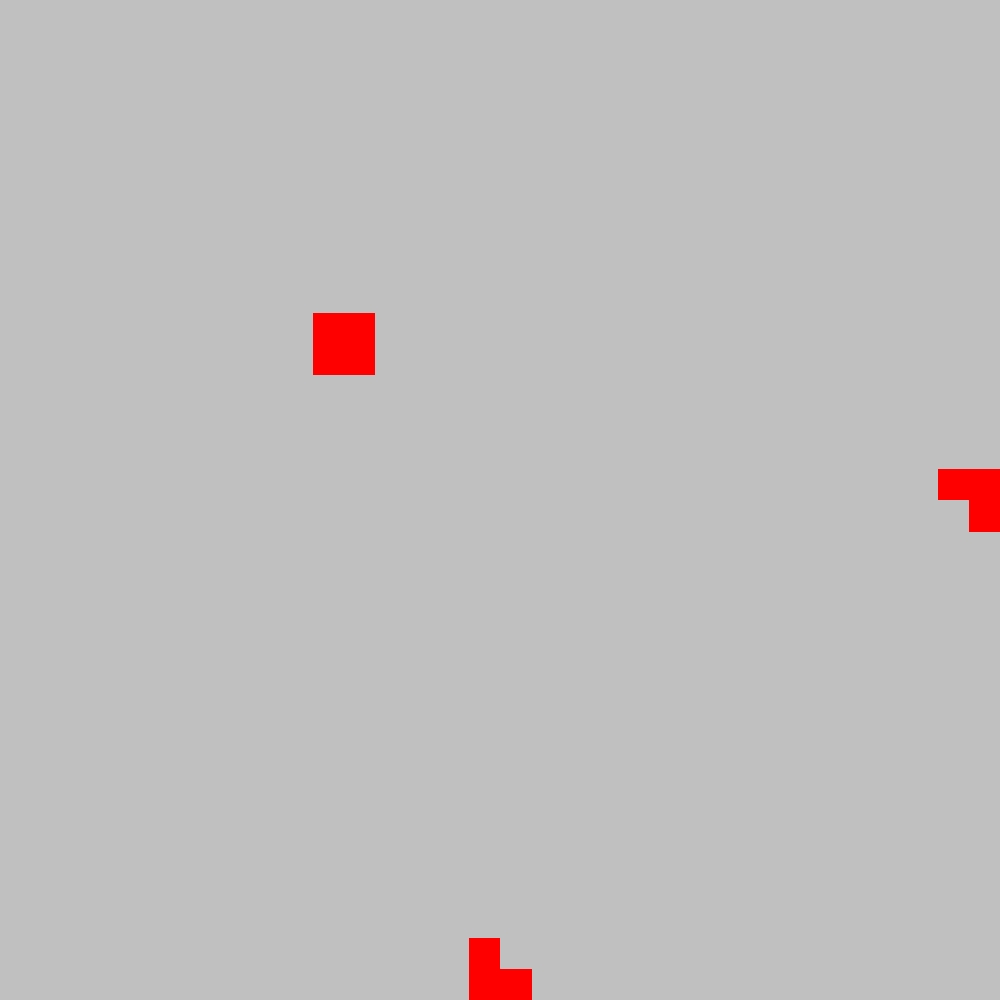}}}
&
\rotatebox{180}{\reflectbox{\includegraphics[scale=0.06]{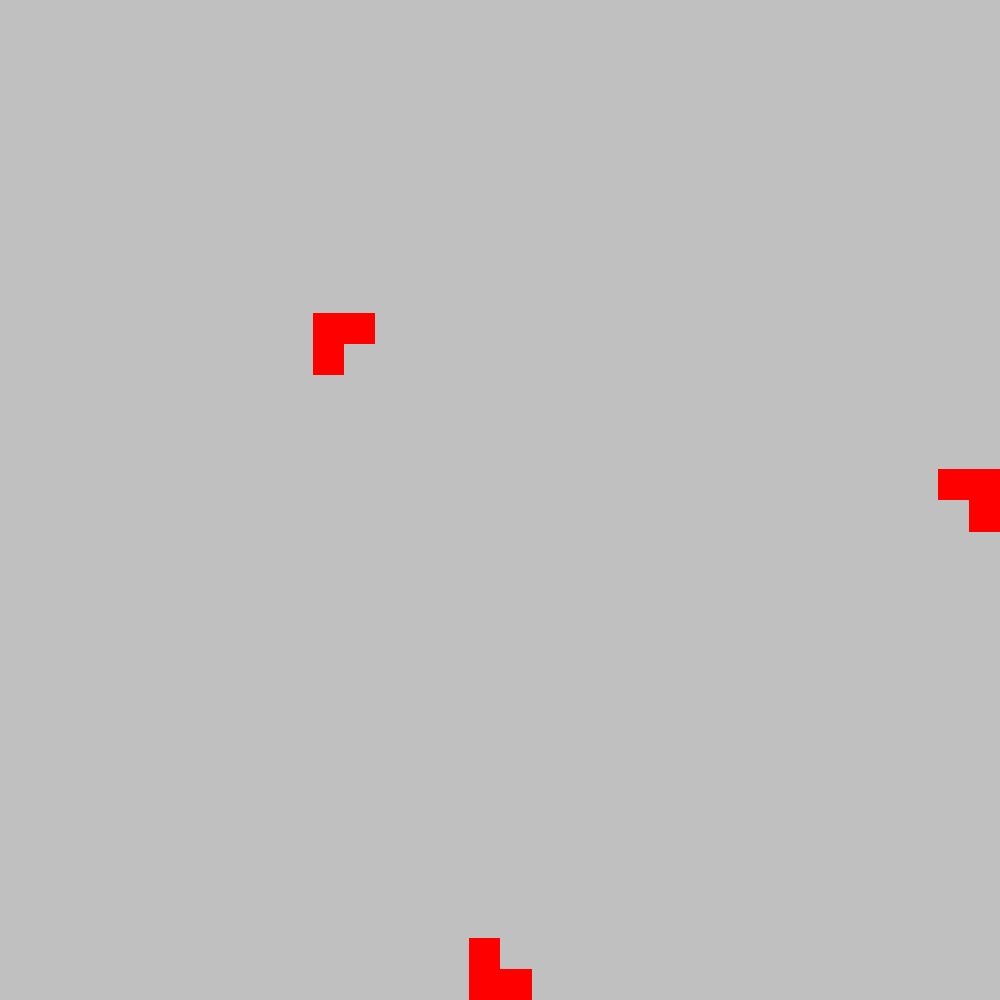}}}
&
\rotatebox{180}{\reflectbox{\includegraphics[scale=0.06]{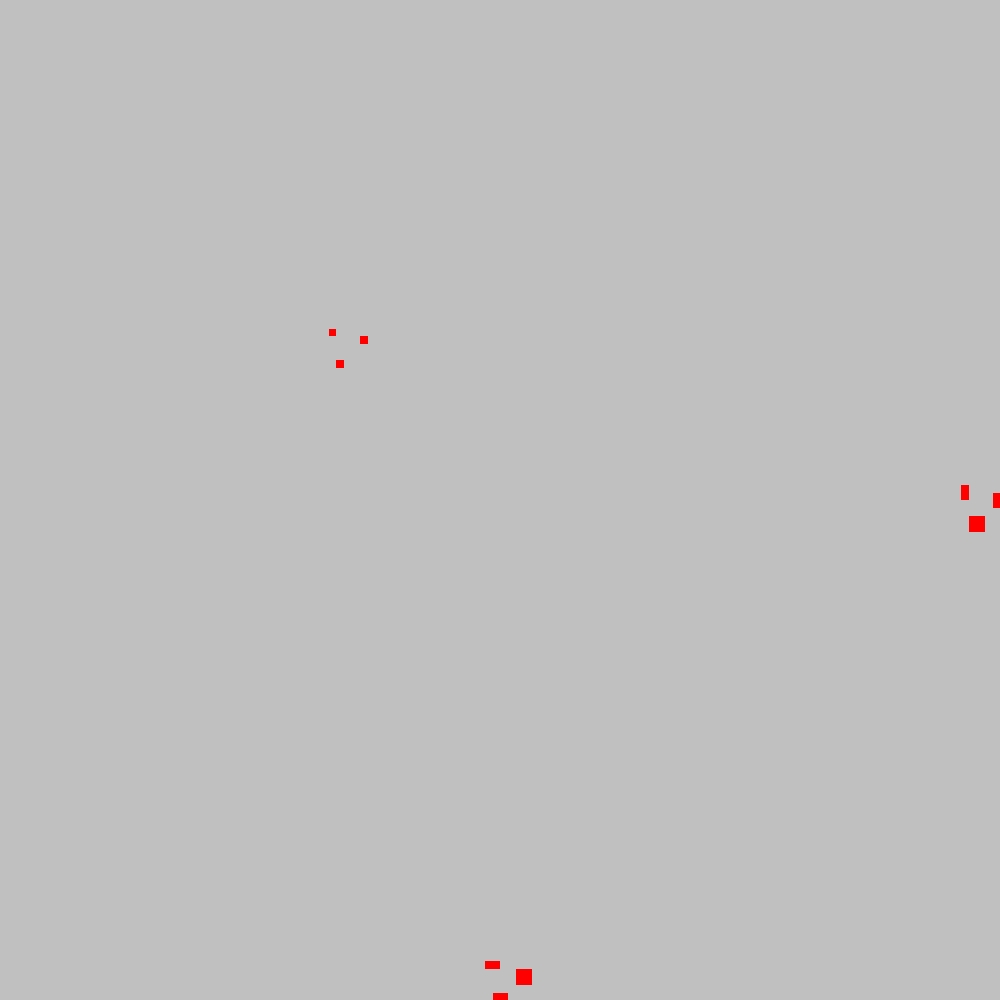}}}\\
  2 & 4 & 6 & 8 & 10 & 15 \\
  \multicolumn{6}{c}{(b) $\gamma=0.05$, $\epsilon=0.01$}
\end{tabular}
  \caption{The evolution of the set of SPE payoff profiles computed by Algorithm~\ref{alg:basic} for mixed actions without public correlation in the repeated Battle of the Sexes.}
  \label{fig:Results_BS_SPE}
\end{figure}

Battle of the Sexes (BoS) is the game that has two pure action
stage-game equilibria, $(O,O)$ and $(F,F)$, with payoff profiles
respectively $(1,2)$ and $(2,1)$. The game also has one mixed action
stage-game equilibrium with payoff profile $(2/3,2/3)$. When
$\gamma$ is sufficiently close to~$0$, the set of SPE payoff
profiles computed by Algorithm~\ref{alg:basic} converges towards
these three points (Figure~\ref{fig:Results_BS_SPE}b), which is the
expected behavior. As $\gamma$ grows, the set of SPE payoff profiles
becomes larger (Figure~\ref{fig:Results_BS_SPE}a). We also
ascertained that when the value of $\gamma$ becomes sufficiently
close to~$1$, the set of SPE payoff profiles converges towards $F^*$
and eventually includes the point $(3/2,3/2)$. The latter point is
interesting in that it maximizes the Nash
product~\citep{nashjr1950bp}.

\begin{figure}
\center
\begin{tabular}{cc}
\begin{tabular}{ccc}
\\[-67 mm]
\rotatebox{180}{\reflectbox{\includegraphics[scale=0.06]{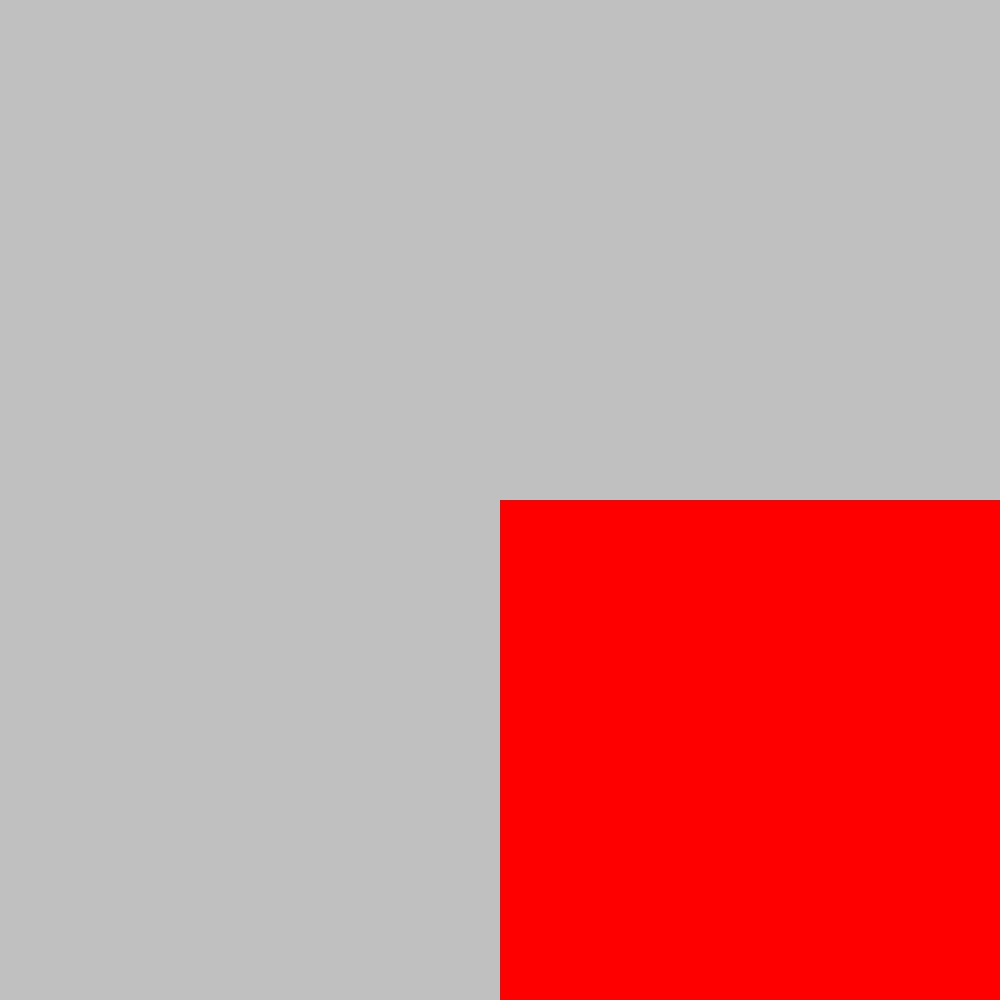}}}
&
\rotatebox{180}{\reflectbox{\includegraphics[scale=0.06]{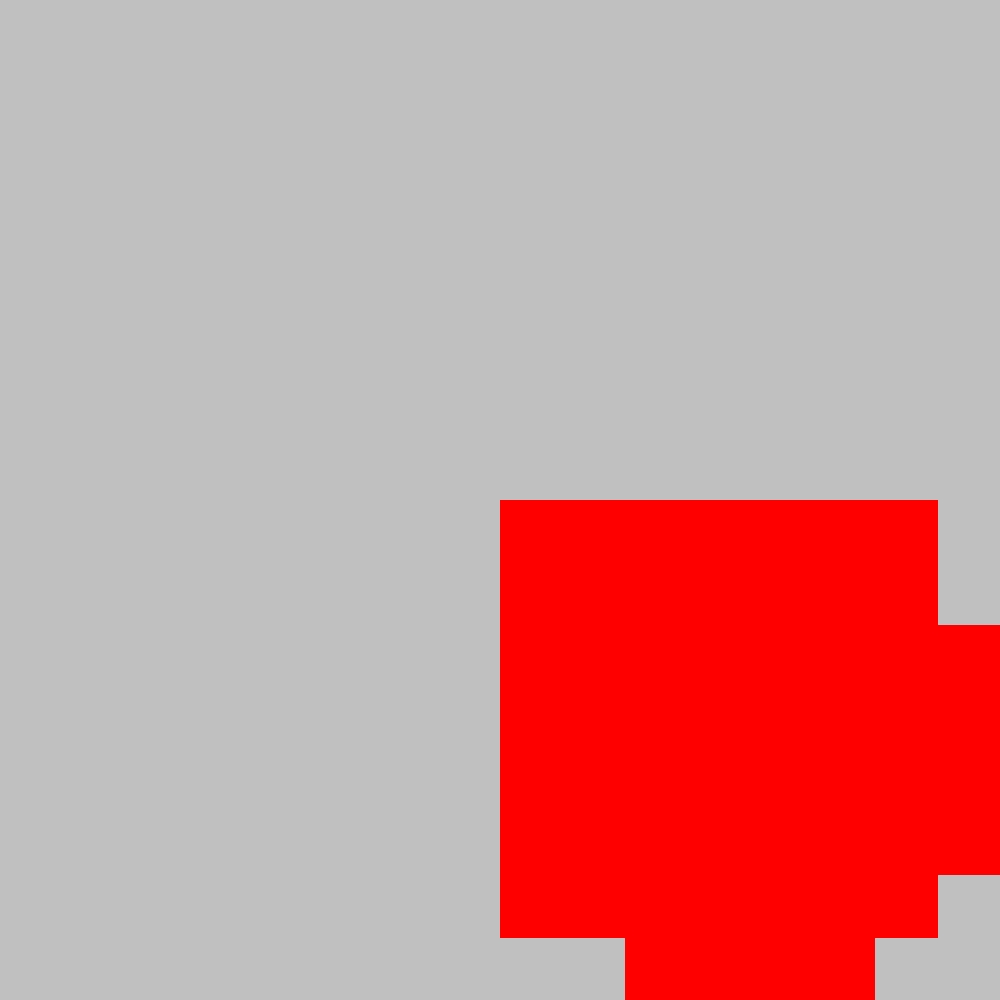}}}
&
\rotatebox{180}{\reflectbox{\includegraphics[scale=0.06]{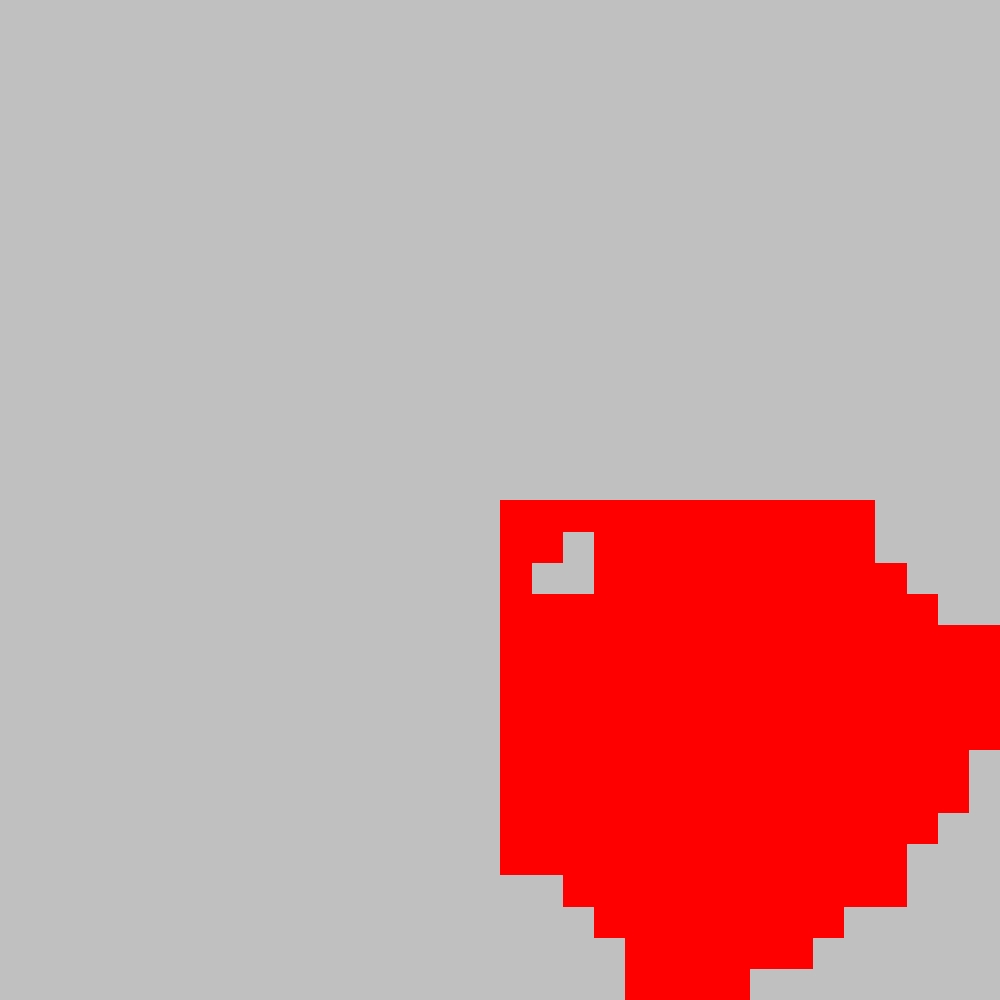}}}\\
2 & 5 & 10\\
\rotatebox{180}{\reflectbox{\includegraphics[scale=0.06]{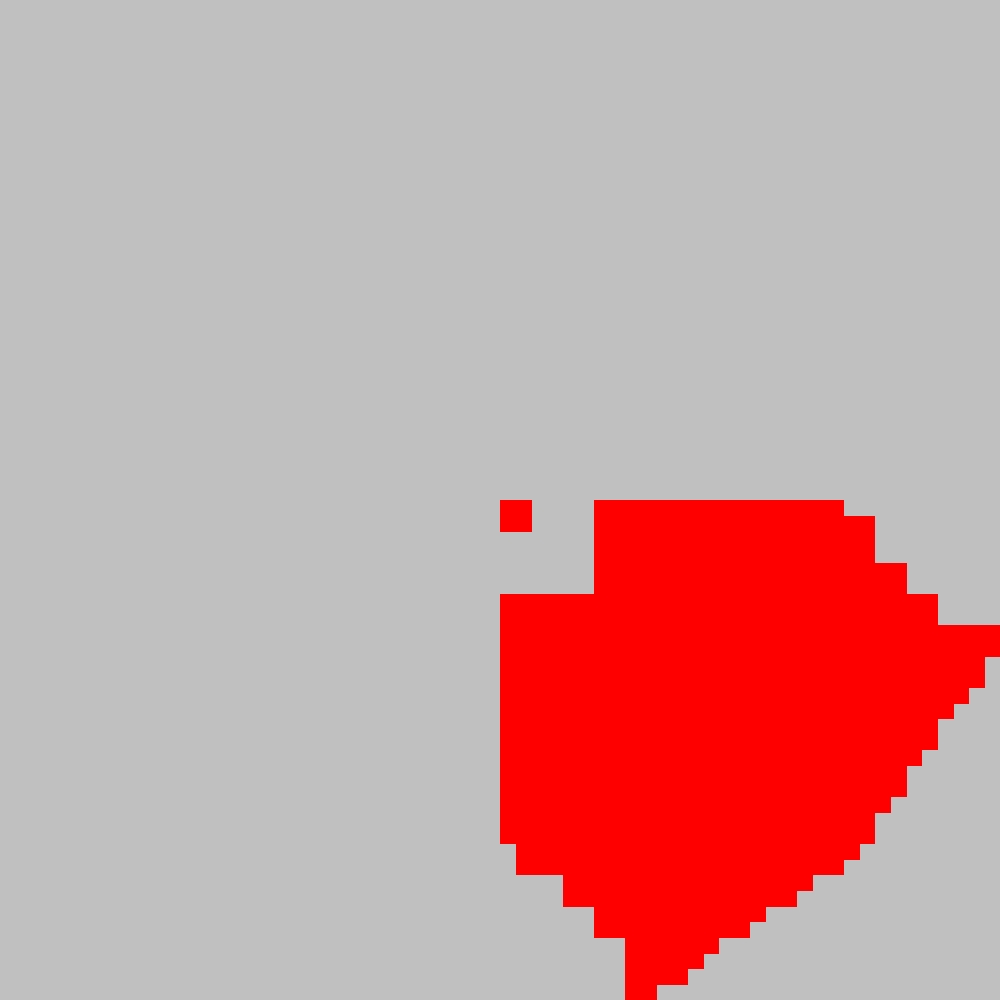}}}
&
\rotatebox{180}{\reflectbox{\includegraphics[scale=0.06]{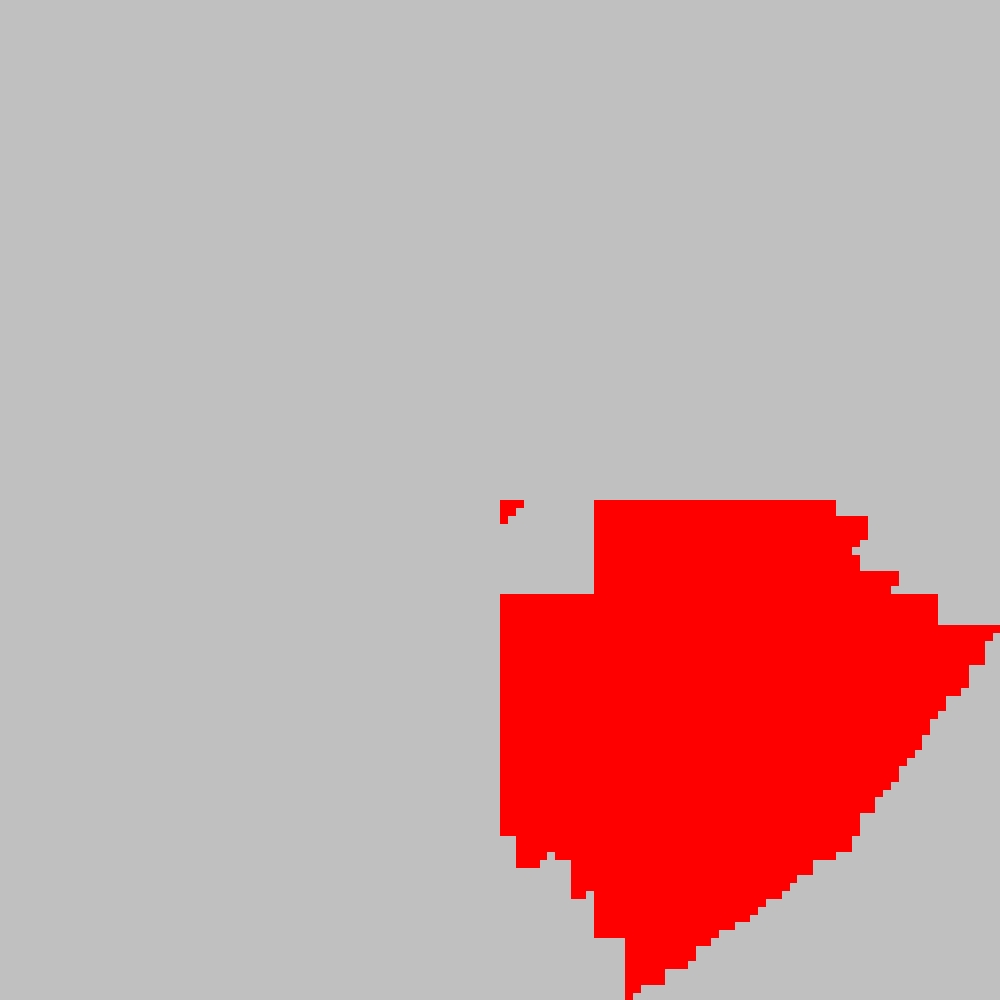}}}
&
\rotatebox{180}{\reflectbox{\includegraphics[scale=0.06]{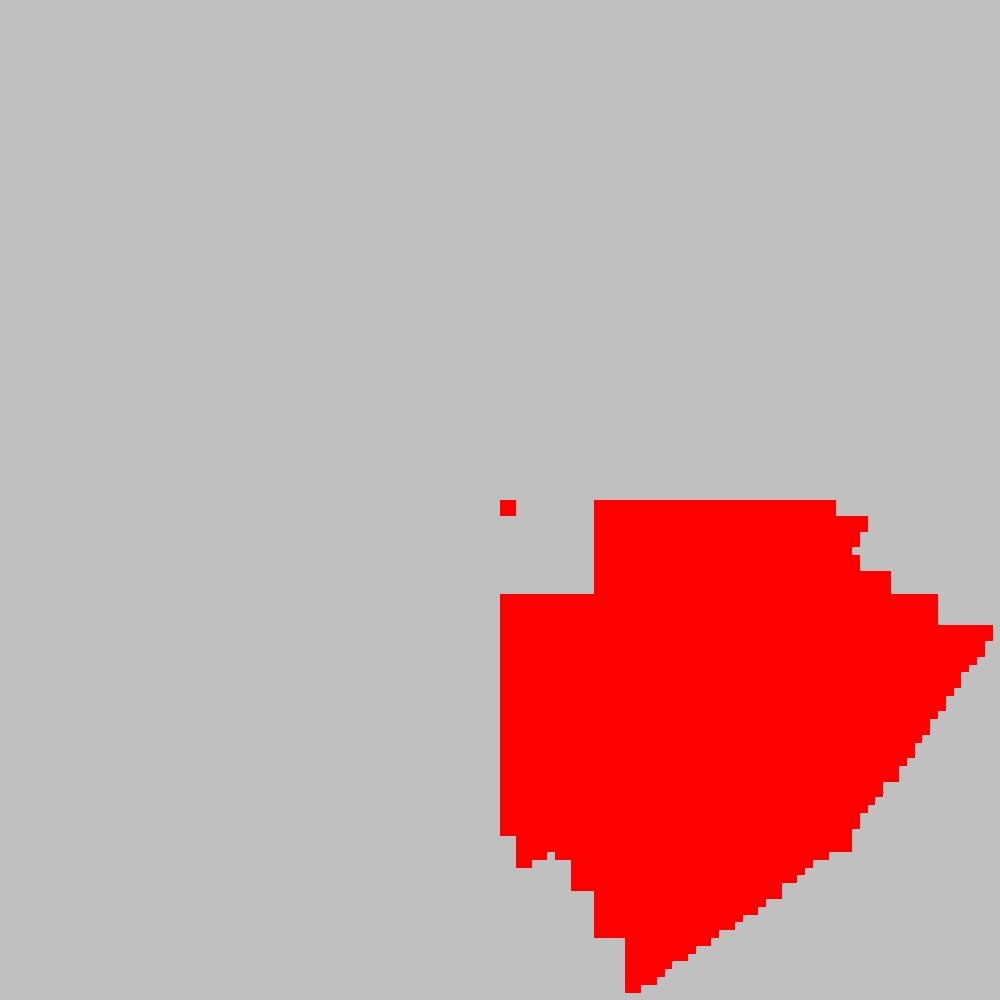}}}\\
15 & 20 & 25
\end{tabular}
&
\includegraphics[scale=0.83]{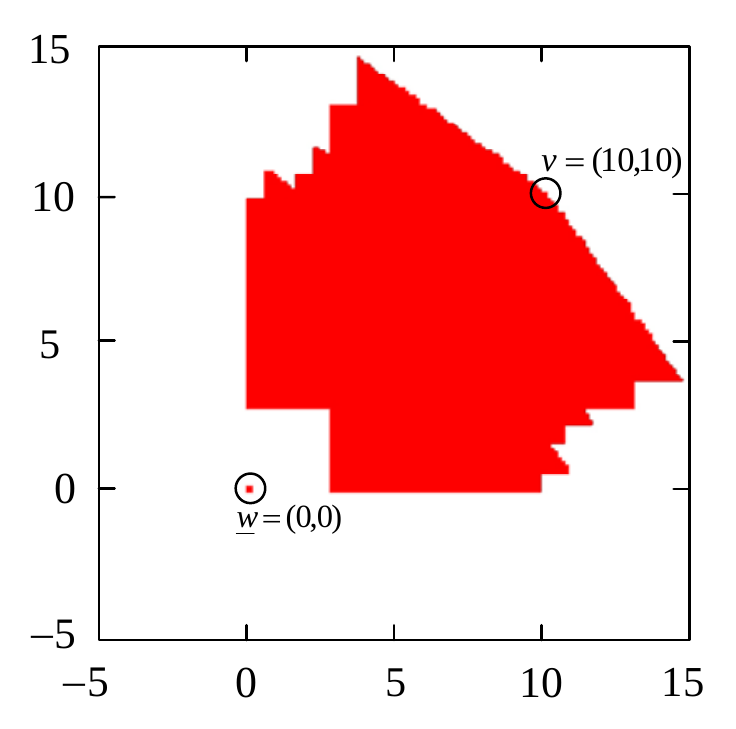}\\
(a) & (b)
\end{tabular}
  \caption{SPE payoff profiles in repeated Duopoly game computed by Algorithm~\ref{alg:basic} for pure strategies with $\gamma=0.6$ and $\epsilon=0.01$. (a)~The evolution of the set of SPE payoff profiles through different algorithm's iterations. (b) Abreu's optimal penal code solution is contained within the set of SPE payoff profiles.}
  \label{fig:Results_Abreu_SPE}
\end{figure}

It was particularly interesting for us to see whether, when applied
to the repeated Duopoly game (Figure~\ref{fig:matrix_games}a),
Algorithm~\ref{alg:basic} for pure strategies preserves the point
$(10,10)$ in the set of SPE payoff profiles. \citet{abreu1988theory}
showed that this point can only make part of the set of SPE payoff
profiles, if $\gamma>4/7$. In our experiments, we observed that for
$\epsilon=0.01$, the point $(10,10)$ indeed remains in the set of
SPE payoff profiles, when $4/7<\gamma<1$. Moreover, the payoff
profile $(0,0)$ of the \emph{optimal penal code}, which was proposed
by \citet{abreu1988theory} as the profile of punishment strategies,
does also remain there (Figures~\ref{fig:Results_Abreu_SPE}a and b).
Algorithm~\ref{alg:basic} also returns an automaton that induces a
strategy profile that generates the payoff profile $(10,10)$.
Interestingly, this automaton induces a strategy profile, which is
equivalent to the optimal penal code based strategy profile proposed
by \citet{abreu1988theory}. To the best of our knowledge, this the
first time that optimal penal code based strategies, which so far
were only proven to exist (in the general case), were
algorithmically computed.

\begin{figure}
\center
\begin{tabular}{ccccc}
\rotatebox{180}{\reflectbox{\includegraphics[scale=0.07]{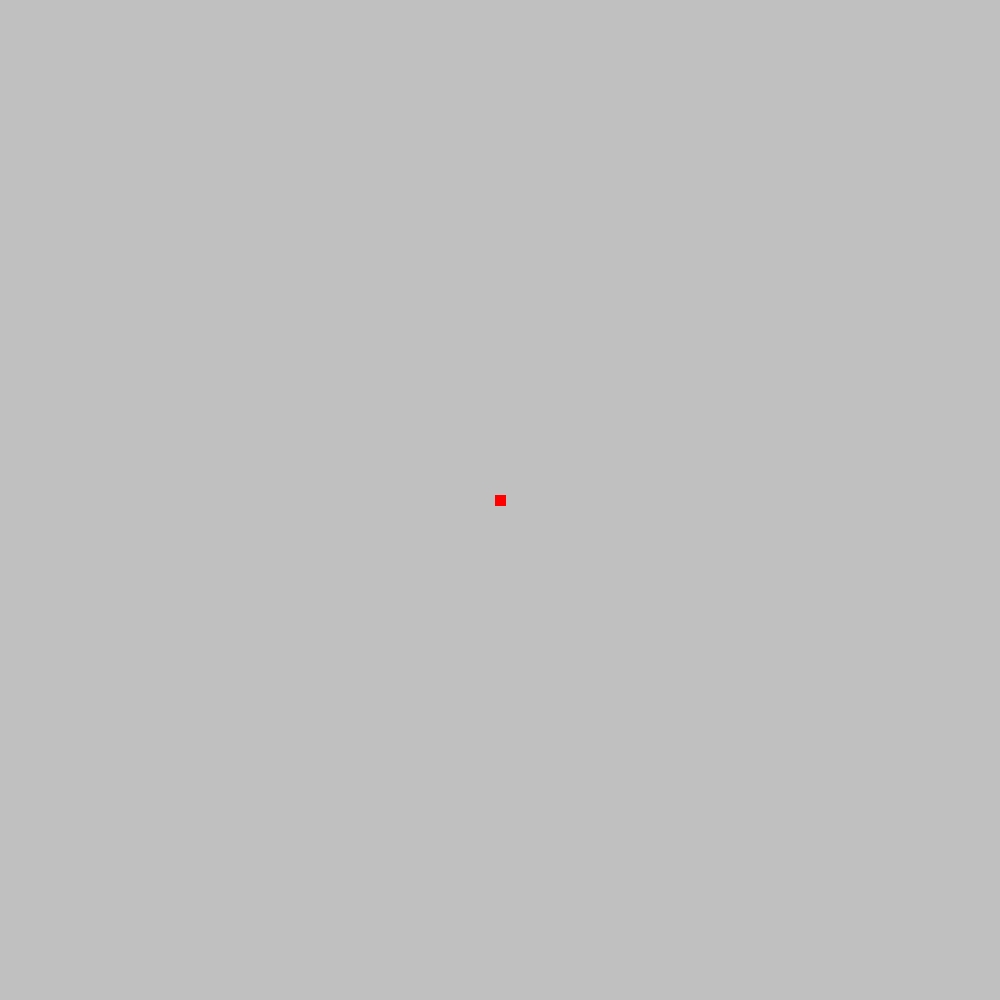}}}
&
\rotatebox{180}{\reflectbox{\includegraphics[scale=0.07]{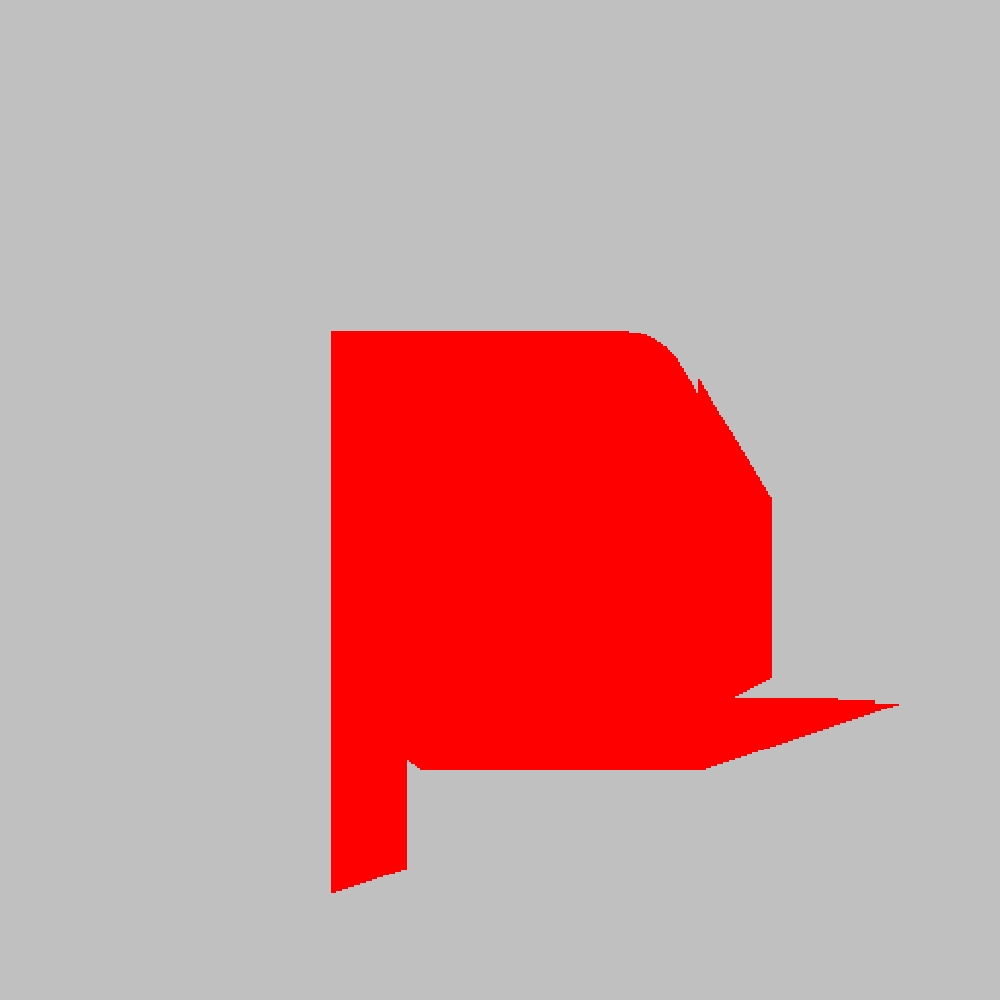}}}
&
\rotatebox{180}{\reflectbox{\includegraphics[scale=0.07]{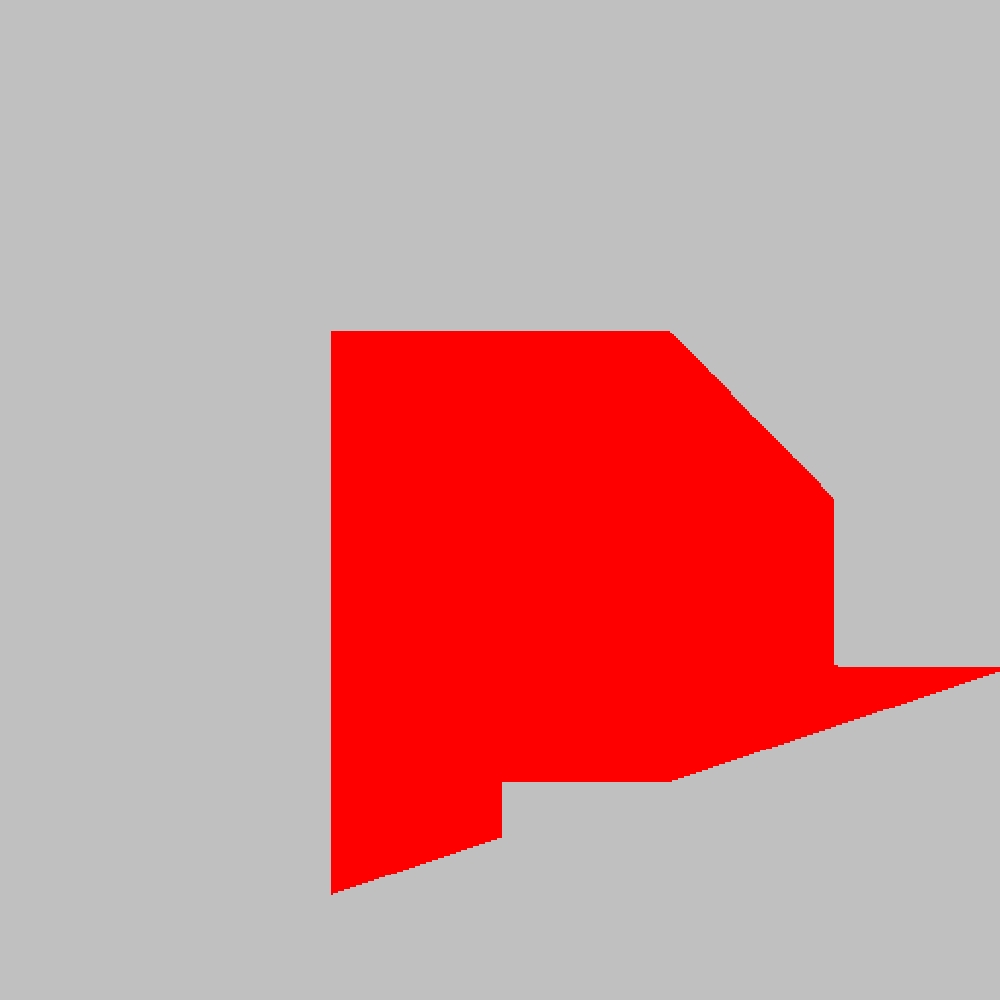}}}
&
\rotatebox{180}{\reflectbox{\includegraphics[scale=0.07]{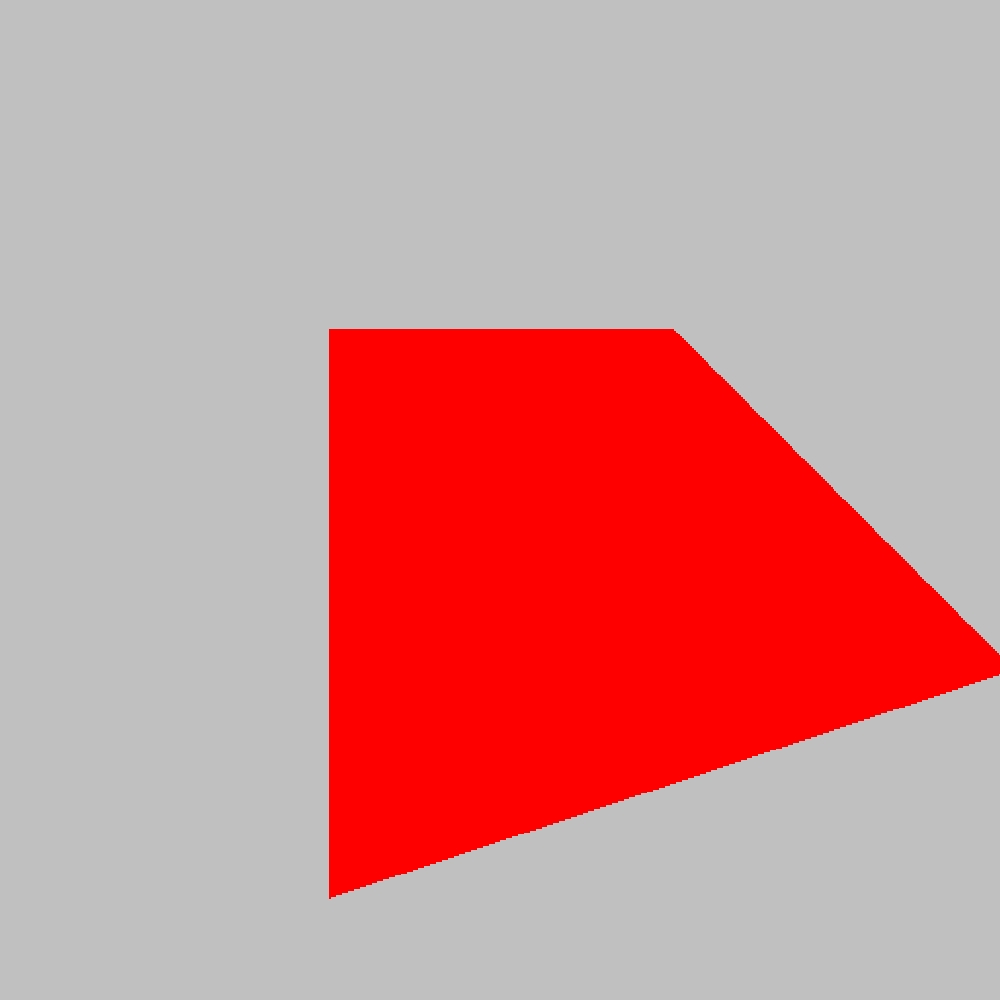}}}
&
\rotatebox{180}{\reflectbox{\includegraphics[scale=0.07]{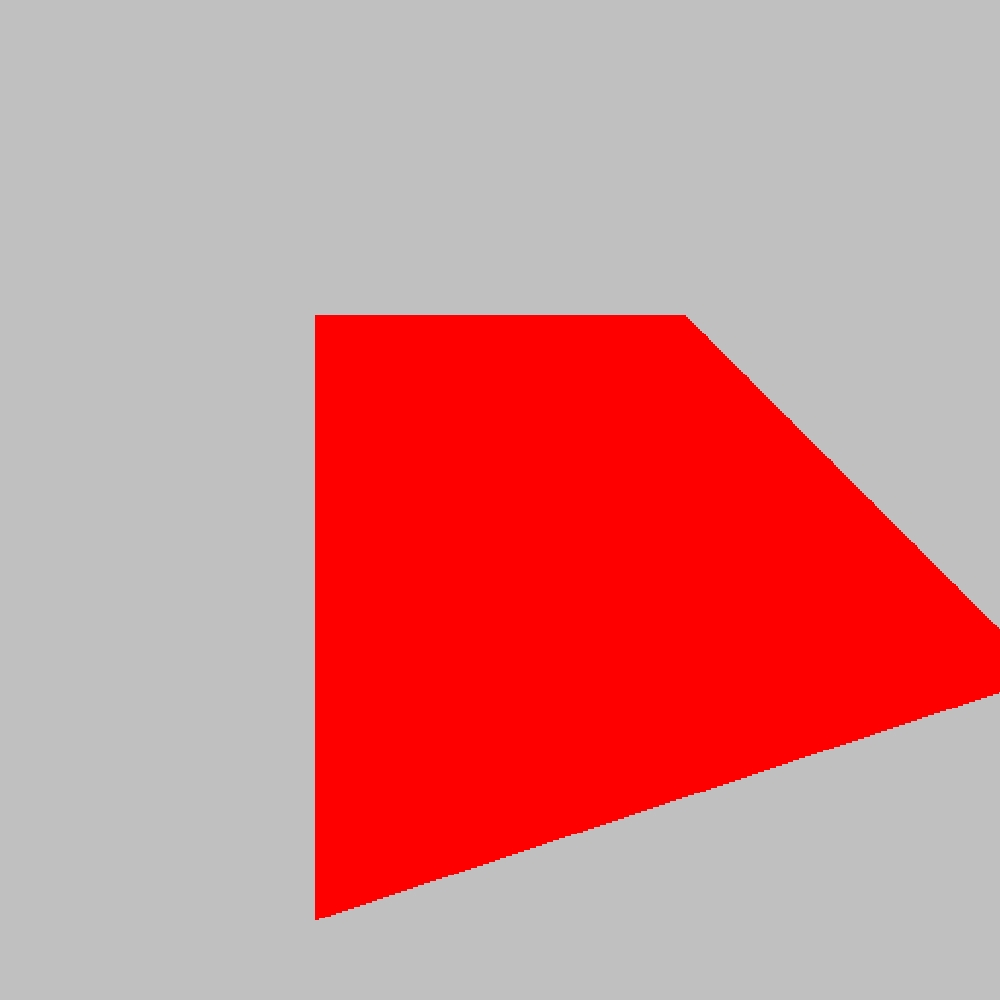}}}\\
  $\gamma \in [0.01,0.4]$ & $\gamma = 0.45$ & $\gamma = 0.5$ & $\gamma = 0.7$ & $\gamma = 0.9$
\end{tabular}
  \caption{The sets of SPE payoff profiles computed in the repeated game from Figure~\ref{fig:matrix_games}d with $\epsilon=0.01$ for different values of the discount factor.}
  \label{fig:Results_NoPure_SPE}
\end{figure}

Another experiment was conducted with the game that does not possess
any pure action stationary equilibrium
(Figure~\ref{fig:matrix_games}d). In such games, for lower discount
factors, the algorithm of~\citet{judd2003computing}, that can only
compute pure action strategy and payoff profiles, is incapable of
returning any SPE point. On the other hand,
Algorithm~\ref{alg:basic} does return a non-empty SPE set for the
whole range of values of the discount factor
(Figure~\ref{fig:Results_NoPure_SPE}).

Finally, the numbers in Table~\ref{tab:performance} demonstrate how
different values of the approximation factor $\epsilon$ impact the
performance of Algorithm~\ref{alg:basic} (with clusters) in terms of
(\emph{i})~number of iterations until convergence and
(\emph{ii})~time spent by the algorithm to compute a solution. The
game used in this experiment is the repeated Battle of the Sexes
from Figure~\ref{fig:matrix_games}c.
\begin{table}
  \centering
  \renewcommand{\arraystretch}{1.5}
  \renewcommand{\tabcolsep}{3.5mm}
  \begin{tabular}{cccc}
    \hline\hline
    $\epsilon$ & $l$ & Iterations & Time\\
    \hline
    0.025&0.008&55&1750\\
    0.050&0.016&41&770\\
    0.100&0.031&28&165\\
    0.200&0.063&19&55\\
    0.300&0.125&10&19\\
    0.500&0.250&5&15\\
    \hline
  \end{tabular}
  \caption{The performance of Algorithm~\ref{alg:basic} in the
  repeated Battle of the Sexes for different values of the approximation factor~$\epsilon$.
  The second column represents the hypercube side length~$l$ at the end of the algorithm's execution; the third column contains
  the number of iterations until convergence; the last column
  contains the overall execution time in seconds.}\label{tab:performance}
\end{table}

\section{Discussion}\label{sec:discussion}
We have presented an approach for approximately computing the set of
subgame-perfect equilibrium (SPE) payoff profiles in repeated games
and for deriving strategies implementable as finite automata and
capable of approximately inducing those payoff profiles. To our
knowledge, this is the first time that both these goals are achieved
simultaneously.

Furthermore, for the setting where no coordination during the
game-play is possible, our algorithm returns the richest set of SPE
payoff profiles among all existing algorithms for repeated games.
More precisely, it returns a set that contains all stationary Nash
equilibrium payoff profiles, all non-stationary pure SPE payoff
profiles, and a subset of non-stationary mixed SPE payoff profiles.
In a case where a certain level of coordination can be assumed, such
as the availability of a public correlating device, our algorithm
returns a set containing all SPE payoff profiles, while satisfying
the necessary approximation properties.

In this paper, we adopted a usual assumption that the discount
factor,~$\gamma$, is the same for all players. However, our
algorithms can readily be modified to incorporate player specific
discount factors. Furthermore, for simplicity of presentation, we
assumed that the hypercube side length,~$l$, is the same for all
players. This is also not a strict requirement; it is
straightforward to generalize all algorithms and theoretical results
to the case of player specific hypercube side lengths.

One formulation of the procedure \textsc{CubeSupported} assumes the
presence of a source of a commonly observed random signal (public
correlating device). A natural question would be why not aiming, in
that case, at computing a richer set of \emph{subgame perfect
correlated equilibrium} (SPCE) payoff
profiles~\citep{aumann1987correlated}. Indeed, several algorithms
for computing the set of SPCE payoff profiles and the strategies to
achieve them have recently been
proposed~\citep{murray2007finding,dermed2009solving}. Our algorithm
can also be transformed into one for approximating the set of SPCE
payoff profiles. Indeed, in that case, the mathematical programming
problem for the \textsc{CubeSupported} procedure will be even
simpler than that for SPE. This is due to the fact that for
computing a correlated equilibrium, one has to find a \emph{unique}
probability distribution for players' action profiles, and not a
profile of probability distributions whose product enforces
equilibrium.

However, in order to implement correlated equilibria in practice,
one has to have a reliable third-party mediator that can send
\emph{private signals} to the players before \emph{every} repeated
game stage. Furthermore, at every period, the signals coming to the
players have to be thrown from a specific distribution, different at
different repeated game stages. In the presence of communication,
the mediator can be replaced by a special communication
protocol~\citep{dodis2000cryptographic}. Nevertheless, each stage of
the repeated game has to be preceded by a round of communication in
order to simulate the mediator.

On the other hand, SPE equilibria computed using
Algorithm~\ref{alg:basic} with the \textsc{CubeSupported} procedure
given by Algorithms~\ref{alg:CubeSupportedPure}
or~\ref{alg:CubeSupportedMixedClusters} neither require a mediating
party nor a communication. Furthermore, the assumption of public
correlation, adopted in order to implement
Algorithm~\ref{alg:CubeSupportedMixedConvex}, only requires the
presence of a source of a (constant) uniformly distributed signal
that has to be observed by all players only at certain repeated game
periods. This is a significantly less restrictive assumption than
the one that has to be satisfied for implementing correlated
equilibria in practice.

Algorithm~\ref{alg:basic} with the \textsc{CubeSupported} procedure
given by Algorithm~\ref{alg:CubeSupportedPure} can be
straightforwardly extended to stochastic games while preserving the
linearity of the mathematical programming problem of the
\textsc{CubeSupported} procedure. In more general cases, however,
the existence of multiple states in the environment is a source of
non-linearity. The latter property, together with the presence of
integer variables, require special techniques to solve the problem;
this constitutes subject for future research.




\vskip 0.2in
\bibliographystyle{apalike}
\bibliography{burkov}

\end{document}